\documentclass[preprint,11pt]{elsarticle}
\usepackage{graphicx}

\usepackage{amssymb}
\usepackage{amsmath}
\usepackage{textcomp}  

\usepackage{lineno}
\usepackage[margin=2cm]{geometry}
\usepackage{doi}
\usepackage{hyperref}
\hypersetup{
    colorlinks,
    linkcolor={red!50!black},
    citecolor={blue!50!black},
    urlcolor={blue!80!black}
}

\usepackage[table]{xcolor}
\usepackage{enumerate}
\usepackage[scriptsize,tight,figbotcap]{subfigure}

\biboptions{compress}

\journal{Journal of Computational Physics}

\begin{document}

\begin{frontmatter}



\title{Numerical experiments on the efficiency of local grid refinement based
on truncation error estimates}


\author[uowm]{Alexandros Syrakos\corref{cor}}
\ead{asirakos@uowm.gr, alexandros.syrakos@gmail.com}
\author[uowm]{Georgios Efthimiou}
\author[uowm]{John G. Bartzis}
\author[auth]{Apostolos Goulas}

\cortext[cor]{Corresponding author. Tel.: +30 24610 56711; Fax.: +30 24610
21730}

\address[uowm]{Environmental Technology Laboratory, Department of Mechanical
Engineering, University of West Macedonia, Bakola \& Sialvera, Kozani 50100,
West Macedonia, Greece}

\address[auth]{Laboratory of Fluid Mechanics and Turbomachinery, Department of
Mechanical Engineering, Aristotle University of Thessaloniki, Thessaloniki
54124, Central Macedonia, Greece}

\begin{abstract}
Local grid refinement aims to optimise the relationship between accuracy of the
results and number of grid nodes. In the context of the finite volume method no
single local refinement criterion has been globally established as optimum for
the selection of the control volumes to subdivide, since it is not easy to
associate the discretisation error with an easily computable quantity in each
control volume. Often the grid refinement criterion is based on an estimate of
the truncation error in each control volume, because the truncation error is a
natural measure of the discrepancy between the algebraic finite-volume equations
and the original differential equations. However, it is not a straightforward
task to associate the truncation error with the optimum grid density because of
the complexity of the relationship between truncation and discretisation errors.
In the present work several criteria based on a truncation error estimate are
tested and compared on a regularised lid-driven cavity case at various Reynolds
numbers. It is shown that criteria where the truncation error is weighted by
the volume of the grid cells perform better than using just the truncation
error as the criterion. Also it is observed that the efficiency of local
refinement increases with the Reynolds number. The truncation error is estimated
by restricting the solution to a coarser grid and applying the coarse grid
discrete operator. The complication that high truncation error develops at grid
level interfaces is also investigated and several treatments are tested.
\end{abstract}

\begin{keyword}
finite volume method \sep local grid refinement \sep adaptivity \sep truncation
error \sep regularised lid driven cavity
\end{keyword}

\end{frontmatter}

This is the accepted version of the article published in: Journal of Computational Physics 231 (2012) 6725--6753, 
\doi{10.1016/j.jcp.2012.06.023}

\textcopyright 2015. This manuscript version is made available under the CC-BY-NC-ND 4.0 license 
\url{http://creativecommons.org/licenses/by-nc-nd/4.0/}


\section{Introduction}
\label{sec:intro}

The finite volume method is a popular method in Computational Fluid Dynamics
(CFD): The domain is divided into a number of Control Volumes (CVs) using a
grid, the differential equations are integrated over each CV, and
the integrals are approximated by algebraic expressions involving the values of
the unknown variables at specific discrete locations (e.g. the CV centres).
This results in a system of algebraic equations whose solution gives
approximate values for the unknowns at these locations. In general, the finer
the grid the smaller the discrepancy between the algebraic and the differential
equations, and the better the results. On the other hand, using a higher grid
density also increases the computational effort required. Therefore, for given
computational resources, it is desirable to have a way of determining the
optimal grid density distribution in space which results in the maximum possible
accuracy.

Some methods move the existing nodes of the grid to new locations in order to
improve the accuracy of the results without altering the number of nodes of the
grid (an interesting example is \cite{Yamaleev_01}, based on truncation error).
However, the most popular methods are local refinement methods which are more
easily applicable to more general grids and complex geometries and which locally
increase the grid density by subdividing selected CVs into smaller CVs. Various
grid refinement criteria have been proposed which try to identify the CVs whose
subdivision would be most beneficial in terms of increase of accuracy, but there
is no global agreement among researchers as to which criterion is optimal.

In the context of the finite volume method, refining CVs where the gradient of a
selected flow variable is large appears to have been popular in the past
(e.g. \cite{Vilsmeier_93, Chen_97}). This simple choice is useful for capturing
flow features such as fronts or shock waves, but otherwise it is not
appropriate; second- or higher-order discretisation schemes should be able to
capture these gradients no matter how high without special grid refinement, as
long as the higher-order derivatives of the flow variables are small. Also, such
a method could potentially not converge to the exact solution of the partial
differential equation as convergence may require also the refinement of CVs where
the gradients are small but the higher-order derivatives are large.

Another popular class includes local refinement criteria which are based on the
truncation error (e.g. \cite{Brandt, Berger_84, Berger_89, Thompson_89,
Trottenberg, Aftosmis02, Brown_05, Syrakos06b, Lee_07}). This has a more solid
theoretical background since the truncation error acts as the source of the
discretisation error (see Section \ref{sec:tau}). An interesting alternative to
the use of the truncation error is the use of the \textit{residual} (e.g.
\cite{Muzaferija_97, Trottenberg, Jasak_01, Jasak_03, Hay_06, Hay_07, Ganesh_09, Roy_09}):
While the truncation error is obtained by applying the discrete finite volume
operator to the exact solution, the residual is obtained by applying the
exact differential operator to the approximate finite volume solution. These two
quantities have similar properties. Yet another method is to base the local
refinement criterion on an estimate or indicator of the discretisation error
itself (e.g. \cite{Pretorius_06, Shen_11}), but as discussed in (\cite{Hay_06,
Ganesh_09}) this is not efficient since the discretisation error is convected
and diffused in the domain and may be high in regions which do not really need a
higher grid density. Grid refinement indirectly reduces the discretisation error
by reducing the \textit{source} of this error (which can be viewed to be either
the truncation error or the residual) - and so it is more appropriate to base
the refinement criterion on the source.

The present study attempts a detailed evaluation of the efficiency of truncation
error-based local refinement for the Navier-Stokes equations, by testing and
comparing various truncation error-based criteria on a lid-driven
cavity problem. Although the aforementioned studies demonstrate that truncation
error-guided local refinement can be beneficial, such a detailed study is
missing, to the authors' knowledge. A possible explanation for this is that it
is difficult to quantify the efficiency (the ratio of the discretisation error
reduction to the number of control volumes added to the grid), as the Navier -
Stokes equations do not have analytic solutions except for very simple cases,
and so it is difficult to calculate the discretisation error accurately. In the
present work a variant of the lid-driven cavity test case is selected (by far
the most popular test case for incompressible flows), and the problem is first
solved on very fine structured grids for various Reynolds numbers. Using
Richardson extrapolation ``exact'' solutions are obtained, and the various local
refinement techniques are evaluated by comparing the discretisation error against
the number of CVs of the locally refined grids in each case. Sections \ref{sec:tau} and
\ref{sec:tau_estimate} present some theory concerning the truncation error, its
use as a refinement criterion, and a method to estimate it \textit{a posteriori}.
Section \ref{sec:eqns_and_discr} describes the specific finite volume method used
to solve the incompressible Navier - Stokes equations. Finally, Sections
\ref{sec:experiments_setup} and \ref{sec:results} present the lid-driven cavity
test case and the results.

\section{The truncation error as a grid refinement criterion}
\label{sec:tau}

The truncation error quantifies the discrepancy between the differential
equation integrated over a CV and the approximate algebraic equation which is
obtained for that CV via the finite volume method. This makes it appear to be
the right quantity to use as a refinement criterion, and indeed many researchers
have used it as such (e.g. \cite{Brandt, Berger_84, Berger_89, Thompson_89,
Trottenberg, Aftosmis02, Brown_05, Syrakos06b, Lee_07}), but there are some
complexities. At this point it is appropriate to include a brief description of
the truncation error and of some of its properties that are of interest.

Suppose a finite volume method is used to solve the (possibly non-linear) partial
differential equation (PDE) $N(u)=b$ where $N$ is the differential operator, $u$
is the unknown function, and $b$ is a known function independent of $u$. This
equation is integrated over each CV and the result is approximated by an algebraic
equation. For the whole grid one obtains a system of algebraic equations:

\begin{equation} \label{eq:tau}
 \left[ \frac{1}{\delta\Omega_P} \int_{\delta\Omega_P} N(u) d\Omega
 \right]_{P=1}^K =\; N_h(u_h) \;+\; \tau_h
\end{equation}
Both sides of (\ref{eq:tau}) are vectors with $K$ components, where $K$ is the
number of CVs in the grid. On the left-hand side, each component $P$ is the
integral of the image $N(u)$ over CV $P$ divided by the CV volume
$\delta\Omega_P$. On the right-hand side the term $N_h(u_h)$ is a vector which
is obtained by applying the algebraic operator $N_h$ to the vector $u_h$ of the
values of the exact solution function $u$ at the CV centres. The symbol $h$ is
sometimes used to denote a particular grid, and sometimes used to denote the
spacing between successive grid lines of that grid - this will be clear from the
context. $N_h$ is constructed by discretising the continuous differential operator
$N$ according to the finite volume methodology using selected discretisation schemes.
No matter how accurate these discretisation schemes are, it is in general not possible
for the term $N_h(u_h)$ to equal the left hand side, but they will differ by the
truncation error $\tau_h$. The truncation error consists of all the terms that
were truncated from the Taylor series used in converting the differential
equations to algebraic form (see e.g. \cite{Peric}).

The PDE $N(u)=b$ implies that both sides of (\ref{eq:tau}) should equal the
vector $b_h$ of the integrals of $b$ over each CV of the grid $h$ divided by
the corresponding volume. As the truncation error is unknown, the finite volume
method drops it under the assumption that it is small enough such that its
omission does not have a significant impact on the results. Therefore, the
system $N_h(\tilde{u}_h)=b_h$ is solved instead of $N_h(u_h)+\tau_h=b_h$, where
$\tilde{u}_h$ is an approximation to the exact solution $u_h$, and their
difference is called the discretisation error $\epsilon_h = u_h - \tilde{u}_h$.

The truncation error acts as the source of the discretisation error
(\cite{Peric}), which is distributed in space according to the same laws
described by the differential equation - e.g., convection and diffusion. This
can be seen from the following equation which is straightforward to derive:

\begin{equation} \label{eq:eps_tau}
 N_h(\tilde{u}_h + \epsilon_h) \;-\; N_h(\tilde{u}_h) \;=\; -\tau_h
\end{equation}

In the linear case this simplifies to $N_h\epsilon_h = -\tau_h$ where it is
evident that the discretisation error obeys the same PDE with the truncation
error acting as the source. In the non-linear case for small enough $\epsilon_h$
equation (\ref{eq:eps_tau}) becomes $N'_h\epsilon_h \approx -\tau_h$ where
$N'_h$ is the Jacobian matrix of the operator $N_h$. This makes it a reasonable
choice to base the refinement criterion on the truncation error. Local grid
refinement normally leads to a local truncation error reduction, and according
to the above discussion this should also reduce the discretisation error, provided
that $N_h$ is continuous and $\epsilon_h$ is small enough.

On the other hand there are some complexities. The first is that $\tau_h$ is
unknown; however, there are ways to estimate it, so this is not a major problem
in many cases. Also, the relationship between $\epsilon_h$ and $\tau_h$ is not
simple, as can be seen from (\ref{eq:eps_tau}). If the truncation error in any
two CVs is the same, then this does not necessarily mean that subdivision of
either of them will bring about the same reduction in discretisation error.
Therefore to use the truncation error as a local refinement criterion, its value
in each CV should be multiplied by a weighting factor which should come from
equation (\ref{eq:eps_tau}). One of the simplest options is to divide the
truncation error in each CV by the corresponding main diagonal coefficient of
the $N_h$ operator after it is linearised. This is equivalent to estimating
$\epsilon_h$ by performing a single Jacobi iteration on (\ref{eq:eps_tau})
(after linearising $N_h$) with zero initial guess, and then using this estimate
of $\epsilon_h$ as the refinement criterion. This sort of weighting has been used
e.g. by \cite{Muzaferija_97, Jasak_01}.

There exists another difficulty in using the truncation error as the refinement
criterion. When the grid is smooth then the rate of convergence of the
truncation and discretisation errors is the same - e.g. $O(h^2)$ for second
order methods. But when the grid is not smooth, e.g. it exhibits significant
skewness or other geometric irregularities, then it is usual that the
truncation error converges more slowly than the discretisation error. For
example, it may be the case that $\epsilon_h \in O(h^2)$ while $\tau_h \in O(h)$
or even $\tau_h \in O(1)$ meaning that the truncation error does not reduce with
grid refinement. This has been observed by several researchers, e.g.
\cite{Manteuffel86, Jones2000, Syrakos06a, Syrakos06b, diskin_nia_07_08,
Thomas_08, Eriksson09}. Unfortunately, such large truncation errors appear not
only on irregular grids but also on smooth grids under the usual discretisation
schemes near the domain boundaries \cite{Syrakos06a} and near the interfaces
between different levels on locally refined grids \cite{Syrakos06b}. In all
these cases, when second order accurate discretisation schemes are used,
numerical experiments show that the high truncation errors of order $O(1)$ do
not cause degradation of the convergence rate of the discretisation error which
remains $O(h^2)$. An interesting possible explanation for this based on
arguments from potential theory is sketched in \cite{johansen98}. More rigorous
mathematical arguments can be found for the domain boundaries e.g. in
\cite{Matsunaga2000} and \cite{Svard_06}. This phenomenon requires some
investigation in relation to the use of the truncation error as a local
refinement criterion, because in the regions of high truncation error it seems
likely that there will result perpetual refinement without actual benefit in
terms of reduction of the discretisation error.

\section{Truncation error estimate} \label{sec:tau_estimate}

A popular method for estimating the truncation error works by transferring
(restricting) the solution obtained on a given grid $h$ to a coarser but
geometrically similar grid $H$, and applying the discrete operator of the
coarse grid (constructed using the exact same discretisation schemes as the
operator of the fine grid $h$) to this restricted solution \cite{Brandt,
Bernert97, Trottenberg, Fulton03, Brown_05, Syrakos06a}. This concept arises
naturally in FAS multigrid methods (Full Approximation Storage schemes - see
any of \cite{Brandt, Briggs, Trottenberg}), from which the estimate originates.
The method is valid under the assumption that the truncation and discretisation
errors vary smoothly within the domain, and there exist continuous functions
$\tau$ and $\epsilon$ such that:

\begin{align}
 \tau_h \;&=\; I_0^h \left[ h^p \cdot \tau \;+\; O(h^q) \right]
\label{eq:tau_leading_term}
\\
 \epsilon_h \;&=\; I_0^h \left[ h^p \cdot \epsilon \;+\; O(h^q) \right]
\label{eq:epsilon_leading_term}
\end{align}
where $p$ is the order of the finite volume method and $q>p$, with both $p$ and
$q$ being integers. $I_0^h$ is an operator that discretises a continuous
function by keeping only the values of the function at the centres of the CVs of
grid $h$. 

Suppose as discussed in Section \ref{sec:tau} that to solve the differential
equation $N(u)=b$ the problem is discretised on a grid $h$ using the finite
volume method to obtain the algebraic system $N_h (\tilde{u}_h) = b_h$ by
neglecting the truncation error, which is solved to obtain an approximate
solution $\tilde{u}_h$. Suppose also that there is available a grid $2h$ which
is geometrically similar to grid $h$ but which has everywhere twice the grid
spacing. Consider the following systems on grid $2h$, whose solutions are
increasingly more accurate:

\begin{align}
  N_{2h}(\tilde{u}_{2h}) \;&=\; b_{2h}
\label{eq:problem_2h}
\\
  N_{2h}(\tilde{u}_{2h}^h) \;&=\; b_{2h} \;-\; \tau_{2h}^h 
                          \;\;=\; N_{2h}(\tilde{u}_{2h}) \;-\; \tau_{2h}^h
\label{eq:problem_2h_w_t_rel}
\\
  N_{2h}(u_{2h}) \;&=\; b_{2h} \;-\; \tau_{2h}
                \;\;=\; N_{2h}(\tilde{u}_{2h}) \;-\; \tau_{2h}
\label{eq:problem_2h_w_t}
\end{align}
where
\begin{equation} \label{eq:tau_2h_h}
  \tau_{2h}^h \;=\; b_{2h} \;-\; N_{2h} \left( I_h^{2h}\tilde{u}_h \right)
\end{equation}
is called the \textit{relative truncation error} on grid $2h$ with respect to
grid $h$. In (\ref{eq:tau_2h_h}) $\tilde{u}_h$ is the approximate solution
obtained on grid $h$ by the finite volume method, and $I_h^{2h}$ is a transfer
(restriction) operator which transfers the function $\tilde{u}_h$ to grid $2h$.
For the time being it will be assumed that this transfer does not introduce any
errors, in the sense that, for example, $I_h^{2h}u_h = u_{2h}$ ($u_h = I_0^h u$
and $u_{2h}=I_0^{2h}u$ being the values of the exact solution $u$ of the
differential equation at the centres of the CVs of grids $h$ and $2h$
respectively). This is a natural assumption if the CV centres of grid $2h$ are a
subset of the CV centres of grid $h$, as in node-centred grids, and direct
injection is used (i.e. the values of $u_{2h}$ are simply copies of the corresponding
values of $u_h$ at the coincident points). But in cell-centred grids like
those used in the present study this assumption is not valid, and the implications
will be discussed later on. The finite volume operator $N_{2h}$ is constructed
on grid $2h$ using the exact same discretisation schemes as $N_h$ on grid $h$.

The following can be noted about the solutions of systems (\ref{eq:problem_2h})
- (\ref{eq:problem_2h_w_t}): $\tilde{u}_{2h}$ is the approximate solution on
grid $2h$ obtained by solving (\ref{eq:problem_2h}). By substituting
(\ref{eq:tau_2h_h}) into (\ref{eq:problem_2h_w_t_rel}) one can see that the
solution of (\ref{eq:problem_2h_w_t_rel}) is $\,\tilde{u}_{2h}^h = I_h^{2h}
\tilde{u}_h$, i.e. the solution of grid $h$ restricted to grid $2h$. So, by
subtracting $\tau_{2h}^h$ from the right hand side of (\ref{eq:problem_2h}) one
gains the accuracy of grid $h$ on grid $2h$. Finally, by subtracting the actual
truncation error $\tau_{2h}$ from the right hand side of (\ref{eq:problem_2h})
one obtains the exact solution of the PDE ($u_{2h}$) by solving
(\ref{eq:problem_2h_w_t}).

Let $N'_{2h}$ be the Jacobian matrix of $N_{2h}$ evaluated at $\tilde{u}_{2h}$
($N'_{2h}\equiv N_{2h}$ if $N_{2h}$ is linear), and assume that the functions
$\tilde{u}_{2h}$, $I_h^{2h}\tilde{u}_h$ and $u_{2h}$ are close enough such that
the following hold:

\begin{align}
  N_{2h}\left( I_h^{2h} \tilde{u}_h \right) \;&=\; N_{2h}(\tilde{u}_{2h})
\;+\; N'_{2h}\cdot \left( I_h^{2h} \tilde{u}_h - \tilde{u}_{2h} \right)
\;+\; O\left( \delta_1^2 \right)
\label{eq:jacobian_u_h}
\\
  N_{2h}(u_{2h}) \;&=\; N_{2h}(\tilde{u}_{2h}) \;+\; N'_{2h}\cdot \left(
u_{2h} - \tilde{u}_{2h} \right) \;+\; O\left(\delta_2^2 \right)
\label{eq:jacobian_u}
\end{align}
where $\delta_1=I_h^{2h} \tilde{u}_h-\tilde{u}_{2h}$ and $\delta_2=u_{2h} -
\tilde{u}_{2h}$. Now, $\delta_1$ and $\delta_2$ are of the order of the
discretisation error (see also equation (\ref{eq:delta_1}) below) and so
$O(\delta_1^2),O(\delta_2^2) \in O(h^{2p})$ due to
(\ref{eq:epsilon_leading_term}). By rearranging (\ref{eq:jacobian_u_h}) -
(\ref{eq:jacobian_u}) and comparing to (\ref{eq:problem_2h_w_t_rel}) -
(\ref{eq:problem_2h_w_t}) respectively one obtains that:

\begin{align}
  N'_{2h}\cdot \left( I_h^{2h} \tilde{u}_h - \tilde{u}_{2h} \right)
\;&=\; -\tau_{2h}^h \;+\; O(h^{2p})
\label{eq:jacobial_u_h_2}
\\
  N'_{2h}\cdot \left( u_{2h} - \tilde{u}_{2h} \right) \;&=\; -\tau_{2h}
\label{eq:jacobial_u_2} \;+\; O(h^{2p})
\end{align}

Now, $u_{2h}-\tilde{u}_{2h}=\epsilon_{2h}$, and also:

\begin{align}
  I_h^{2h} \tilde{u}_h - \tilde{u}_{2h} \;&=\; \left(u_{2h} - \tilde{u}_{2h}
\right) \;-\; \left( u_{2h} - I_h^{2h} \tilde{u}_h \right) \nonumber
  \;=\; \epsilon_{2h} \;-\; I_h^{2h} \epsilon_h 
\\
  &=\; I_0^{2h} \left[ (2h)^p \epsilon + O(h^q) \right] \;-\;
       I_h^{2h} I_0^h \left[ h^p \epsilon + O(h^q) \right] \nonumber
\\
  &=\; I_0^{2h} \left[ 2^p h^p \epsilon - h^p \epsilon + O(h^q) \right] 
  \nonumber
\\
  &=\; \left( 1-\frac{1}{2^p} \right) I_0^{2h} \left[ 2^ph^p\epsilon +
       O(h^q) \right] \nonumber
\\ \label{eq:delta_1}
  &=\; \left( \frac{2^p-1}{2^p} \right) \epsilon_{2h} \;+\; O(h^q)
\end{align}
where (\ref{eq:epsilon_leading_term}) has been used for both $\epsilon_h$ and
$\epsilon_{2h}$, and also it has been assumed that $I_h^{2h}$ has not
introduced any errors ($I_h^{2h} I_0^h = I_0^{2h}$), as stated above. Therefore,
(\ref{eq:jacobial_u_h_2}) and (\ref{eq:jacobial_u_2}) become:

\begin{align}
  \left( \frac{2^p-1}{2^p} \right) &N'_{2h} \cdot \epsilon_{2h} \;+\;
  N'_{2h} \cdot O(h^q) \;=\; -\tau_{2h}^h \;+\; O(h^{2p})
\label{eq:jacobial_tau_2h_h}
\\
  &N'_{2h}\cdot \epsilon_{2h} \;=\; -\tau_{2h} \;+\; O(h^{2p})
\label{eq:jacobial_tau_2h}
\end{align}
Usually it will be the case that $2p\geq q$. Also, it may be assumed that ${N'_{2h}
\cdot O(h^q)} \in O(h^q)$ because neither $N'_{2h}$ nor the vector $O(h^q)$
change arbitrarily as the grid spacing, $h$, decreases, but they approximate
specific continuous counterparts. The eigenvalues of $N'_{2h}$ are therefore
expected to not change much with grid refinement. So, substituting
(\ref{eq:jacobial_tau_2h}) into (\ref{eq:jacobial_tau_2h_h}) and rearranging one
gets:

\begin{equation} \label{eq:tau_2h_estimate}
  \tau_{2h} \;=\; \frac{2^p}{2^p-1}\tau_{2h}^h \;+\; O(h^q)
\end{equation}
Finally, (\ref{eq:tau_leading_term}) implies that:
\begin{equation} \label{eq:tau_h_from_tau_2h}
  \tau_h \;=\; \frac{1}{2^p} I_{2h}^h \tau_{2h} \;+\; O(h^q)
\end{equation}
where $I_{2h}^h$ is a transfer (prolongation) operator which transfers a
function from grid $2h$ to grid $h$. This, combined with
(\ref{eq:tau_2h_estimate}) and the definition of $\tau_{2h}^h$
(\ref{eq:tau_2h_h}) gives the final result:

\begin{align} \label{eq:tau_h_estimate}
  \tau_h \;&=\; \frac{1}{2^p-1} I_{2h}^h \left[ b_{2h} \;-\; N_{2h} \left(
I_h^{2h}\tilde{u}_h \right) \right] \;+\; O(h^q)
\\
  &=\; \tilde{\tau}_h \;+\; O(h^q) \nonumber
\end{align}

The above estimate $\tilde{\tau}_h$ differs from the actual truncation error
$\tau_h$ by $O(h^q)$. Since the truncation error itself has a magnitude of
$O(h^p)$, the relative error in the approximation is $(\tau_h - \tilde{\tau}_h)
/ \tau_h \in O(h^{q-p})$. In most cases it will be the case that $q-p=1$, but
there are some central difference approximations where due to cancellation of
every second term in the corresponding Taylor series it happens that $q-p=2$.
So, it may be stated that the approximation $\tilde{\tau}_h$
(\ref{eq:tau_h_estimate}) is either first- or second-order accurate.

The analysis so far has assumed that the transfer operators $I_h^{2h}$ and
$I_{2h}^h$ are ``perfect'' in the sense that they introduce no errors. In
practice this is not possible on cell-centred grids, and so an attempt will now
be made to examine the effect of the order of accuracy of these operators on the
order of accuracy of the estimate (\ref{eq:tau_h_estimate}). Starting with the
prolongation operator, suppose an $r$-th order accurate operator
$\tilde{I}_{2h}^h$ such that $\tilde{I}_{2h}^h u_{2h} = I_{2h}^h u_{2h} +
O(h^r)$, $I_{2h}^h$ being again the ``perfect'' operator which introduces no
errors. Then $\tilde{I}_{2h}^h \tau_{2h} = I_{2h}^h \tau_{2h} + O(h^{p+r})$
because $\tau_{2h} \in O(h^p)$. So (\ref{eq:tau_h_from_tau_2h}) becomes $\tau_h
= \frac{1}{2^p} \tilde{I}_{2h}^h \tau_{2h} + O(h^{p+r}) + O(h^q)$. This means
that the order of accuracy of the estimate (\ref{eq:tau_h_estimate}) will
degrade only if $p+r<q$, or $r<q-p$. If $q-p=1$ then using a first-order
accurate operator ($r=1$) suffices. In the present study a first-order accurate
prolongation operator is used for this purpose which sets the value at each CV
of grid $h$ equal to the value of its parent CV of grid $2h$. This causes the
estimate (\ref{eq:tau_h_estimate}) to predict the same truncation error for all
sibling CVs, i.e. CVs which share the same parent. In general therefore all
siblings are refined together. But this is already a requirement of the
refinement procedure in order to ensure that the underlying grid $2h$ always
exists (see Section \ref{sec:eqns_and_discr}).

Now, suppose an $r$-th order accurate restriction operator $\tilde{I}_h^{2h}$
such that $\tilde{I}_h^{2h} u_h = I_h^{2h} u_h + O(h^r)$. Then
(\ref{eq:tau_h_estimate}) can be written as:

\begin{align}
  \tau_h \;&=\; \frac{1}{2^p-1} I_{2h}^h \left[ b_{2h} \;-\; N_{2h} \left(
\tilde{I}_h^{2h}\tilde{u}_h + O(h^r) \right) \right] \;+\; O(h^q) \nonumber
\\
  &=\; \frac{1}{2^p-1} I_{2h}^h \left[ b_{2h} \;-\; N_{2h} \left(
  \tilde{I}_h^{2h}\tilde{u}_h \right) \;-\; N'_{2h} \cdot O(h^r)
  \;+\; O(h^{2r}) \right] \;+\; O(h^q) \nonumber
\\
  &=\; \frac{1}{2^p-1} I_{2h}^h \left[ b_{2h} \;-\; N_{2h} \left(
  \tilde{I}_h^{2h}\tilde{u}_h \right) \right] \;+\; O(h^r) \;+\; O(h^q)
\nonumber
\\
  &=\; \tilde{\tau}_h \;+\; O(h^{\min(r,q)})
\end{align}
So now $(\tau_h - \tilde{\tau}_h)/\tau_h \in O(h^{\min(r,q)-p})$ so that in
order for the estimate $\tilde{\tau}_h$ to converge to the actual $\tau_h$
it is necessary that $r>p$. The maximum accuracy is obtained for $r \geq q$.
These theoretical findings are demonstrated in numerical experiments in
\cite{Syrakos06a}. For the present study, the better of the two restriction
operators tested in \cite{Syrakos06a} is used, which is third-order accurate on
smooth grids and second order accurate on irregular grids (including near domain
boundaries and near interfaces between different local refinement levels).

In the case of non-smooth grids, the accuracy of the restriction operator is
not the only issue of concern. Indeed, in this case (\ref{eq:tau_leading_term})
does not hold (although (\ref{eq:epsilon_leading_term}) does, in general, hold)
so (\ref{eq:tau_h_estimate}) does not hold because $p$ is wrong or varies
across the domain. Yet even in this case (\ref{eq:tau_h_estimate}) should be
able to roughly predict the order of magnitude of the truncation error, due to
the validity of (\ref{eq:epsilon_leading_term}), which could be sufficient for
local refinement purposes. In this case one can regard (\ref{eq:tau_h_estimate})
as calculating the truncation error on grid $2h$ by applying the coarse grid operator
$N_{2h}$ to the fine-grid solution instead of the exact solution which would give the
exact truncation error. The fine grid solution may be regarded as a good approximation
to the exact solution on grid $2h$ due to (\ref{eq:epsilon_leading_term}).

Finally we note that the estimate (\ref{eq:tau_h_estimate}) can also be used
with any other coarse-to-fine grid ratio, $H=\rho \cdot h$, by substituting
$\rho$ instead of $2$ in the denominator.

\section{Equations and discretisation} \label{sec:eqns_and_discr}

For simplicity, the 2-dimensional steady-state incompressible Navier - Stokes
equations with constant density and viscosity are solved in this work.
Integrated over a CV $P$, the equations solved are:

\textit{$x$-momentum:}
\begin{equation}
 \int_{S_P} \rho u \vec{V} \cdot d\vec{S} \;=\; 
 \int_{S_P} \mu \nabla u \cdot d\vec{S} \;+\;
 \int_{S_P} \mu \left( \frac{\partial u}{\partial x}\vec{i} +\frac{\partial
v}{\partial x}\vec{j} \right) \cdot d\vec{S} \;-\; 
 \int_{S_P} p \vec{i} \cdot d\vec{S} \label{eq:mom_x}
\end{equation}

\textit{$y$-momentum:}
\begin{equation}
 \int_{S_P} \rho v \vec{V} \cdot d\vec{S} \;=\; 
 \int_{S_P} \mu \nabla v \cdot d\vec{S} \;+\;
 \int_{S_P} \mu \left( \frac{\partial u}{\partial y}\vec{i} +\frac{\partial
v}{\partial y}\vec{j} \right) \cdot d\vec{S} \;-\; 
 \int_{S_P} p \vec{j} \cdot d\vec{S} \label{eq:mom_y}
\end{equation}

\textit{continuity:}
\begin{equation}
 \int_{S_P} \rho \vec{V} \cdot d\vec{S} \;=\; 0 \label{eq:continuity}
\end{equation}
where $S_P$ is the boundary surface of CV $P$, $d\vec{S}$ is an infinitesimal
element of this area pointing out of the CV, $\vec{V}=(u,v)$ is the fluid
velocity, $\rho$, $\mu$ and $p$ are the density, viscosity and pressure
respectively, and $\vec{i}$, $\vec{j}$ are the unit vectors in the $x$ and $y$
directions respectively. The second terms on the right-hand sides of the
momentum equations are zero for constant density and viscosity, yet they have
been retained because their discrete counterparts were present in the Fortran
code that was used for this study (in their discrete form these terms are not
zero but they add to the truncation error, tending to zero as the grid is
refined; cf. end of Sec. \ref{sec:boundary treatment}).

\begin{figure}[tpb]
 \centering
 \noindent\makebox[\textwidth]{
 \includegraphics[width=1.0\textwidth]{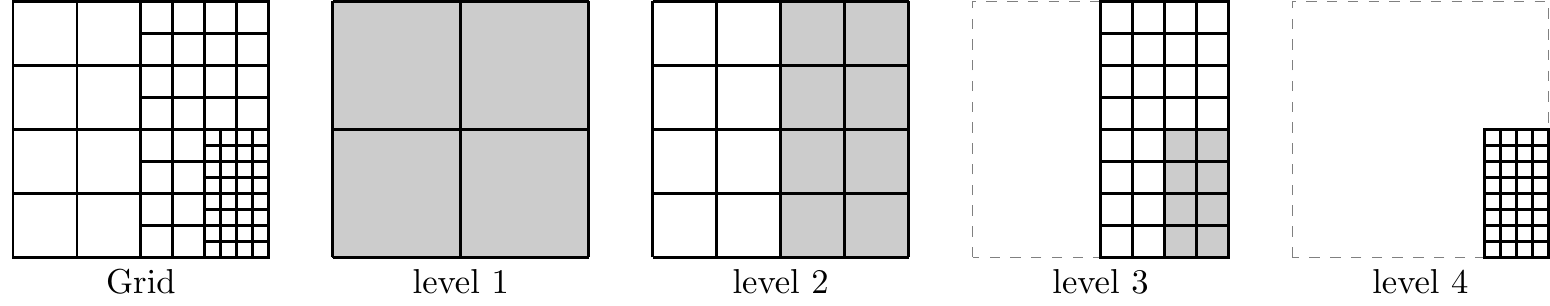}}
 \caption{A composite grid and its decomposition into levels. The local CVs of
each level are marked in grey.}
 \label{fig:grid}
\end{figure}

\begin{figure}[tpb]
 \centering
 \includegraphics[]{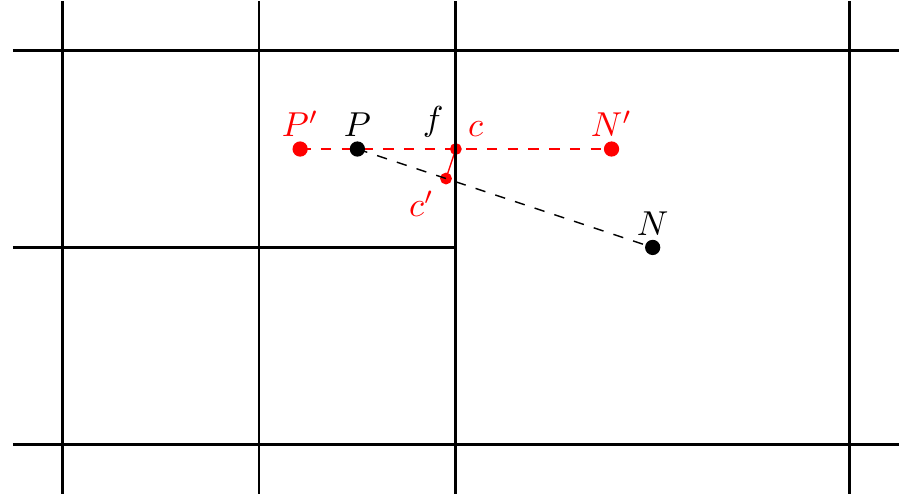}
 \caption{Two adjacent CVs of different level.}
 \label{fig:schemes}
\end{figure}

The in-house code used for the present study can handle grids composed of
quadrilateral CVs. A structured or block-structured grid is used as a starting
point, and local refinement levels are added on top of that by the adaptive
procedure. Refinement of a CV takes place by splitting it along the line
segments which join its centre with the centres of its faces, into four
\textit{child} CVs. The original CV, called the \textit{parent} CV, is retained
in the data structure, and is used by a multigrid algorithm to accelerate
algebraic convergence. Each CV belongs to a certain \textit{grid level}: these
levels are numbered, and the children of a CV of level $l$ belong to level
$l+1$. Each level consists of \textit{global} and \textit{local} CVs: Global are
those which do not have children, and local are those which do have (see Figure
\ref{fig:grid}). The actual grid on which the problem is discretised and solved
at any one time (\textit{composite grid}) is the set of all global CVs of all
levels. The CVs of the composite grid, although geometrically quadrilateral, are
treated as logically polygonal because one or more sides of a CV may actually
consist of two faces, if the CV borders a finer level on that side. A face is
defined as the piece of the boundary of a CV which separates it from another
single CV. For example, the left side of the CV whose centre is $N$ in Figure
\ref{fig:schemes} consists of two faces, one of which is face $f$ separating it
from the CV whose centre is $P$, which belongs to a finer level. Neighbouring
CVs are not allowed to be more than one level apart. For more information on the
grid structure see \cite{Syrakos06b}.

The discretisation schemes are those used in \cite{Syrakos06a} and are similar
to those described in \cite{Peric}. They are central difference schemes with
correction terms for grid non-orthogonality, skewness etc. Suppose a face $f$
of area $S_f$ and centre $c$ separating two CVs whose centres have position
vectors $P$ and $N$ as shown in Figure \ref{fig:schemes}. Suppose also $c'$
is the point on line $PN$ which is closest to $c$. Also let points $P'$, $N'$
be such that the line segment $P'N'$ is perpendicular to $f$, its length is
equal to the length of $PN$, and its centre is at $c$. The convective and
viscous terms of (\ref{eq:mom_x}) are discretised thus:

\begin{equation} \label{eq:CDS_conv}
 \int_{S_f} \rho u \vec{V} \cdot d\vec{S} \;\approx\; F_{h,f}u_{h,c} 
\end{equation}

\begin{equation} \label{eq:CDS_visc}
 \int_{S_f} \mu \nabla u \cdot d\vec{S} \;\approx\;
\mu S_f \frac{u_{h,N'}-u_{h,P'}}{|N'-P'|}
\end{equation}
where:
\begin{align}
 u_{h,c} \; &=\; (\overline{u_h})_{c'} \;+\; (\overline{\nabla_h u_h})_{c'}
  \cdot (c-c')
\label{eq:u_c}
\\
 u_{h,P'}\; &=\; u_{h,P} \;+\; (\nabla_h u_h)_P \cdot (P'-P)
\label{eq:u_P_prime}
\end{align}
In the above equations, $u_h$ is the grid function of the values of $u$ at the
CV centres of grid $h$ and $u_{h,P}$ is the value of that function at the centre
of CV $P$. The overline denotes linear interpolation at point $c'$ from the
values at points $P$ and $N$. The discrete gradient operator $\nabla_h$ is based
on a weighted least squares method that is second-order accurate on smooth
grids and first-order accurate on irregular grids (such as the one shown in Figure
\ref{fig:schemes}), described in \cite{Muzaferija_97, Syrakos}. The mass flux
$F_{h,f}$ through face $f$ is calculated using a variant of momentum interpolation
to avoid pressure oscillations which may otherwise be exhibited by the solution when
both velocity and pressure are stored at CV centres \cite{Peric}:

\begin{equation} \label{eq:mom_interp}
 F_{h,f} \;=\; \rho \vec{V}_{h,c} \cdot \vec{S}_f \;+\; \frac{\rho S_f^2}{A_f}
\left[ \left( p_{h,P} - p_{h,N} \right) - \left(\overline{\nabla_h p_h}\right)_
{\frac{P+N}{2}}  \cdot (P-N) \right]
\end{equation}

Here the overline denotes interpolation of the gradient at the middle of the
line segment $PN$. The original concept of momentum interpolation was proposed
in \cite{Rhie_Chow}, but it has the shortcoming that $A_f$ is calculated in the
context of the SIMPLE solution method \cite{Patankar_72}. To separate completely
the discretisation scheme from the algebraic solution method, a modified momentum
interpolation was proposed in \cite{Syrakos06a}, where $A_f$ is calculated directly
from flow variables and grid geometry:

\begin{equation} \label{eq:a_f}
 A_f \;=\; \rho S_f \,|\vec{V}_{h,c}\cdot \vec{n}_f| \;+\;
           \rho S_d \,|\vec{V}_{h,c}\cdot \text{ROT}(\vec{n}_f)| \;+\;
           2\mu \left[ \frac{S_f}{S_d} + \frac{S_d}{S_f} \right]
\end{equation}
Here $\vec{n}_f$ is the unit vector normal to face $f$, ROT is a function which
rotates a vector by 90 degrees, and $S_d=(N-P)\cdot \vec{n}_f$. The pressure
term of (\ref{eq:mom_interp}) does not allow the development of pressure
oscillations in the solution. It is an artificial term totally belonging to the
truncation error, and, as discussed in \cite{Syrakos06a}, its contribution to the
truncation error is $O(h^4)$ on smooth grids and $O(h^3)$ on irregular grids.
Therefore it does not affect the overall accuracy of the method.

The convection approximations (\ref{eq:CDS_conv}) of all faces of a CV
contribute $O(h)$ to the truncation error on non-smooth grids, and $O(h^2)$ on
smooth grids because in the latter case parts of the contributions cancel out
between opposite faces of the CV, as shown e.g. in \cite{Syrakos}. Also shown in
\cite{Syrakos} is that the diffusion approximations (\ref{eq:CDS_visc})
contribute $O(1)$ to the truncation error on non-smooth grids and $O(h^2)$ on
smooth grids. Nevertheless, as discussed in Section \ref{sec:tau}, the
discretisation error remains $O(h^2)$ under these discretisation schemes, even
when the truncation error is $O(h)$ or $O(1)$.

The discretisation of the pressure terms of (\ref{eq:mom_x}) - (\ref{eq:mom_y})
is similar to that of the convection terms. As for the second viscous terms
(second terms on the right hand sides of (\ref{eq:mom_x}) and (\ref{eq:mom_y})),
they are discretised using the midpoint rule approximation, using the value of
the gradients at point $c'$ as obtained by linear interpolation.

\begin{figure}[tpb]
 \centering
 \includegraphics[]{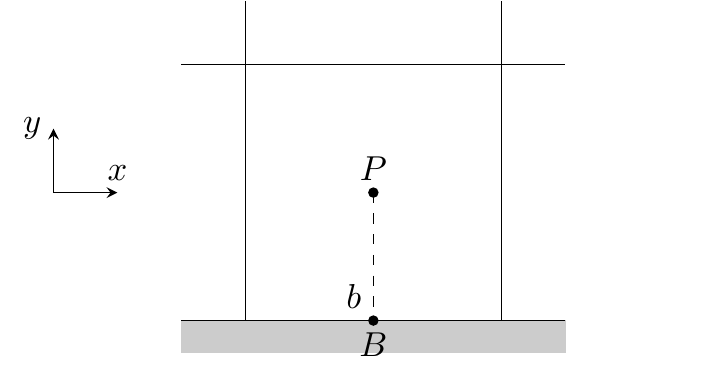}
 \caption{A boundary CV with its boundary face $b$.}
 \label{fig:boundary}
\end{figure}

It is appropriate also to discuss briefly the implementation of the wall
boundary condition, which is the only boundary condition used in the test case
of Section \ref{sec:experiments_setup}. The original version of the CFD code
used for the present study followed the popular approach described in
\cite{Peric} where instead of just implementing a Dirichlet boundary condition
for the momentum equations, the continuity equation is also used to further
refine the boundary condition. For example, Figure \ref{fig:boundary} shows a
boundary CV with a boundary face $f$ oriented along the $x$ direction. If the
wall is rigid and moves with a constant tangential velocity then $\partial u/
\partial x =0$ implies that also $\partial v / \partial y =0$. Therefore the
normal viscous stress is zero at a wall, and so only a tangential viscous force
is implemented on the boundary, discretised as $\int_{S_b}-\mu\frac{\partial
u}{\partial y} dS \approx -\mu S_b \frac{u_P - u_B}{y_P-y_B}$. But if the
tangential velocity of the wall is not constant (imagine the wall as being
elastic, stretching at some parts and contracting at others) then $\partial
u/\partial x \neq 0$ implies that also $\partial v/\partial y \neq 0$ and there
is also a normal viscous stress. To allow for this the CFD code was modified to
include also a normal viscous force: $\int_{S_b} -\mu \frac{\partial v}{\partial
y} dS \approx -\mu S_b \frac{v_P - v_B}{y_P - y_B}$ where $v_B=0$ on the wall
(the normal component of velocity is zero, but its normal derivative is not, in
general, zero). The 2$^\text{nd}$ viscous terms on the right hand sides of
(\ref{eq:mom_x}) - (\ref{eq:mom_y}) are also discretised by taking the gradients
$(\nabla u)_B$ and $(\nabla v)_B$ as equal to the respective gradients at the
centre of CV $P$. Preliminary numerical experiments on the classic lid-driven
cavity problem with rigid walls, for which both approaches are valid, showed
very little difference in the results (the method of \cite{Peric} produced very
slightly better results, as evaluated against the benchmark results of \cite{Botella_98}).

The discretisation of (\ref{eq:mom_x}) - (\ref{eq:continuity}) for each CV
results in a system of algebraic equations whose unknowns are the velocity and
pressure at each CV centre. This system is solved using the popular SIMPLE
algorithm \cite{Patankar_72} in a multigrid context which is such that it
allows for the existence of local refinement levels that do not extend throughout
the domain. This multigrid method was proposed in \cite{Syrakos06b}. SIMPLE is a
fixed-point iteration type algorithm, where during each outer iteration a linear
system is solved for each of the velocity components, followed by a linear
system for the pressure. The velocity component linear systems are obtained by
linearising the corresponding momentum equations using values from the previous
SIMPLE iteration in the nonlinear terms, and in terms containing variables other
than that particular velocity component. The velocity field that results from
the solution of these linear systems does not satisfy the continuity equation
in general, and so a Poisson-type linear system is constructed and solved to
calculate corrections to the pressure field that would force the velocity field
to be more mass-conserving. This completes one outer SIMPLE iteration, and the
whole procedure is repeated in the next outer iteration. This brief description
of the SIMPLE algorithm will be useful to the unfamiliar reader as mention of
it is made a few times in this paper, although the methodology and results are
in no way connected to a particular algebraic solver.

The SIMPLE-multigrid procedure solves the algebraic system up to a residual.
To examine the effect of this on the accuracy of the solution, consider again
the abstract problem $N(u)=b$. As already noted, the exact solution
$\tilde{u}_h$ of the discrete system which approximates the continuous problem
satisfies equation (\ref{eq:eps_tau}), repeated here:

\begin{equation} \label{eq:u_tau}
  N_h(\tilde{u}_h) \;=\; N_h(u_h) \;+\; \tau_h
\end{equation}
where $u_h$ is the exact solution of the continuous problem. Now, the iterative
procedure does not solve for $\tilde{u}_h$ exactly, but suppose that after the
$k$-th iteration the algebraic residual is $r_h^k$ and the approximate solution
is $\tilde{u}_h^k$:

\begin{equation} \label{eq:u_tau_res}
  N_h(\tilde{u}_h^k) \;=\; N_h(\tilde{u}_h) \;+\; r_h^k \;=\;
                           N_h(u_h) \;+\; (\tau_h + r_h^k)
\end{equation}
where (\ref{eq:u_tau}) has been used to obtain the second equality. Comparing
with (\ref{eq:u_tau}), this result shows that $\tilde{u}_h^k$ corresponds to a
``truncation error'' $\tau_h + r_h^k$, just like $\tilde{u}_h$ corresponds to
the truncation error $\tau_h$. To obtain the maximum accuracy, iterations must
be carried on until the residual $r_h^k$ has dropped below the truncation error
$\tau_h$. This is important also in order to accurately estimate the truncation
error using (\ref{eq:tau_h_estimate})\footnote{A detailed discussion of the effect
of the iteration error on the truncation error estimate is given in \cite{Fraysse_12},
where it is shown that it is possible, using specialised restriction operators,
to obtain a good estimate of the truncation error even before the algebraic
residuals have dropped to a low level. The results of \cite{Fraysse_12} were
not known to the authors when the present study was conducted.}.

\section{Local refinement} \label{sec:refinement}

\subsection{General procedure} \label{sec:refinement_general_procedure}

In the present work local refinement takes place as follows: After solving the
equations on a given grid (which may be composite), the truncation error is
calculated at every CV using equation (\ref{eq:tau_h_estimate}). This truncation
error estimate is used to calculate another quantity which will be used to
select the CVs to be refined (such quantities will be described shortly).
Then all CVs are sorted in descending order according to the absolute value of
this quantity. A fraction $R$ of the CVs, those with the largest value of this
quantity, are marked for refinement, i.e. the first $R\cdot K$ CVs of this
descending list are marked, where $K$ is the total number of CVs in the grid.
An efficient algorithm should be chosen to perform the sorting because an
inefficient algorithm could require a significant amount of time if the grid is
large in terms of number of CVs. This has been observed in the present work when
the selection sort algorithm was first used due to its simplicity. The situation
improved greatly when the quicksort algorithm was used instead, with the cost
of sorting becoming insignificant compared to the overall computational cost
(see e.g. \cite{Lewis} for a description of the algorithms).

\begin{figure}[tpb]
 \centering
 \noindent\makebox[\textwidth]{
   \includegraphics[width=1.0\textwidth]{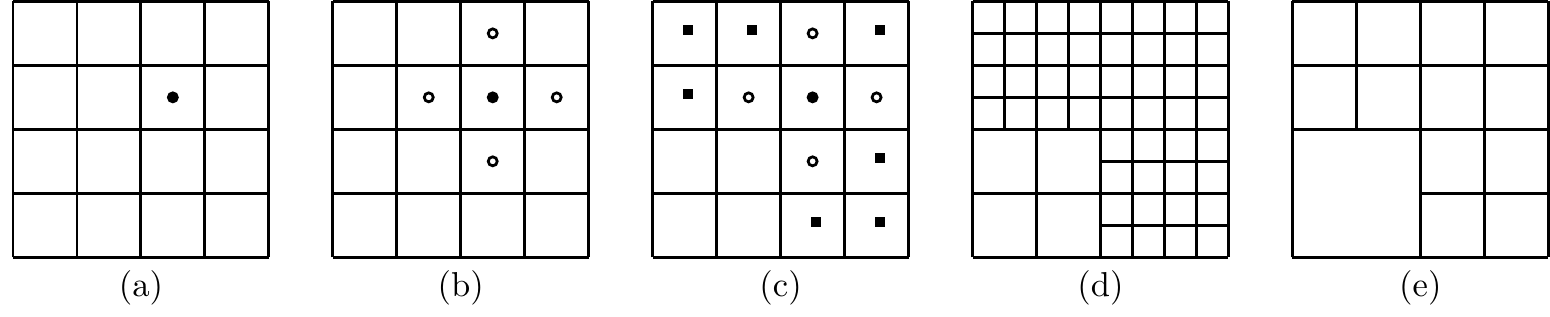}}
 \caption{Local Refinement. (a): Suppose a single CV (\textbullet) is marked for
refinement according to the criterion. (b): its neighbours (\textopenbullet)
are also marked, as a safety margin. (c): The siblings ($\scriptscriptstyle
\blacksquare$) of any marked CVs are also marked, as required for the existence
of the underlying grid. (d): After all CVs have been marked, refinement takes
place to produce the new grid $h$, whose underlying grid $2h$ is (e).}
 \label{fig:refinement}
\end{figure}

Marking of the CVs takes place as shown in Figure \ref{fig:refinement}: When a
CV is marked, its neighbours are also marked as a safety margin (Figure
\ref{fig:refinement} (b)). Also, as shown in Figure \ref{fig:refinement} (c),
for each CV marked its siblings are also marked (i.e. all CVs which share the
same parent CV), which is necessary so that the new grid $h$ which will result
when refinement takes place will have an underlying grid $2h$ on which the new
truncation error will be calculated. This marking procedure results in more
than $R\cdot K$ CVs being marked overall. Furthermore, if a system of PDEs is
solved, like in the present case where there are 3 equations for each CV (2
momentum equations (\ref{eq:mom_x}) - (\ref{eq:mom_y}) and the continuity
equation (\ref{eq:continuity})), the marking procedure can be repeated for as
many of these equations as desired, and the results are logically OR-ed to
establish the CVs that are to be refined.

After all relevant CVs have been marked, refinement takes place by splitting
each marked CV into four children, at the line segments joining their centres
with the centres of their four sides, as shown in Figure \ref{fig:refinement}
(d). By marking all siblings together, this procedures ensures the existence of
the coarse grid $2h$ (Figure \ref{fig:refinement} (e)) which consists of the
parents of all the global CVs, or equivalently of all the local CVs whose
children are global. The equations are then discretised and solved on the new
grid $h$, the truncation error is calculated using the coarse grid $2h$, and
the refinement procedure is repeated to produce yet another grid. This cycle is
repeated a fixed number of times.

\subsection{Quantities used as refinement criteria} \label{sec:criteria}

In the present work, the quantities that are used to assess whether a CV $P$
should be refined or not are the absolute values of the following:

\begin{enumerate}
 \item The truncation error itself ($\tau_{h,P}$). \label{q:tau}
 \item The truncation error times the volume of the CV ($\tau_{h,P} \cdot \delta
\Omega_P$). \label{q:tau_vol}
 \item The truncation error divided by the corresponding main diagonal
coefficient of a matrix $A_h$ which comes from linearising the discrete operator
$N_h$ ($\tau_{h,P} / A_{hP,P}$).
\label{q:tau_ap}
\end{enumerate}

The quantity \ref{q:tau} is used for example in \cite{Thompson_89, Aftosmis02,
Brown_05}. The quantity \ref{q:tau_vol} is used e.g. in \cite{Trottenberg}. And
quantities similar to \ref{q:tau_ap} are used e.g. in \cite{Muzaferija_97,
Jasak_01} (except that there the residual is used instead of the truncation
error).

The reasoning behind quantity \ref{q:tau_ap} can be sketched as follows: If
$N_h$ is linear, then the relationship (\ref{eq:eps_tau}) between the truncation
and discretisation errors simplifies to $N_h\epsilon_h = -\tau_h$. Then,
dividing $\tau_h$ by the main diagonal coefficients of $N_h$ is equivalent to
performing a single Jacobi iteration on that system with zero initial guess.
This produces a very rough estimate of the discretisation error $\epsilon_h$
produced by the truncation error in each CV that takes into account the
physical processes (e.g. convection, diffusion) inherent in $N_h$. This rough
estimate of $\epsilon_h$ is actually used as the refinement criterion.

The main diagonal coefficients used in quantity \ref{q:tau_ap} are those of the
velocity linear systems in the SIMPLE solution method, as long as the systems
are expressed per unit volume and the matrices of coefficients are constructed
using the UDS (first order upwind) scheme for convection with any higher-order
schemes being implemented through deferred correction \cite{Khosla_74, Peric}.
The UDS ensures that only the flow out of the CV is taken into account, and not
the flow in that would lead to cancellations and possibly near zero or negative
contribution of convection to the coefficient \cite{Muzaferija_97, Jasak_01}.

If the systems are expressed per unit volume then quantity \ref{q:tau_ap} for a
CV $P$ becomes $|\tau_{h,P}^x/A_{P,P}^u|$ and $|\tau_{h,P}^y/A_{P,P}^v|$ for the
$x$- and $y$- momentum equations respectively, $A^u$, $A^v$ being the matrices
of coefficients of the linear systems for the velocity components. If the
systems are not expressed per unit volume then quantity \ref{q:tau_ap} is
calculated as $|(\tau_{h,P}^x\!\cdot\!\delta\Omega_P) / A_{P,P}^u|$ and
$|(\tau_{h,P}^y\!\cdot\!\delta\Omega_P) /A_{P,P}^v|$. The coefficients are
readily available if the SIMPLE solution method is used, otherwise they have to
be calculated. When not expressed per unit volume their form is similar to
(\ref{eq:a_f}), only that (\ref{eq:a_f}) is staggered, i.e. centred around
a face centre, while $A_{P,P}$ would be centred around the centre of CV $P$.
An examination of (\ref{eq:a_f}) shows that grid refinement causes $A_{P,P}$ to
tend to a constant; its convection part tends to zero because it is
proportional to the CV dimensions $S_f$ and $S_d$, while the viscous part is
constant because the refinement procedure adopted here does not alter the
aspect ratio $S_f/S_d$ of the CVs. If all the CVs of the grid have the same
aspect ratio then as the grid becomes finer $A_{P,P}$ tends to the same value
for all CVs (for constant $\mu$) and the quantities \ref{q:tau_vol}
$(\tau_{h,P}\!\cdot\!\delta\Omega_P)$ and \ref{q:tau_ap}
$(\tau_{h,P}\!\cdot\!\delta\Omega_P / A_{P,P})$ become equivalent. If the CV
aspect ratio varies across the grid then criterion \ref{q:tau_ap} has a
preference for refining CVs with an aspect ratio close to 1.

In the present work quantity \ref{q:tau_ap} is not calculated for the continuity
equation because a corresponding coefficient is not readily available, although
it should be possible to derive an approximate equation relating the truncation
error of the continuity equation to the discretisation error of pressure in the
same way that the ``pressure correction equation'' is derived in the SIMPLE
method.

\subsection{Internal boundary treatment} \label{sec:boundary treatment}

There is another important issue which must be addressed. The derivation of the
truncation error estimate in Section \ref{sec:tau_estimate} has been based on a
number of idealised assumptions, not all of which are valid in a realistic case.
Specifically, as discussed in Section \ref{sec:tau}, the truncation error
becomes $O(1)$ at domain boundaries and at the interior boundaries between
different grid levels. This is because although the grid is generally Cartesian,
it exhibits all kinds of distortions directly on the level interface as shown
in Figure \ref{fig:schemes}. Therefore the truncation error distribution is
discontinuous at these regions, and the index $p$ of equation
(\ref{eq:tau_leading_term}) is zero there, while at the smooth interior regions
of the grid it equals $2$. Yet, as also discussed in Section \ref{sec:tau},
equation (\ref{eq:epsilon_leading_term}) remains valid even on non-smooth grids.
Even though (\ref{eq:tau_leading_term}) is not completely valid on composite
grids, equation (\ref{eq:tau_h_estimate}) should provide a rough estimate of the
truncation error, due to the validity of (\ref{eq:epsilon_leading_term}). That
is, since the truncation error estimate works by assuming that the solution of
the fine grid $h$ is roughly equivalent to the exact solution of the PDE when
considered with respect to grid $2h$, and since this assumption is roughly true 
even on distorted grids due to (\ref{eq:epsilon_leading_term}), $\tilde{\tau}_h$
should provide a rough estimate even at distorted regions of the grid, albeit
not as accurate as at smooth regions where all the assumptions hold.

\begin{figure}[tpb]
 \centering
 \includegraphics[scale=0.45]{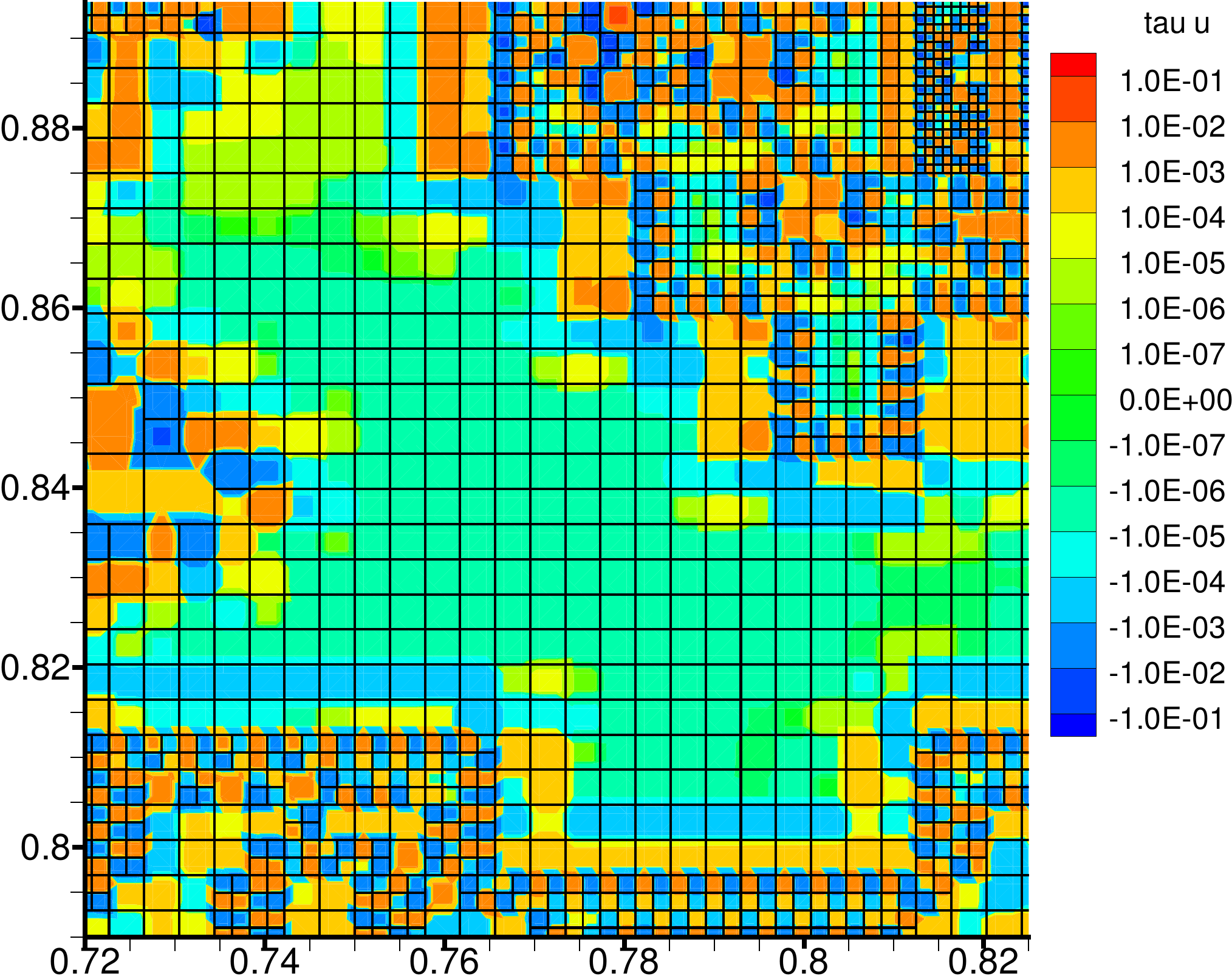}
 \caption{Truncation error (estimate) of the $x$-momentum equation at a region
with grid level interfaces. The underlying grid $2h$ is shown.}
 \label{fig:tau_at_boundaries}
\end{figure}

Figure \ref{fig:tau_at_boundaries} shows an example of the high truncation
errors that occur near grid level boundaries. The $x$-momentum equation
truncation error estimate is displayed along with the underlying grid $2h$ for
one of the test cases which will be presented in Section \ref{sec:results}.
It is a good idea to display the coarse grid $2h$ rather than grid $h$ because
most of the truncation error calculation occurs on grid $2h$, and since an
injection prolongation operator is used all children of a CV of grid $2h$
inherit the same value of truncation error calculated at their parent. It may be
seen that the truncation error is several orders of magnitude higher
near the level interfaces than in the interior of the levels. This is expected
since in the former case $\tau_h \in O(1)$ while in the latter case $\tau_h \in
O(h^2)$. It may also be noticed that the high truncation error regions appear to
extend up to two CVs of grid $2h$ away from the level interface, i.e. up to 4
CVs of grid $h$. This is an artificial result of the truncation error estimate
(\ref{eq:tau_h_estimate}) which calculates the truncation error mostly on grid
$2h$, caused by the discontinuity of the truncation error distribution. In fact
the actual high truncation error occurs only up to two CVs of grid $h$ away from
the interface. A similar complication also seen in the Figure is that in the
high truncation error regions the truncation error often appears to alternate in
sign between neighbouring CVs of grid $2h$. This should be expected to occur in
fact between neighbouring CVs of grid $h$, but a usual prolongation operator
cannot produce this.

Therefore, to transfer correctly the truncation error estimate from grid $2h$ to
grid $h$ a specialised prolongation operator should be used that takes into
account the similarity between CVs of grid $2h$ and CVs of grid $h$. For
example, the truncation error at CVs of grid $h$ that are in the second row
away from the level interface should be based not on the calculations performed
at their parents (which are in direct contact with the level interface) but on
their parents' neighbours that also lie on the second row (of grid $2h$) of CVs
away from the interface. Anyway, in the present work such a specialised
prolongation operator is not used, because the issue is not how to transfer
accurately the $O(1)$ truncation error from grid $2h$ to grid $h$, but how to
deal with it in the context of local refinement, which assumes that refinement of
CVs causes a reduction in truncation error. In this case, of course, such a reduction
does not occur because $\tau_h \in O(1)$, so refinement appears to bring no benefit,
while on the other hand it increases the computational cost by increasing the number
of CVs. Three different strategies to deal with this issue are tested in the
present work:

\begin{enumerate}
 \item Take no particular action. CVs near grid level interfaces may be refined
just as all other CVs of the grid.
 \item Do not allow refinement of CVs that are near level interfaces. This
includes CVs that are up to 4 positions away from the interface, as discussed
above.
 \item Do not allow refinement of CVs that are on the fine side of the
interface, but allow refinement of CVs that are on the coarse side of the
interface. This does not allow the creation of an even finer level, but rather
the extension of the existing fine level towards the coarse side.
\end{enumerate}

These strategies will be tested in Section \ref{sec:results}. Similar problems
occur at the computational domain boundaries where the boundary conditions are
applied, but in the present work no special restrictions on CV refinement have
been applied there.

To check whether a CV is within a 4-CV width zone on either side of the level
interface, it is checked whether its parent or any of the parent's neigbours
of the same level as the parent are in contact with the level interface
(the 4-CV width zone appears as a 2-CV width zone on the underlying grid).
The parent is checked first; since it belongs to the underlying
grid it will be in touch with the interface if any of its neighbours of the same
level do not belong to the underlying grid, i.e. if they either do not have
children, or they have grandchildren. If all the parent's neighbours belong
to the underlying grid then the parent is not on the interface, but the check
must be repeated for its neighbours as well. Alternatively, one could traverse
the linked list of faces which lie on the interface (they separate a fine from
a coarser CV). The data structures are implemented using pointers in Fortran 95.

Closing the present section, we would like to make a couple of comments. The
first concerns the reason why the high truncation error extends also to the second
CV away from the interface. The reason is that, according to formulae
(\ref{eq:CDS_conv}) - (\ref{eq:u_P_prime}), the calculation of the fluxes through
a face that separates such a CV from another CV that touches directly the level
interface involves also the gradients at this neighbour CV, which are only first
order accurate, as mentioned in Section \ref{sec:eqns_and_discr} (because the
grid is distorted directly on the interface, as shown in Figure \ref{fig:schemes}).
On the other hand, one may notice that the gradients in the terms (\ref{eq:CDS_conv}) -
(\ref{eq:u_P_prime}) are used only if the grid exhibits skewness,
non-orthogonality or unequal spacing of the CV centres from that face. However,
even in the absence of these geometric distortions the gradient at the CV which
touches the interface is still used to calculate the second viscosity terms of
(\ref{eq:mom_x}) - (\ref{eq:mom_y}), i.e. the second terms on the right-hand
sides, for the face that separates the two CVs. So even if the grid is smooth,
a CV that has a neighbour in direct contact with the level interface will
exhibit large truncation errors.

The second comment concerns the alternating sign of the truncation error at
neighbouring CVs near the level interfaces. This is due to the fact that when a
conservative discretisation scheme is used, like in the present work, each face
contributes equal and opposite contributions to the equations of the CVs on
either side of the face. This holds also for the truncation error contributions:
the discretisation of the flux through a face contributes equal and opposite
contributions to the truncation errors of the CVs on either side. Therefore the
total truncation error in the domain is determined only by boundary faces and
source terms. Also, if a pair of neighbouring CVs has $O(1)$ truncation errors
of similar magnitude but opposite sign, then the effect of these truncation
errors (in terms of the discretisation errors they produce) at distant locations
will tend to cancel out. Local refinement of these CVs, although it will not
reduce the truncation error since it is of $O(1)$, will produce smaller CVs of
opposite truncation error that are closer together. Bringing the two sources of
opposite strength closer together would cause the overall discretisation error
that they produce at any given location to decrease. So local refinement could
prove beneficial even in cases when $\tau_h \in O(1)$, which is in line with the
observation mentioned in Section \ref{sec:tau} concerning the validity of
equation (\ref{eq:epsilon_leading_term}) even in cases when
(\ref{eq:tau_leading_term}) does not hold. For example, in \cite{Syrakos06b} a
numerical experiment is described where the lid-driven cavity problem is solved
on a series of similar composite grids of increasing fineness (each successive
grid is produced by refining every CV of the previous grid) and although the
truncation error does not decrease near level interfaces, the discretisation error
does decrease as $O(h^2)$.

\section{Description of the test case and of the evaluation methodology}
\label{sec:experiments_setup}

\begin{figure}[tpb]
 \centering
 \includegraphics[scale=0.7]{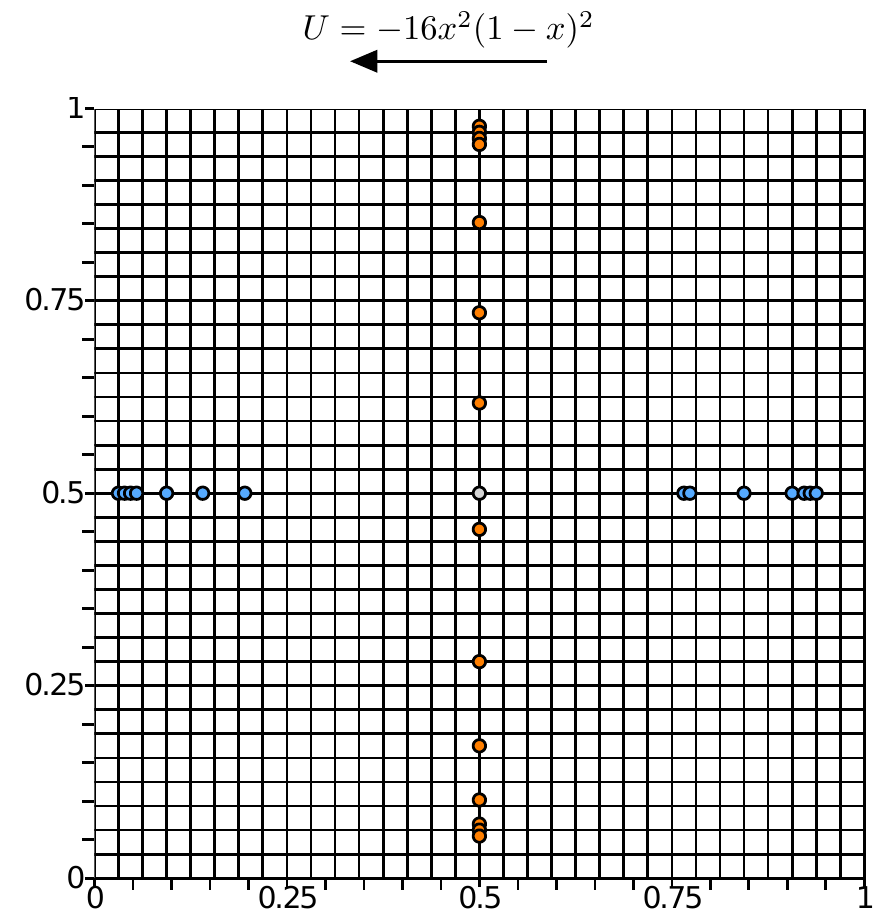}
 \caption{The $32 \times 32$ grid and the location of the probe points for the
lid driven cavity problem.}
 \label{fig:probes}
\end{figure}

\subsection{The regularised lid-driven cavity case}

The refinement strategies previously presented were tested on a lid-driven
cavity problem. The classic lid-driven cavity problem is the most popular
problem for testing and validating new techniques for incompressible flows (see
e.g. \cite{Ghia_82, Botella_98, Bruneau_06, Syrakos06b} and references therein).
It concerns the flow in a square cavity with three stationary sides and one side
(lid) - usually taken as the top side as shown in Figure \ref{fig:probes} -
moving at a constant tangential velocity which drives the flow. Unfortunately
the resulting flow field is discontinuous because the velocity changes abruptly
from a non-zero constant value (top lid) to zero (side walls) at the top corners
of the cavity. This causes very large truncation errors at these two corners,
which increase with refinement (see e.g. \cite{Syrakos06b}). This would pose an
additional complexity for the present investigation of the efficiency of
truncation error-based local refinement, making it more difficult to interpret
the results. So, to keep things simple, the so-called regularized lid-driven
cavity problem \cite{Shen_89, Shen_91} was selected instead, where the
tangential velocity of the lid is not constant (the lid may be viewed as made of
an elastic material that can stretch and contract), but is given by a
polynomial that is such that the velocity and its first derivative are zero at
the edges of the lid:

\begin{equation} \label{eq:lid_velocity}
 U(x) \;=\; -16x^2(1-x)^2
\end{equation}

It is necessary that $dU/dx$ be zero at the lid edges so that the continuity
equation is satisfied: At the lid corners it holds that $\partial v/\partial y =
0$ because $v=0$ at the side walls, so it must also hold that $\partial
u/\partial x = dU/dx = 0$. Equation (\ref{eq:lid_velocity}) sets the maximum
velocity at the centre of the lid with $|U_{max}|=1\;m/s$. There are no
Dirichlet boundary conditions for pressure but instead pressure is linearly
extrapolated to the boundaries from the interior \cite{Peric}. After solution,
the whole pressure field is adjusted by a constant so that $p=0$ at the centre
of the cavity.

First, the problem was solved on a series of structured grids with uniform grid
line spacing: $32\times 32$ CVs (shown in Figure \ref{fig:probes}), $64\times
64$ CVs, etc. up to $2048\times 2048$ CVs. These will be referred to as
\textit{structured} grids in the following, while the locally refined grids will
also be referred to as \textit{composite} grids. Then using Richardson
extrapolation (see e.g. \cite{Peric}) a very accurate solution was obtained
which was treated as the exact solution. This was repeated for a range of
Reynolds numbers: $Re$ = 100, 1000 and 10000 where $Re = \frac{\rho
U_{max}L}{\mu}$. The values of the constants are $\rho=1\;kg/m^3$ (density),
$U_{max}=1\;m/s$ (max. lid velocity) and $L=1\;m$ (length of cavity side). The
viscosity $\mu$ was varied to obtain the desired Reynolds number.

For the classic lid-driven cavity problem, published results (e.g.
\cite{Bruneau_06}) indicate that the flow is steady for $Re=$ 100 and 1000, but
not for 10000 where it becomes periodic in time. For the regularised lid-driven
cavity problem it is suggested in \cite{Shen_89, Shen_91} that the flow is
steady at $Re=$ 10000, although it becomes unsteady at slightly higher Reynolds
numbers. The present work deals only with steady-state flows, so the flow at $Re=$
10000 is considered to be steady (which is most probably the case as suggested in
\cite{Shen_89, Shen_91}).

Multigrid cycles were carried out on each grid until the residuals of all
equations in every CV had dropped below $10^{-8}$ to ensure that they were well
below the truncation error (see eqn. (\ref{eq:u_tau_res}))\footnote{Actually it
suffices to ensure that the residuals have dropped below the truncation error by,
say, an order of magnitude, according to equation (\ref{eq:u_tau_res}). Any further
reduction would be a waste of computational effort because the accuracy would not
increase. In this respect the $10^{-8}$ threshold that we used throughout the domain
is supererogatory.}. For $Re$=100, iterations converged on each grid (including
composite grids) in about 13 V(2,2) cycles. For $Re$ = 1000, converge was obtained
with around 12 W(2,2) cycles (more on coarse grids). And for $Re$ = 10000 using
again W(2,2) cycles, convergence was obtained with 12 and 10 cycles for the
1024$\times$1024 and 2048$\times$2048 grids, respectively, while for coarser grids
more cycles were necessary. Finally we note that for $Re$=1000 and $Re$=10000 within
each cycle the corrections were smoothed before being prolonged to the finer
grid according to the procedure described in \cite{Syrakos06b}.

Tables \ref{table:vert_centreline} and \ref{table:horiz_centreline} present the
``exact'' solution as obtained using Richardson extrapolation at a number of
points along the vertical and horizontal centrelines respectively. The points
are marked on Figure \ref{fig:probes}, and they are the same points used in
numerous studies on the classic lid-driven cavity problem; they were first used
in \cite{Ghia_82}, and also later by other authors such as \cite{Botella_98,
Bruneau_06} who compared their results against those of \cite{Ghia_82}. So we
present here the respective results for the regularised cavity case, to be
available as benchmark results for future studies. The present results were
obtained by first linearly interpolating the solution of the 1024$\times$1024
and 2048$\times$2048 grids at these points and then performing Richardson
extrapolation, assuming 2$^{\text{nd}}$ order convergence. Repeating the
procedure using grids 512$\times$512 and 1024$\times$1024 instead showed no
difference up to the 7 digits shown in Tables \ref{table:vert_centreline} and
\ref{table:horiz_centreline} for Re=100 and Re=1000 except for a couple of
points near the lid. For Re=10000 the last digit (or the last couple of digits
near the lid) may not be accurate.

\begin{table}[htpb]
\caption{Numerical solutions obtained along the vertical centreline.}
\label{table:vert_centreline}
\begin{center}
\begin{scriptsize}   
\rowcolors{3}{black!20}{white}
\begin{tabular}{ r | r r | r r | r r }
 \hline
    \multicolumn{1}{|c|}{}
  & \multicolumn{2}{ c|}{\textbf{Re = 100}} 
  & \multicolumn{2}{ c|}{\textbf{Re = 1000}}
  & \multicolumn{2}{ c|}{\textbf{Re = 10000}} \\
    \multicolumn{1}{|c|}{\textbf{\textit{y}}}
  & \multicolumn{1}{ c }{\textbf{\textit{u}}}
  & \multicolumn{1}{ c|}{\textbf{\textit{p}}}
  & \multicolumn{1}{ c }{\textbf{\textit{u}}}
  & \multicolumn{1}{ c|}{\textbf{\textit{p}}}
  & \multicolumn{1}{ c }{\textbf{\textit{u}}}
  & \multicolumn{1}{ c|}{\textbf{\textit{p}}} \\
 \hline
0.0000 & 0.0000000 & 0.0250098 & 0.0000000 & 0.0607925 & 0.0000000 & 0.0605268\\
0.0547 & 0.0306724 & 0.0250583 & 0.1054923 & 0.0605163 & 0.3178112 & 0.0557021\\
0.0625 & 0.0345378 & 0.0250546 & 0.1184943 & 0.0603569 & 0.3252990 & 0.0539831\\
0.0703 & 0.0383053 & 0.0250475 & 0.1312872 & 0.0601478 & 0.3243764 & 0.0521744\\
0.1016 & 0.0525963 & 0.0249789 & 0.1813140 & 0.0586509 & 0.2920120 & 0.0451693\\
0.1719 & 0.0815204 & 0.0244648 & 0.2668322 & 0.0496716 & 0.2427313 & 0.0316099\\
0.2813 & 0.1222629 & 0.0217104 & 0.2303403 & 0.0257505 & 0.1690492 & 0.0151767\\
0.4531 & 0.1630748 & 0.0070318 & 0.0873616 & 0.0028316 & 0.0529534 & 0.0012714\\
0.5000 & 0.1612522 & 0.0000000 & 0.0519093 & 0.0000000 & 0.0209067 & 0.0000000\\
0.6172 & 0.1165433 &-0.0211446 &-0.0397520 &-0.0010470 &-0.0612785 & 0.0014953\\
0.7344 & 0.0140787 &-0.0389617 &-0.1399778 & 0.0063314 &-0.1496488 & 0.0092355\\
0.8516 &-0.1591120 &-0.0426021 &-0.2387008 & 0.0201424 &-0.2503312 & 0.0216989\\
0.9531 &-0.5752639 &-0.0250529 &-0.3131419 & 0.0304935 &-0.3470263 & 0.0327463\\
0.9609 &-0.6336426 &-0.0223860 &-0.3464474 & 0.0310851 &-0.3473972 & 0.0333507\\
0.9688 &-0.6980380 &-0.0194587 &-0.4019992 & 0.0317165 &-0.3435578 & 0.0338717\\
0.9766 &-0.7668059 &-0.0163600 &-0.4880689 & 0.0324146 &-0.3367190 & 0.0342828\\
1.0000 &-1.0000000 &-0.0063047 &-1.0000000 & 0.0353790 &-1.0000000 & 0.0351764\\
 \hline
\end{tabular}
\end{scriptsize}
\end{center}
\end{table}

\begin{table}[htpb]
\caption{Numerical solutions obtained along the horizontal centreline.}
\label{table:horiz_centreline}
\begin{center}
\begin{scriptsize}   
\rowcolors{3}{black!20}{white}
\begin{tabular}{ r | r r | r r | r r }
 \hline
    \multicolumn{1}{|c|}{}
  & \multicolumn{2}{ c|}{\textbf{Re = 100}} 
  & \multicolumn{2}{ c|}{\textbf{Re = 1000}}
  & \multicolumn{2}{ c|}{\textbf{Re = 10000}} \\
    \multicolumn{1}{|c|}{\textbf{\textit{x}}}
  & \multicolumn{1}{ c }{\textbf{\textit{v}}}
  & \multicolumn{1}{ c|}{\textbf{\textit{p}}}
  & \multicolumn{1}{ c }{\textbf{\textit{v}}}
  & \multicolumn{1}{ c|}{\textbf{\textit{p}}}
  & \multicolumn{1}{ c }{\textbf{\textit{v}}}
  & \multicolumn{1}{ c|}{\textbf{\textit{p}}} \\
 \hline
0.0000 & 0.0000000 & 0.0194764 & 0.0000000 & 0.0422581 & 0.0000000 & 0.0458727\\
0.0312 &-0.0522265 & 0.0225569 &-0.1460808 & 0.0432335 &-0.3915446 & 0.0441632\\
0.0391 &-0.0648647 & 0.0232348 &-0.1877457 & 0.0432677 &-0.4113208 & 0.0425346\\
0.0469 &-0.0770068 & 0.0238568 &-0.2277794 & 0.0431372 &-0.3964249 & 0.0408660\\
0.0547 &-0.0887644 & 0.0244285 &-0.2651327 & 0.0428066 &-0.3704853 & 0.0393322\\
0.0937 &-0.1401717 & 0.0264122 &-0.3742712 & 0.0378014 &-0.3110442 & 0.0330918\\
0.1406 &-0.1809713 & 0.0265146 &-0.3396911 & 0.0280398 &-0.2709560 & 0.0261939\\
0.1953 &-0.1951736 & 0.0233699 &-0.2491894 & 0.0186821 &-0.2239381 & 0.0189608\\
0.5000 & 0.0501305 & 0.0000000 & 0.0276485 & 0.0000000 & 0.0077100 & 0.0000000\\
0.7656 & 0.1382907 & 0.0082429 & 0.2488983 & 0.0295021 & 0.2001465 & 0.0187390\\
0.7734 & 0.1375760 & 0.0084751 & 0.2527025 & 0.0309354 & 0.2059849 & 0.0197800\\
0.8437 & 0.1216119 & 0.0102571 & 0.2558328 & 0.0427657 & 0.2596504 & 0.0302368\\
0.9062 & 0.0900690 & 0.0115459 & 0.2121234 & 0.0493963 & 0.3135677 & 0.0408176\\
0.9219 & 0.0789643 & 0.0118740 & 0.1947518 & 0.0503543 & 0.3254261 & 0.0435931\\
0.9297 & 0.0729086 & 0.0120435 & 0.1846614 & 0.0507312 & 0.3275184 & 0.0449388\\
0.9375 & 0.0664792 & 0.0122188 & 0.1732839 & 0.0510457 & 0.3253029 & 0.0462217\\
1.0000 & 0.0000000 & 0.0139673 & 0.0000000 & 0.0518188 & 0.0000000 & 0.0508029\\
 \hline
\end{tabular}
\end{scriptsize}
\end{center}
\end{table}

\begin{figure}[tpb]
 \centering
 \noindent\makebox[\textwidth]{
 \includegraphics[width=1.0\textwidth]{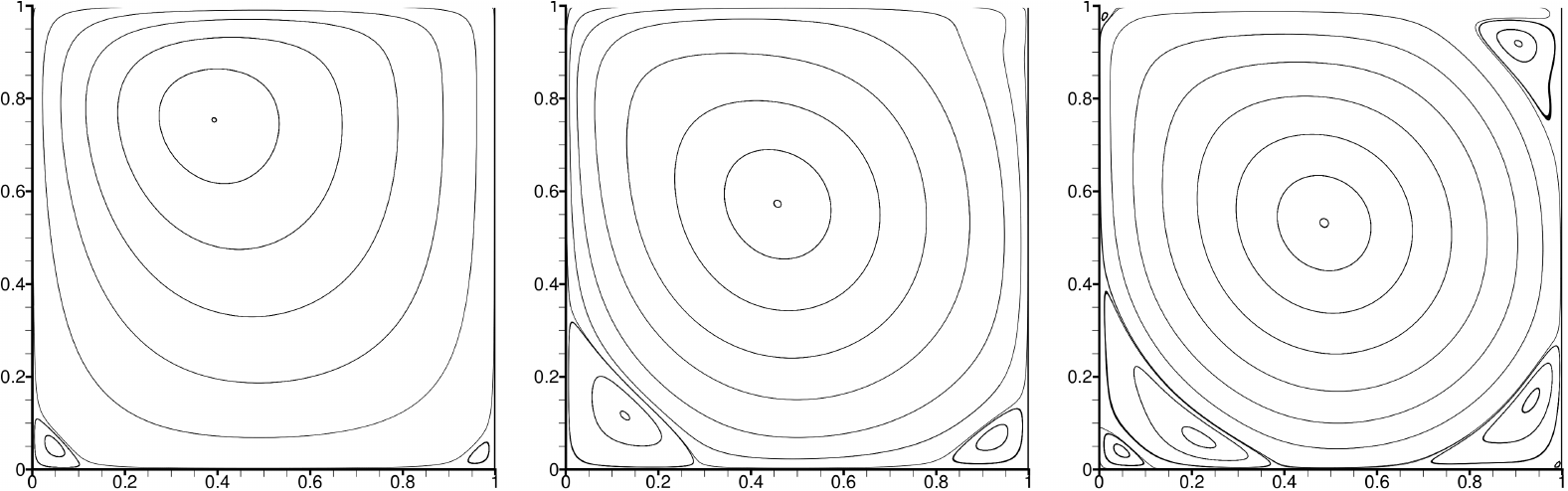}}
 \caption{Particle trajectories: Re=100 (left), Re=1000 (middle) and Re=10000
(right).}
 \label{fig:trajectories}
\end{figure}

Figure \ref{fig:trajectories} visualises the flow field by means of particle
trajectories. The vortex dynamics is similar to the classic lid-driven cavity
case, except that there is a small tertiary vortex at the top left corner for
Re=10000, in agreement with \cite{Shen_91}.

\begin{figure}[htpb]
 \centering
 \noindent\makebox[\textwidth]{
 \includegraphics[width=1.1\textwidth]{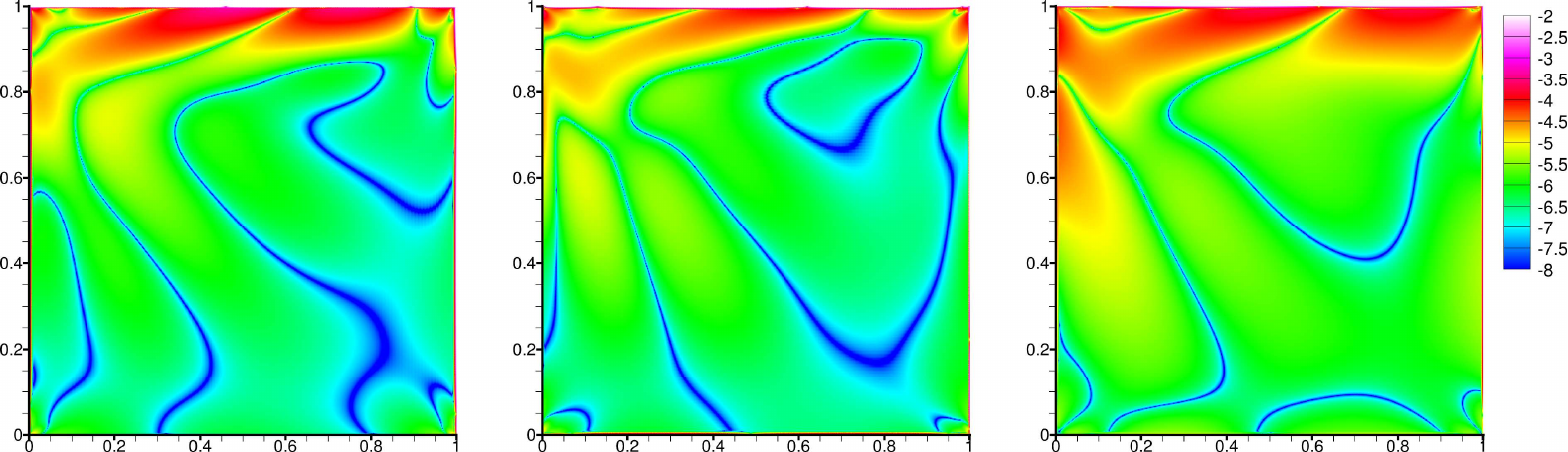}}
 \noindent\makebox[\textwidth]{
 \includegraphics[width=1.1\textwidth]{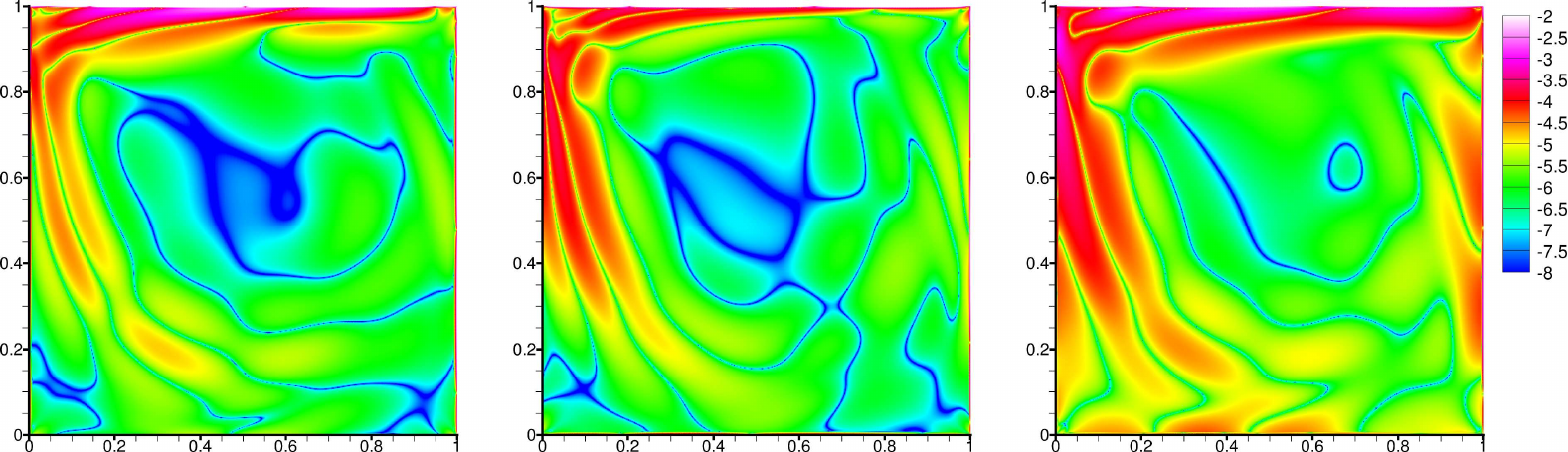}}
 \noindent\makebox[\textwidth]{
 \includegraphics[width=1.1\textwidth]{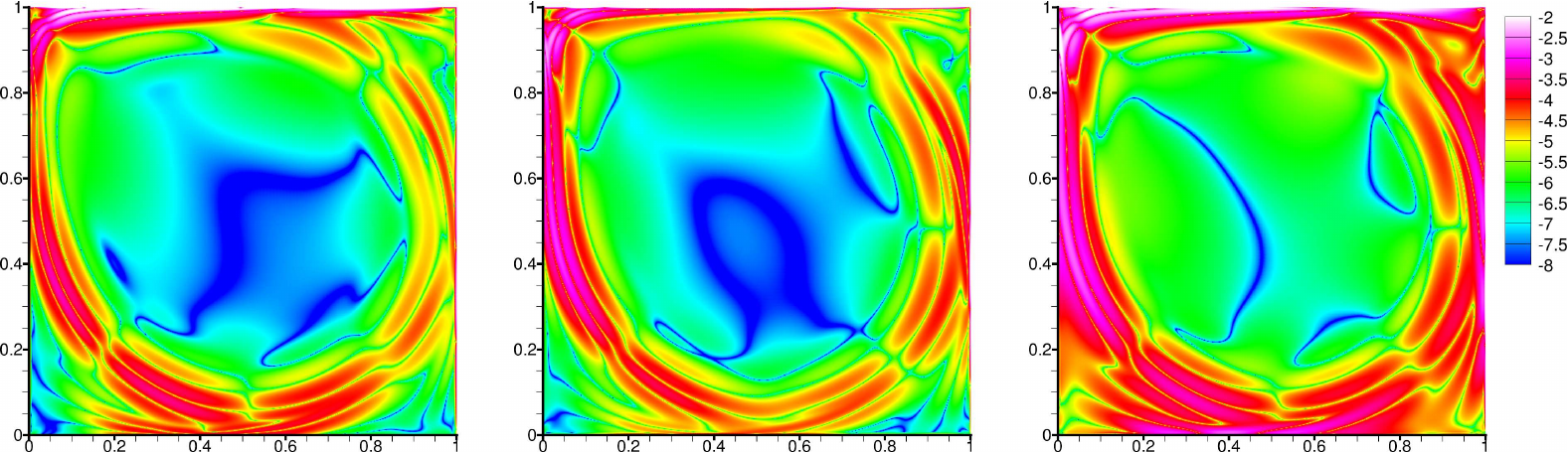}}
 \caption{Base-10 logarithm of the absolute value of the truncation errors of
the $x-$momentum equation (left), of the $y-$momentum equation
(middle) and of the continuity equation (right), for Re=100 (top), Re=1000
(middle) and Re=10000 (bottom).}
 \label{fig:tau}
\end{figure}

Figure \ref{fig:tau} shows the truncation errors (calculated via
(\ref{eq:tau_h_estimate})) of each equation on the 1024$\times$1024 grid for the
various Reynolds numbers. These images will be helpful later on in order to
interpret why each refinement criterion chooses to refine the specific areas
that it does.

\subsection{Assessment of the efficiency of local refinement in reducing the
discretisation error}

To assess the various local refinement strategies, some measure of the
discretisation error must be plotted against the number of CVs of the grid in
each case. The best strategy would be that which achieves the greatest
discretisation error reduction with the smallest number of CVs. When the
discretisation error is plotted against the number of CVs, the curve with the
steepest slope corresponds to the most efficient refinement strategy. Three
kinds of such plots are included in Section \ref{sec:results}:

\begin{enumerate}
 \item The discretisation error is plotted against the total number of CVs at a
point where the grid is locally dense. \label{enum:1a}
 \item The same as above, but for a point where the grid is locally coarse.
\label{enum:1b}
 \item The norm $\lVert\epsilon_h\rVert_1$ of the whole discretisation error
in the domain is plotted against the number of CVs. This norm is defined as:
\label{enum:1c}
\end{enumerate}

\begin{equation} \label{eq:e_norm_1}
 \lVert\epsilon_h\rVert_1 \;=\; \frac{1}{\Omega} \sum_{i=1}^K |\epsilon_{h,i}|
\cdot \delta\Omega_{h,i}
\end{equation}
where $\Omega$ is the total volume of the computational domain, and
$\epsilon_{h,i}$ and $\delta\Omega_{h,i}$ are the discretisation error and
volume of CV $i$ of grid $h$.

In case (\ref{enum:1a}) the discretisation error reduction is partly or mostly
due to grid refinement in the vicinity of the particular point. In case
(\ref{enum:1b}) the reduction is mostly due to grid refinement at distant
regions from where the discretisation error is transported to that particular
point. In case (\ref{enum:1c}) the effect on the whole domain is investigated.
Depending on the actual application, one may either be interested in achieving
high accuracy in a limited region of interest, or in globally achieving high
accuracy throughout the domain.

For cases (\ref{enum:1a}) and (\ref{enum:1b}) the points are selected among
those listed in Tables \ref{table:vert_centreline} and
\ref{table:horiz_centreline}, and the discretisation error is calculated based
on the values displayed in the tables. For case (\ref{enum:1c}) things are more
complex because the ``exact'' solution must be calculated at the centre of each
CV of the locally refined grid, as required by equation (\ref{eq:e_norm_1}).
The following approach was used: First, the 2048$\times$2048 solution was
restricted to the 1024$\times$1024 grid using a third-order accurate
restriction operator which will be described below. This restricted solution
together with the 1024$\times$1024 solution were used to obtain the ``exact''
solution on grid 1024$\times$1024 via Richardson extrapolation assuming
second-order convergence. The solution obtained on each composite grid was
compared against this ``exact'' solution to calculate the discretisation error
on each CV of the composite grid. Since on cell-centred grids, such as those
used in the present study, the CV centres of different levels do not coincide,
the ``exact'' values at the centres of the CVs of the composite grids had to be
interpolated from the values at the 1024$\times$1024 grid in general. The error
introduced by this interpolation should be smaller than the discretisation
error, otherwise the interpolated values cannot be considered as ``exact''.
Therefore, since a second-order accurate finite volume method was used here, a
third-order accurate interpolation operator was chosen which will now be
described.

\begin{figure}[tpb]
 \centering
 \includegraphics[scale=1]{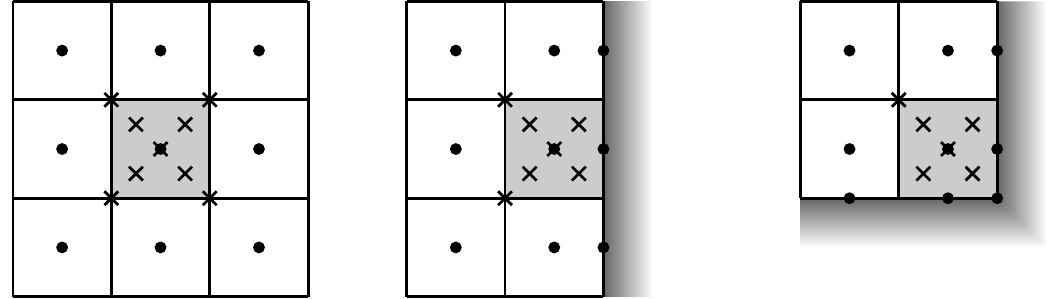}
 \caption{Interpolation stencils. Possible interpolation locations are marked
with $\times$. CV centres or boundary face centres or corner points of the
1024$\times$1024 grid are marked with \textbullet. The CV which contains the
interpolation point is shown shaded. See text for more details.}
 \label{fig:interpolation_stencils}
\end{figure}

Suppose the centre of a CV of a composite grid, where the ``exact'' solution is
to be interpolated (one of the $\times$ points shown in Figure
\ref{fig:interpolation_stencils}). First, a CV of the 1024$\times$1024 grid
which contains the given point is identified (shown shaded in Figure
\ref{fig:interpolation_stencils}). Actually, the following three situations are
possible: If the composite grid CV is of a level coarser than the finest level 
of the 1024$\times$1024 grid then its centre will coincide with one of the corners of
the containing CV (such a point also has alternative container CVs); if it is of
the same level as the finest level of the 1024$\times$1024 grid then its centre will
coincide with the centre of the containing CV; and if it is of a finer level (equal
to the finest level of the 2048$\times$2048 grid) then it will lie in the interior
of the containing CV. The number of CVs of the composite grid may be large, but
locating the container CV for each of them is not expensive because it is performed
using a tree algorithm: First a container CV of the coarsest level is found, then
the container CV of the immediately finer level is sought among that CVs children,
and so on until a container CV of the 1024$\times$1024 level is obtained.

The neighbours of the containing CV are also identified and their centres are
recorded (shown as black dots in the left part of Figure
\ref{fig:interpolation_stencils}). If the containing CV is a boundary CV or a
corner CV then some of the neighbours will be missing, in which case the
corresponding boundary face centres or corner point are selected instead (also
marked as black dots in Figure \ref{fig:interpolation_stencils} middle and
right parts respectively). In any case, we have recorded the centre of
the containing CV $(x_1,y_1)$ and the 8 neighbouring points $(x_2,y_2)$ to
$(x_9,y_9)$. The ``exact'' values $\phi_1, \phi_2, \ldots, \phi_9$ are also
known at these points, where $\phi$ is any of the flow variables - either by
Richardson extrapolation or from the boundary conditions. Suppose also
that $(x_0,y_0)$ is the point where we want to interpolate the function.

The exact solution is reconstructed from these values in the vicinity of these
points by a quadratic polynomial of the form:

\begin{equation} \label{eq:interpolation_o3}
 \phi(\check{x}, \check{y}) \;=\; c_1 \;+\; c_2\check{x} \;+\; c_3\check{y}
\;+\; c_4\check{x}^2 \;+\; c_5\check{x}\check{y} \;+\; c_6\check{y}^2
\end{equation}
where $\check{x}=x-x_0$ and $\check{y}=y-y_0$. The coefficients $c_i$ are
selected such that the values of the polynomial at the 9 neighbouring points
are as close as possible to the ``exact`` values $\phi_1$, $\phi_2$ etc.
Therefore the coefficients $c_i$ should satisfy the following linear system
$A\cdot c = b$ if possible:

\begin{equation}
 \begin{bmatrix}
  \;\; 1 \;\; & \;\;\check{x}_1\;\; & \;\;\check{y}_1\;\; &
\;\;\check{x}_1^2\;\; &
\check{x}_1\check{y}_1 & \;\;\check{y}_1^2\;\; \\
   1 & \check{x}_2 & \check{y}_2 & \check{x}_2^2 & \check{x}_2\check{y}_2 & 
\check{y}_2^2 \\
   \vdots & \vdots & \vdots & \vdots & \vdots & \vdots \\
   \vdots & \vdots & \vdots & \vdots & \vdots & \vdots \\
  1 & \check{x}_9 & \check{y}_9 & \check{x}_9^2 & \check{x}_9\check{y}_9 &
\check{y}_9^2
 \end{bmatrix}
\cdot \begin{matrix} \begin{bmatrix} c_1 \\ c_2 \\ \vdots \\ c_6 \end{bmatrix}
\\ {} \end{matrix} \;=\;
\begin{bmatrix}
 \phi_1 \\ \phi_2 \\ \vdots \\ \vdots \\ \phi_9
\end{bmatrix}
\end{equation}
\smallskip

This system has 9 equations but only 6 unknowns, and is therefore
overdetermined. It is therefore not possible in general for the polynomial to
return exact values at the 9 neighbouring points, but it is possible to minimise
the error by solving the above system in the least squares sense, i.e. to find
$c$ for which the norm $\lVert b-Ac \rVert_2$ is minimised. This is achieved by
computing the reduced QR factorisation of A, $A=QR$ (we used the modified
Gram-Schmidt algorithm) and then solving the 6$\times$6 upper triangular system
$Rc=Q^Tb$ for $c$. For more details see \cite{Trefethen}. The relative
coordinates $\check{x}$, $\check{y}$ are used rather than the actual
coordinates $x$, $y$, because otherwise the column vectors of $A$ are nearly
parallel as the 8 points are very close to each other, and the $QR$
factorisation is not accurate. This interpolation operator introduces an error
of the order $O(h^3)$ as we have verified with simple numerical experiments (not
included).

This interpolation operator was also used for the restriction of the
2048 $\times$ 2048 solution to the 1024 $\times$ 1024 grid, in order to perform
the Richardson extrapolation. In this case, the roles of the 2048 $\times$ 2048
and 1024 $\times$ 1024 grids were interchanged, and the grid shown in Figure
\ref{fig:interpolation_stencils} would be the 2048 $\times$ 2048 grid. The
centres of the 1024 $\times$ 1024 CVs would be those points marked with
$\times$ that lie at the corners of the CVs of the 2048 $\times$ 2048 grid.
Therefore, each 1024 $\times$ 1024 CV centre has 4 possible container CVs on
the 2048 $\times$ 2048 grid. For maximum accuracy, the restriction of the
variables was repeated for each of these 4 possible container CVs, and the
``exact'' value was taken as the average of the four interpolated values.

\subsection{Assessment of the effect of local refinement on the discretisation
error distribution}

Even if local refinement is capable of efficiently reducing the mean
discretisation error this does not necessarily mean that this error may not be
large locally. Therefore, another useful property of local refinement would be
its ability to smooth out the discretisation error distribution. Figure
\ref{fig:tau} shows that on structured grids there are huge variations in
truncation error, by several orders of magnitude. Local refinement focuses on
regions of high truncation error in an effort to reduce it and therefore the
truncation error distributions become smoother on composite grids. It would be
interesting to examine if a similar effect takes place as far as the
discretisation error is concerned.

Discretisation error distributions cannot be compared directly but they have to
be normalised first by dividing them by their mean (given by
(\ref{eq:e_norm_1})). So we define the normalised discretisation error as:

\begin{equation} \label{eq:e_normalised}
 \epsilon_h^* \;=\; \frac{1}{\lVert\epsilon_h\rVert_1} \cdot \epsilon_h
\end{equation}

The main property of $\epsilon_h^*$ which is of interest here is that the
integral of its absolute value in the domain is independent of the order of
magnitude of the actual discretisation error:

\begin{align}
 \int_{\Omega} |\epsilon^*| \,d\Omega \;&=\;
 \sum_{i=1}^K |\epsilon_{h,i}^*| \!\cdot\! \delta\Omega_{h,i} \;=\;
 \sum_{i=1}^K \frac{|\epsilon_{h,i}|}{\lVert \epsilon_h \rVert_1} \!\cdot\!
   \delta\Omega_{h,i} \;=\; \nonumber
\\ \label{eq:e_normalised_integrated} &=\;
 \frac{1}{\lVert \epsilon_h \rVert_1} \sum_{i=1}^K |\epsilon_{h,i}| \!\cdot\!
   \delta\Omega_{h,i} \;=\;
 \frac{1}{\lVert \epsilon_h \rVert_1} \!\cdot\! \left( \Omega \!\cdot\! \lVert
\epsilon_h \rVert_1 \right) \;=\; \Omega
\end{align}
due to (\ref{eq:e_norm_1}). In other words if we draw the $\epsilon_h^*(x,y)$
distribution as a surface in 3D space, the volume enclosed between this surface
and the $x-y$ plane always equals the area $\Omega$ of the domain (which equals
1 in the present test case). This makes $\epsilon_h^*$ distributions comparable
to each other not only among different refinement schemes but also between
different Reynolds numbers and different flow variables ($u$, $v$, $p$).

\begin{figure}[tpb]
 \centering
\noindent\makebox[\textwidth]{
\includegraphics[scale=1]{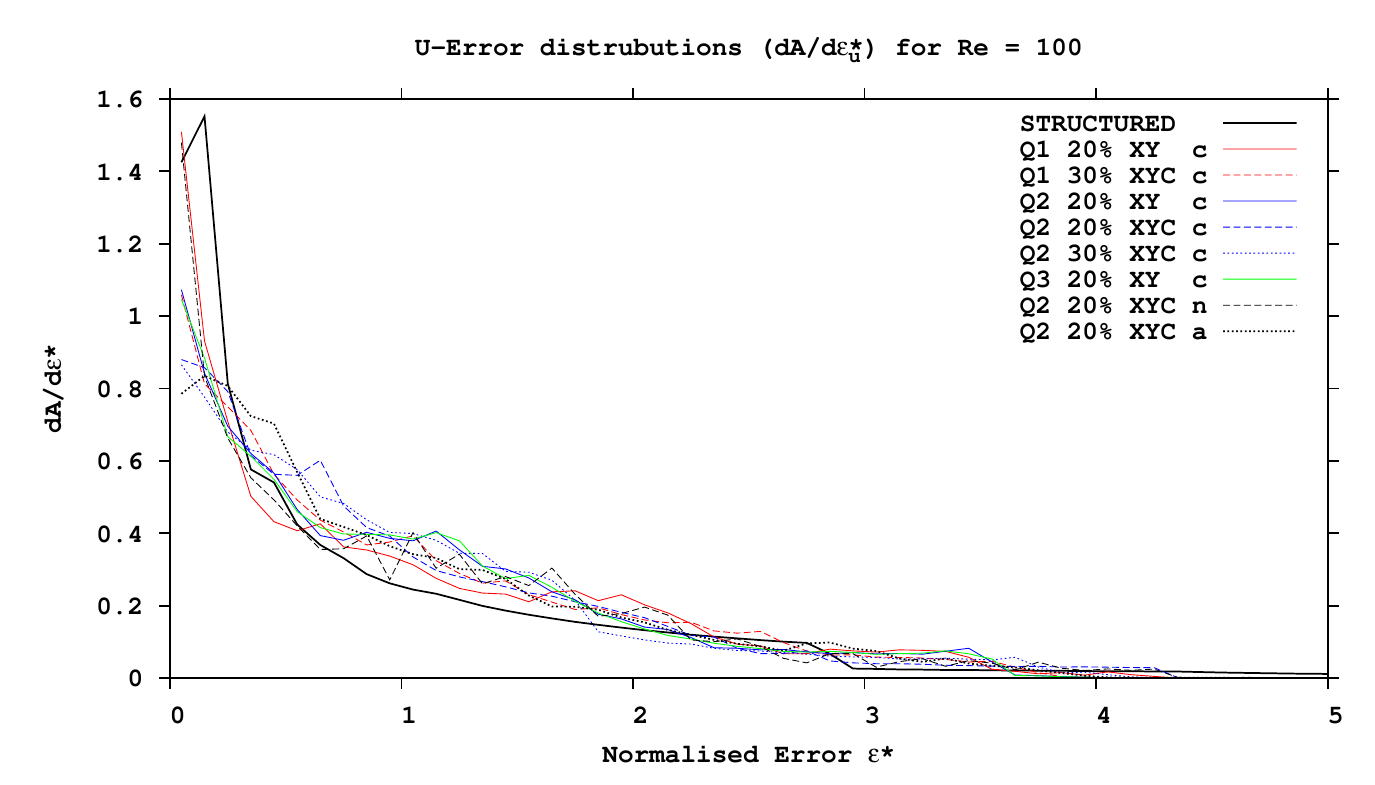}}
 \caption{Distributions of absolute normalised $u$-error on the finest grids, Re
= 100.}
 \label{fig:eun_distributions_100}
\end{figure}

Different $\epsilon_h^*$ distributions are compared graphically by plotting
$\frac{dA(\epsilon^*)}{d\epsilon^*}$ against $\epsilon^*$ (see Figures
\ref{fig:eun_distributions_100}, \ref{fig:eun_distributions_1000} and
\ref{fig:eun_distributions_10000}, which are discussed in the next Section). Here
$A(\epsilon^*)$ is the total area of the domain where the absolute value of the
normalised discretisation error is less than or equal to $\epsilon^*$. To
calculate $\frac{dA(\epsilon^*)}{d\epsilon^*}$, first the interval $[0,
\epsilon_{max}^*]$ is split into a number of bins of size 0.1 each,
$\epsilon_{max}^*$ being the maximum absolute value of normalised
discretisation error in the domain (bin 1 is the interval $[0,0.1)$, bin 2 is
the interval $[0.1, 0.2)$ etc.). Then the area of each CV of the grid is added
to the appropriate bin according to the absolute value of the normalised
discretisation error at that CV. Then $\frac{dA(\epsilon^*)}{d\epsilon^*}$ is
calculated at the centre of each bin as the total area of the bin divided by 0.1
(the width of the bin).

The area of the domain where the absolute value of the normalised
discretisation error is in the range $[\epsilon_1, \epsilon_2]$ equals
$\int_{\epsilon_1}^{\epsilon_2} \frac{dA(\epsilon^*)}{d\epsilon^*} d\epsilon^*$.
In a graph such as that of Figure \ref{fig:eun_distributions_100} that would
equal the area below the curve and above the $\epsilon$-axis, between
$\epsilon_1$ and $\epsilon_2$. The area below the whole curve equals the area of
the whole domain ($\int_0^{\epsilon^*_{max}} \frac{dA(\epsilon^*)}{d\epsilon^*}
d\epsilon^* = \Omega = 1$. For a different test case with $\Omega \neq 1$ one
can use $\frac{1}{\Omega} \frac{dA(\epsilon^*)}{d\epsilon^*}$ instead of
$\frac{dA(\epsilon^*)}{d\epsilon^*}$).

The mean value of $|\epsilon^*|$ in the domain, $\frac{1}{\Omega} \int_{\Omega}
|\epsilon^*| \,d\Omega$, equals 1 as can be seen from
(\ref{eq:e_normalised_integrated}). If $\epsilon^*$ is relatively uniform within
the domain then its distribution would be concentrated around $\epsilon^*=1$.
Otherwise, if it exhibits large variations, then its distribution would be more
spread out.

Also, the variation of $|\epsilon^*|$ in each case is quantitatively assessed
using certain metrics (Tables \ref{table:statistics_100},
\ref{table:statistics_1000} and \ref{table:statistics_10000}). The first metric
is $\epsilon^*_{max}$. However, this maximum value may be just an outlier and
not be representative of the variation of the distribution. Therefore the 99-th
percentile $\epsilon^*_{99\%}$ is also used. This is defined here as the upper
endpoint of the bin $i$ which is the lowest bin such that the sum of the areas
of all bins from 1 to $i$ is greater than or equal to 99\% of the total area of
the domain. In other words, $\epsilon^*_{99\%}$ is roughly such that the error
$|\epsilon^*|$ is less than $\epsilon^*_{99\%}$ in 99\% of the domain. Finally,
the standard deviation is also used to quantify the variation of $\epsilon^*$:

\begin{equation}
 \sigma^* \;=\; \sqrt{\frac{1}{\Omega} \sum_{i=1}^{K} \delta \Omega_{h,i}
   \!\cdot\! ( |\epsilon_{h,i}^*| - \epsilon^*_m )^2 } \;=\;
 \sqrt{\frac{1}{\Omega} \sum_{i=1}^{K} \delta \Omega_{h,i} \!\cdot\!
   ( |\epsilon_{h,i}^*| - 1 )^2 }
\end{equation}
where $\epsilon^*_m=1$ is the mean value of $|\epsilon^*|$ in the domain.

\section{Results and discussion} \label{sec:results}

The results of testing eight different local refinement schemes for each
Reynolds number are presented in this Section. These eight schemes will be
referred to under the following descriptions:

\medskip 
\begin{small}   
\renewcommand\arraystretch{0.8}   
\begin{tabular}{ll}
\textbf{1} & \texttt{Q1 20\% XY \ c}\\
\textbf{2} & \texttt{Q1 30\% XYC c}\\
\textbf{3} & \texttt{Q2 20\% XY \ c}\\
\textbf{4} & \texttt{Q2 20\% XYC c}\\
\textbf{5} & \texttt{Q2 30\% XYC c}\\
\textbf{6} & \texttt{Q3 20\% XY \ c}\\
\textbf{7} & \texttt{Q2 20\% XYC n}\\
\textbf{8} & \texttt{Q2 20\% XYC a}
\end{tabular}	
\end{small}
\medskip

The descriptions are interpreted as follows: Q1, Q2 and Q3 denote the three
local refinement criteria defined in Section \ref{sec:criteria} (Q1: truncation
error; Q2: truncation error times CV volume; Q3: truncation error divided by
main diagonal coefficient). Next comes the percentage R\% of CVs that will be
refined. That is, as described in Section
\ref{sec:refinement_general_procedure}, when a grid is to be refined the
criterion quantity is calculated for every CV, the CVs are sorted in descending
order according to the magnitude of the criterion quantity, and the first R\% of
this list are refined. This is repeated for all selected equations, which are
stated next in the above descriptions: XY means that only the momentum equations
are used, while XYC means that the continuity equation is also used. The last
character given in the descriptions describes the treatment at grid level
interfaces: ``c'' means that only coarse CVs may be refined, ``n'' means that no
refinement is allowed at all up to 4 CVs away from such interfaces, and ``a''
means that all CVs may be refined, i.e. there are no restrictions. See Section
\ref{sec:boundary treatment} for more details.

In every case, the problem was first solved on a 32 $\times$ 32 grid (shown in
Figure \ref{fig:probes}). Then the truncation error was calculated, and the
grid was refined according to the selected refinement scheme to produce a
composite grid. Then the refinement cycle was repeated on this composite grid
(solution, truncation error estimate, refinement) to produce yet another
composite grid. In fact this procedure was repeated six times. In every case
the CVs of the finest level of the last (sixth) composite grid had the same size
as those of the 2048 $\times$ 2048 structured grid.

\begin{figure}[tpb]
\centering
\noindent\makebox[\textwidth]{
 \subfigure[{Q1, Re = 100}] {\label{grid_Q1_20_XYc_Re_100}
  \includegraphics{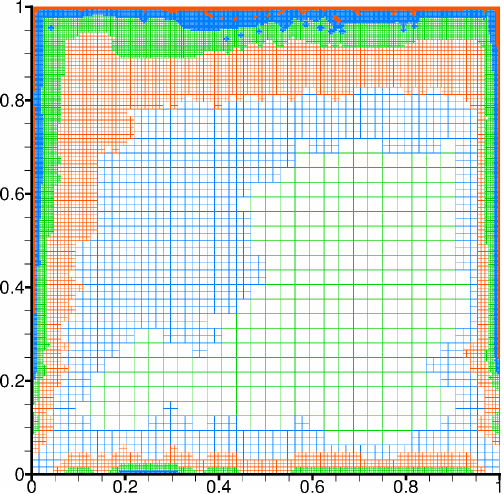}}
 \subfigure[{Q2, Re = 100}] {\label{grid_Q2_20_XYc_Re_100}
  \includegraphics{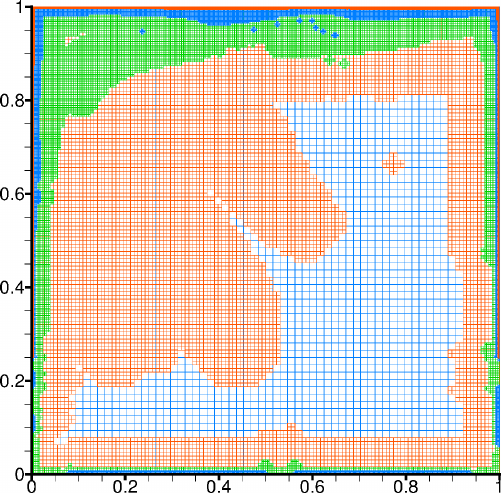}}
 \subfigure[{Q3, Re = 100}] {\label{grid_Q3_20_XYc_Re_100}
  \includegraphics{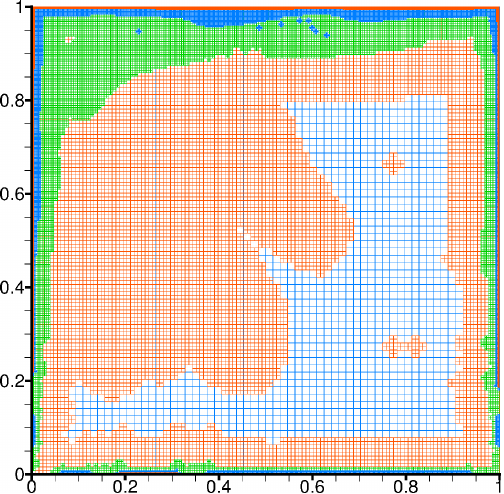}}
}
\noindent\makebox[\textwidth]{
 \subfigure[{Q1, Re = 1000}] {\label{grid_Q1_20_XYc_Re_1000}
  \includegraphics{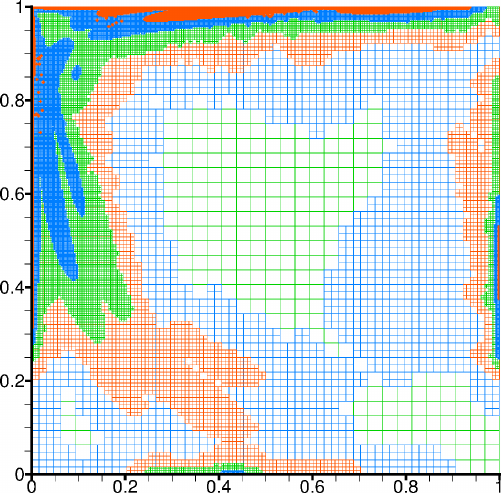}}
 \subfigure[{Q2, Re = 1000}] {\label{grid_Q2_20_XYc_Re_1000}
  \includegraphics{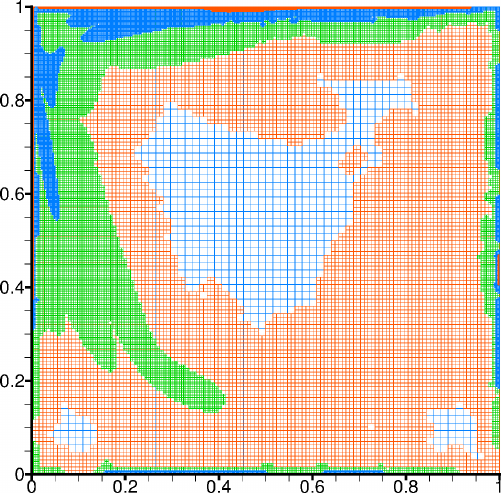}}
 \subfigure[{Q3, Re = 1000}] {\label{grid_Q3_20_XYc_Re_1000}
  \includegraphics{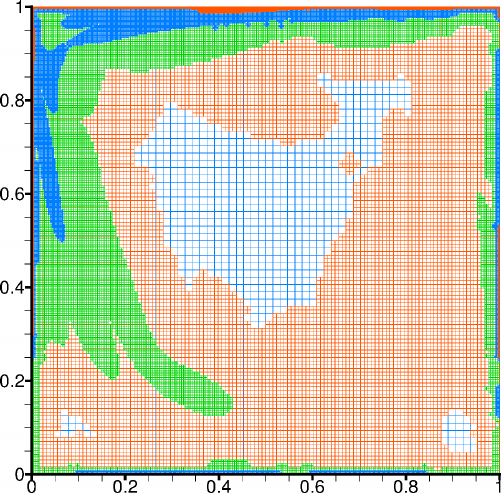}}
}
\noindent\makebox[\textwidth]{
 \subfigure[{Q1, Re = 10000}] {\label{grid_Q1_20_XYc_Re_10000}
  \includegraphics{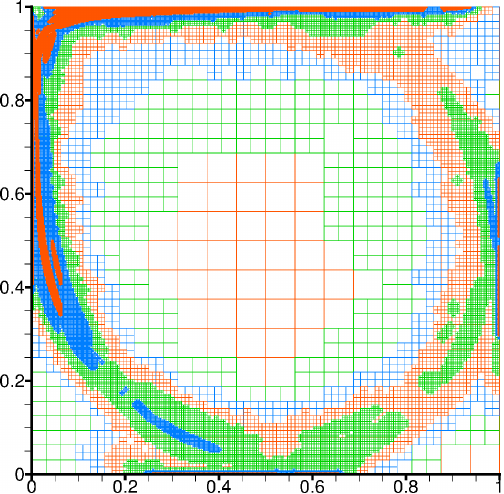}}
 \subfigure[{Q2, Re = 10000}] {\label{grid_Q2_20_XYc_Re_10000}
  \includegraphics{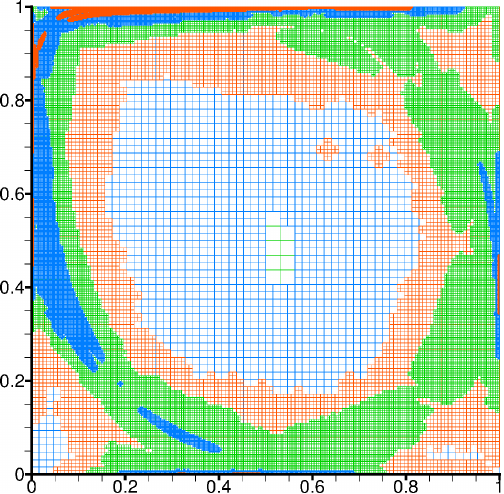}}
 \subfigure[{Q3, Re = 10000}] {\label{grid_Q3_20_XYc_Re_10000}
  \includegraphics{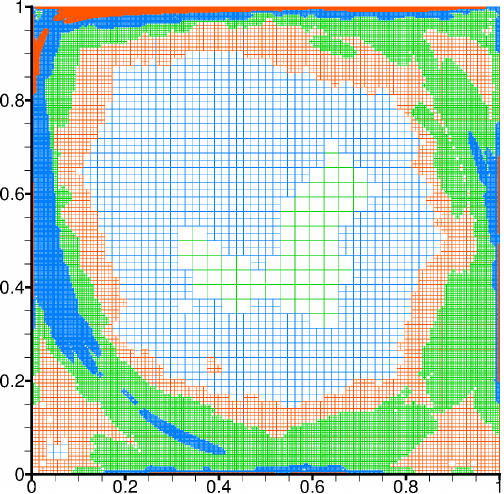}}
}
\caption{The finest grids for cases: Q1 20\% XYc (left column), Q2 20\% XYc
(middle column), and Q3 20\% XYc (right column) (actually the underlying grids
$2h$ are shown). Re = 100 (top), Re = 1000 (middle), Re = 10000 (bottom).
Different colours help to distinguish between levels.}
\label{fig:grids_criteria}
\end{figure}

\begin{figure}[tpb]
\centering
\noindent\makebox[\textwidth]{
 \subfigure[{Re = 100}] {\label{grid_Q2_20_XYCc_Re_100}
  \includegraphics{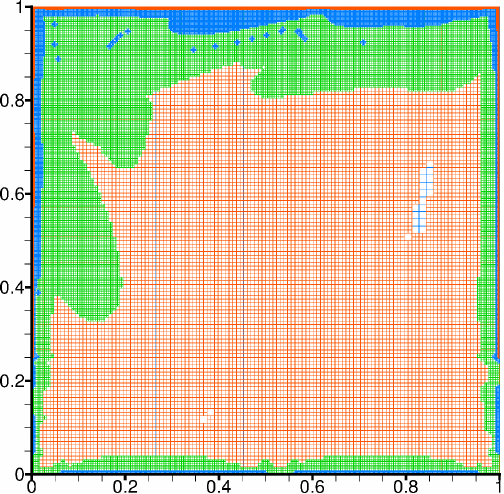}}
 \subfigure[{Re = 1000}] {\label{grid_Q2_20_XYCc_Re_1000}
  \includegraphics{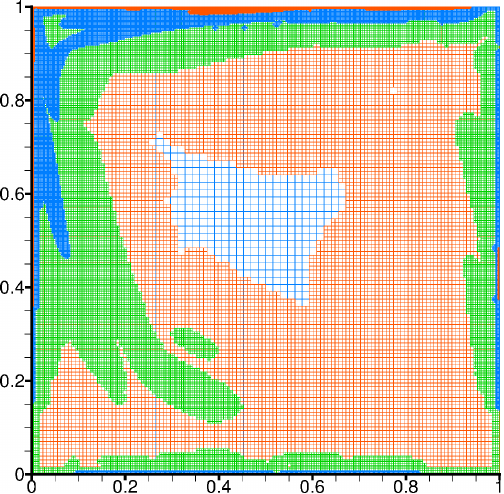}}
 \subfigure[{Re = 10000}] {\label{grid_Q2_20_XYCc_Re_10000}
  \includegraphics{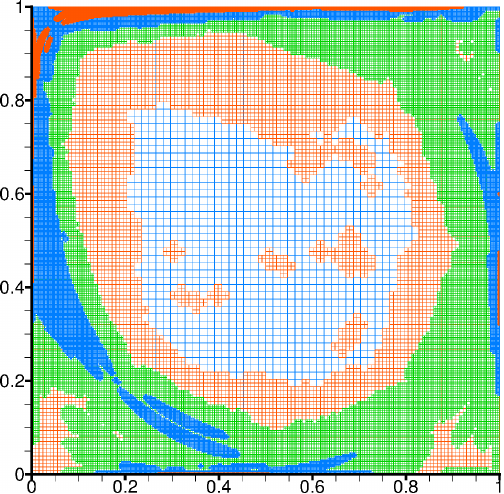}}
}
\caption{The finest grids for case Q2 20\% XYCc (actually the underlying grids
$2h$ are shown). Re =100 (left), Re = 1000 (middle), Re = 10000 (right).}
\label{fig:grids_continuity}
\end{figure}

\begin{figure}[htpb]
\centering
 \subfigure[Q2 20\% XYC n, Re = 1000] {\label{grid_Q2_20_XYCn_Re_1000}
  \includegraphics{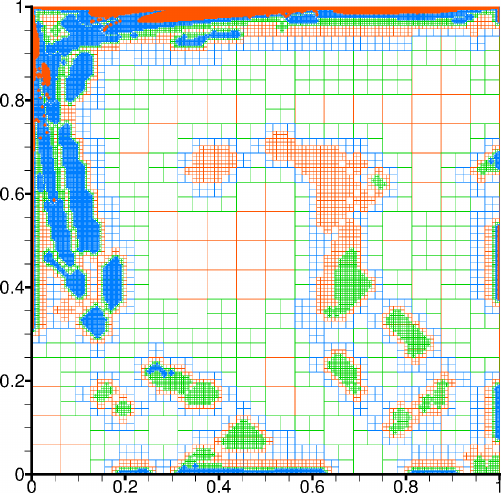}}
 \subfigure[Q2 20\% XYC a, Re = 1000] {\label{grid_Q2_20_XYCa_Re_1000}
  \includegraphics{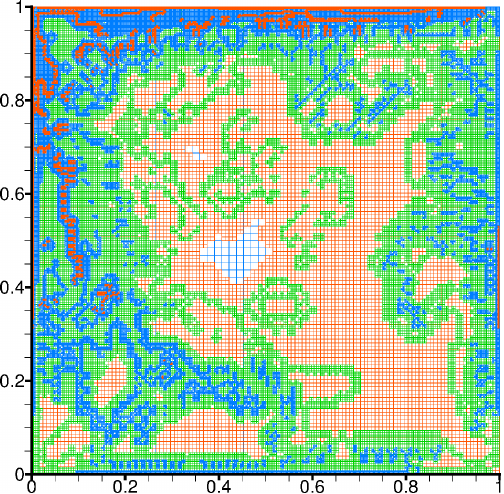}}
\caption{The finest grids for cases Q2 20\% XYCn (left) and Q2 20\% XYCa
(right), Re~=~1000 (actually the underlying grids are shown).}
\label{fig:grids_na}
\end{figure}

Six successive composite grids, times eight refinement schemes, times three
Reynolds numbers produces a large total number of composite grids which are
impossible to display here. Some of the grids are shown in Figures
\ref{fig:grids_criteria}, \ref{fig:grids_continuity} and \ref{fig:grids_na}. In
every case the finest of the series of six grids is shown, and in fact it is the
underlying grid $2h$ which is shown, for clarity.

\begin{figure}[tpb]
 \centering
 \includegraphics[scale=1]{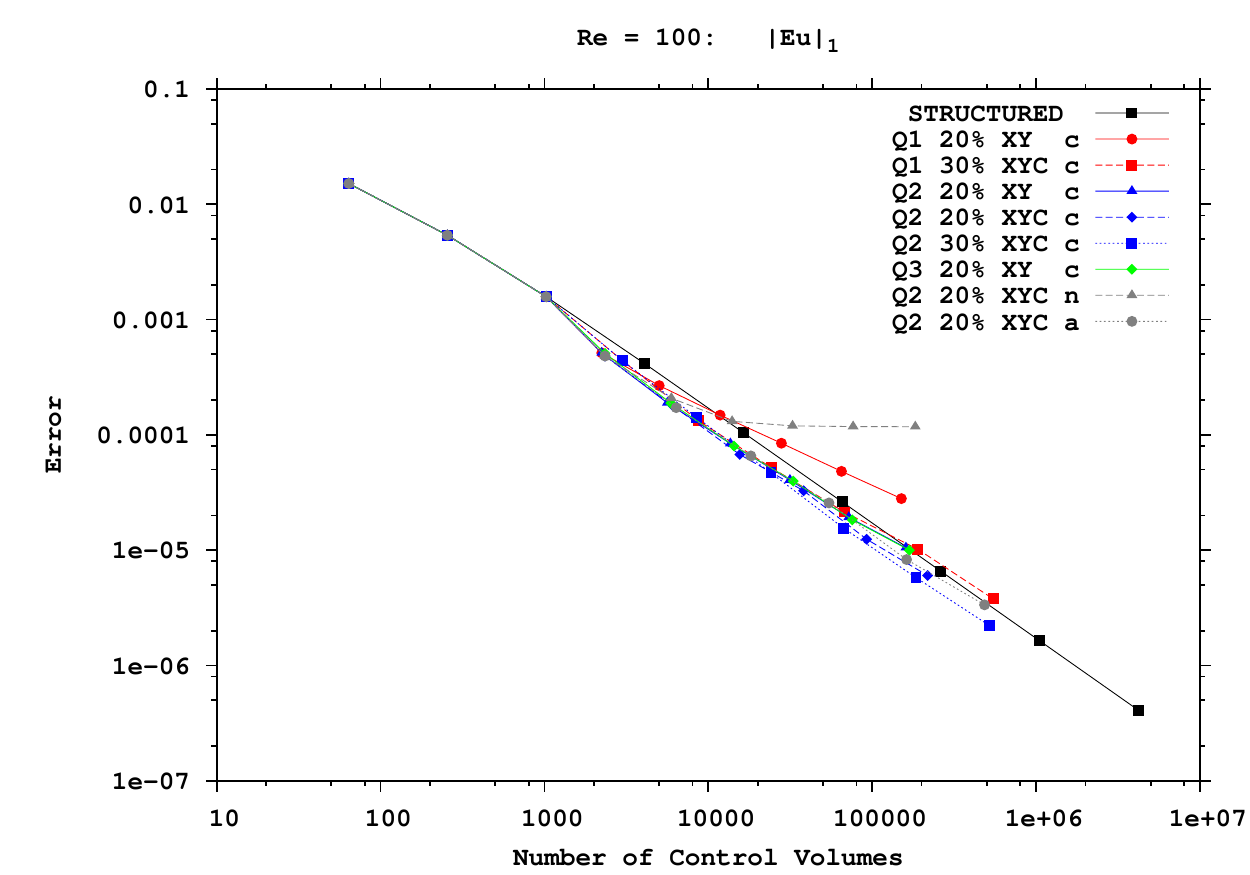}
 \caption{Convergence of the $\lVert . \rVert_1$ norm (\ref{eq:e_norm_1}) of
the $u-$discretisation error with grid refinement for Re = 100.}
 \label{fig:cnv_Re_100_u}
\end{figure}

\begin{figure}[tpb]
 \centering
 \includegraphics[scale=1]{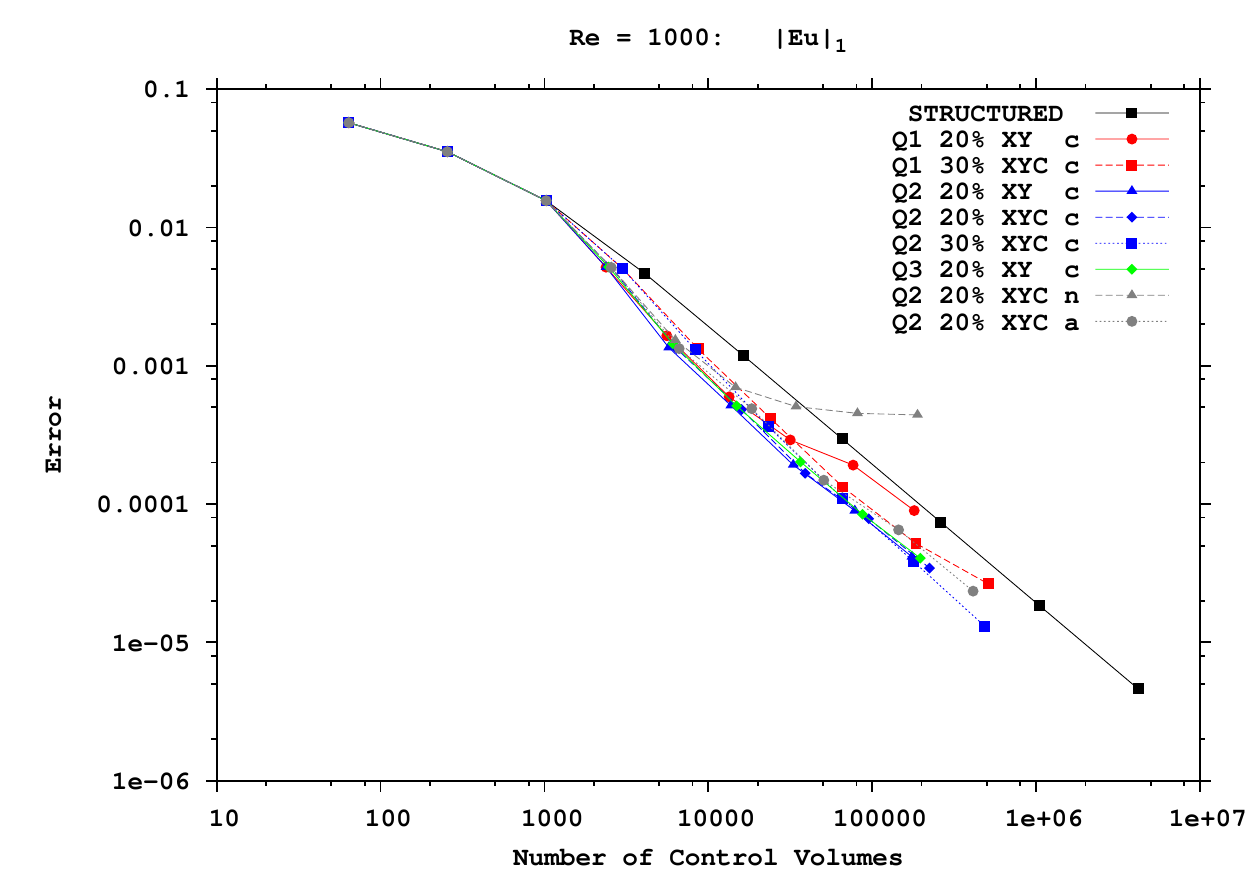}
 \caption{As for Figure \ref{fig:cnv_Re_100_u} but for Re = 1000.}
 \label{fig:cnv_Re_1000_u}
\end{figure}

\begin{figure}[tpb]
 \centering
 \includegraphics[scale=1]{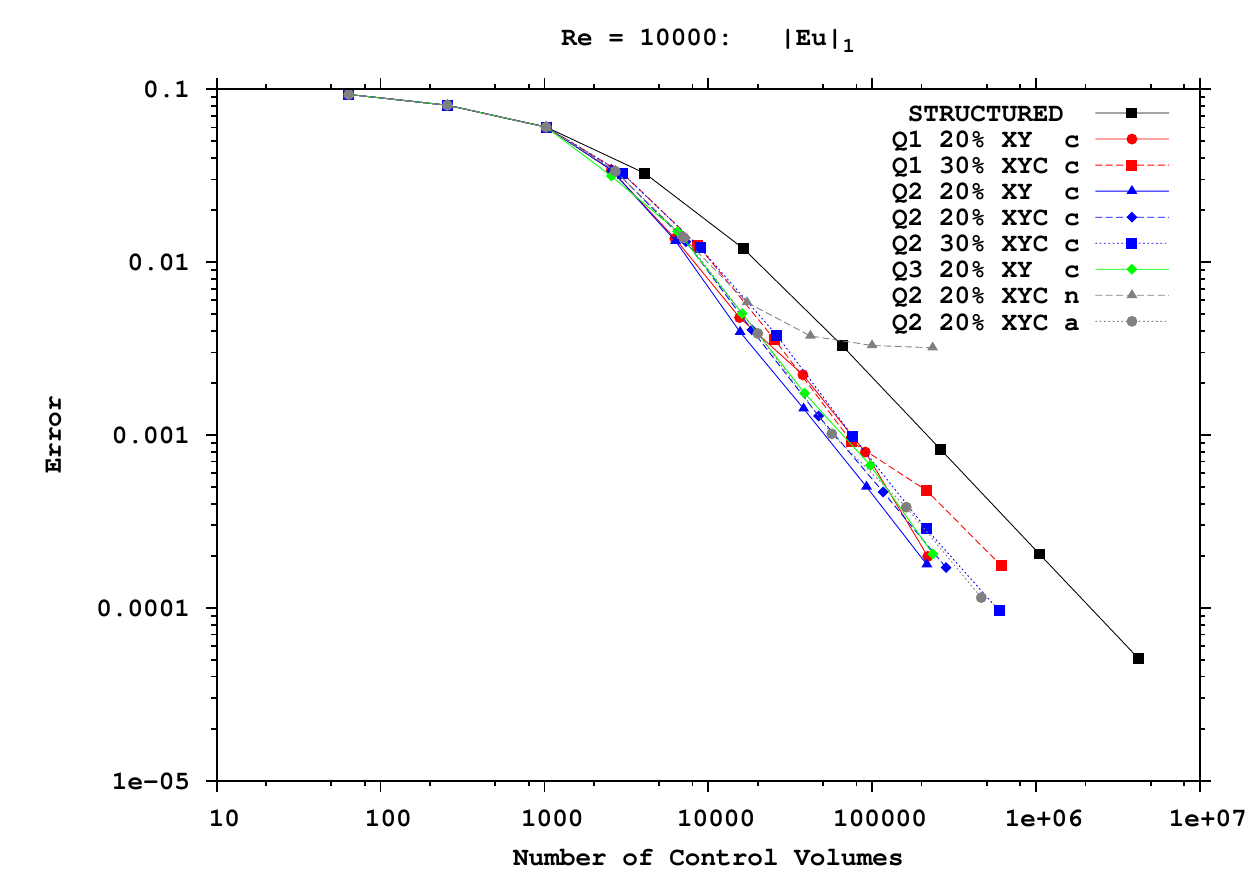}
 \caption{As for Figure \ref{fig:cnv_Re_100_u} but for Re = 10000.}
 \label{fig:cnv_Re_10000_u}
\end{figure}

\begin{figure}[tpb]
\centering
\noindent\makebox[\textwidth]{
 \subfigure {\label{cnv_Re_100_coarse}
  \includegraphics[height=6cm, keepaspectratio=true]
                  {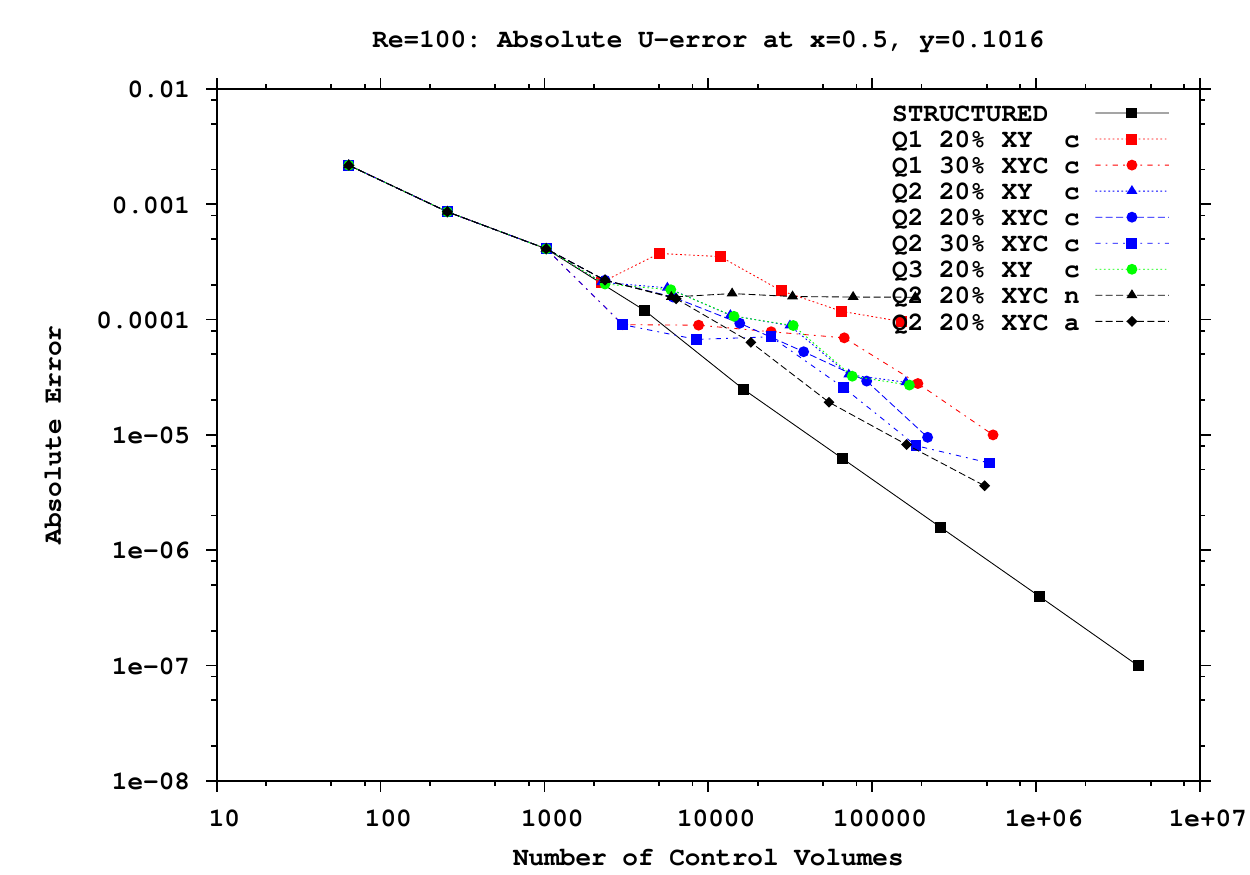}}
 \subfigure {\label{cnv_Re_100_fine}
  \includegraphics[height=6cm, keepaspectratio=true]
                  {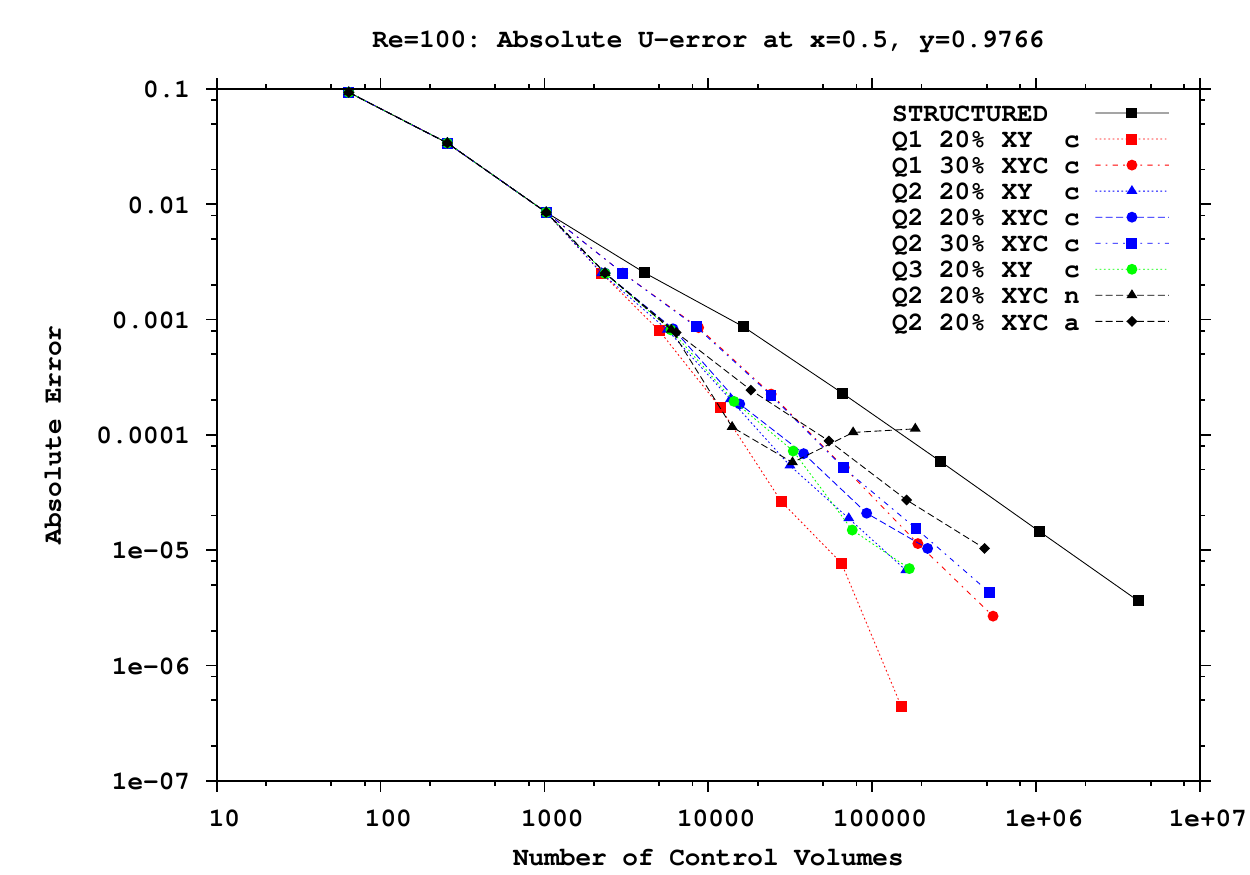}}
}
\noindent\makebox[\textwidth]{
 \subfigure {\label{cnv_Re_1000_coarse}
  \includegraphics[height=6cm, keepaspectratio=true]
                  {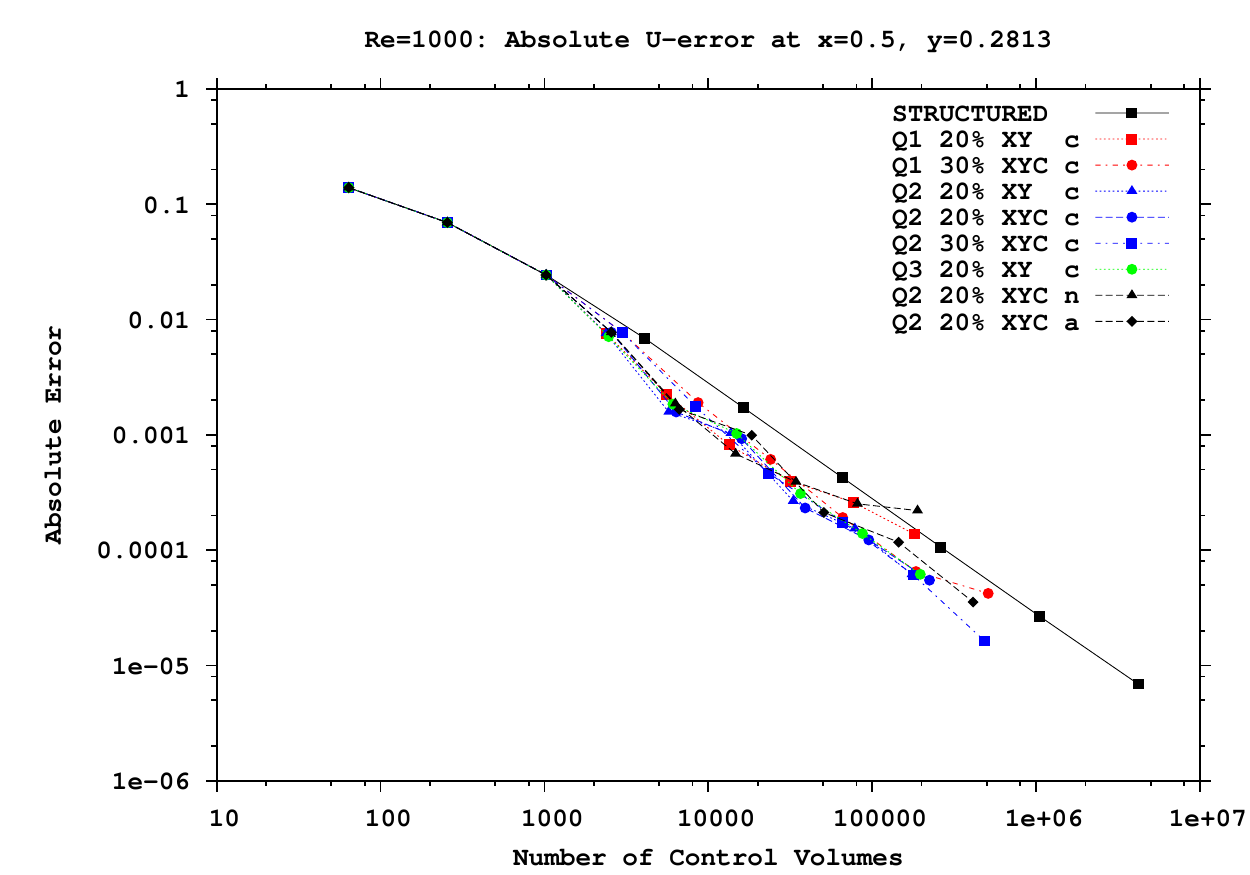}}
 \subfigure {\label{cnv_Re_1000_fine}
  \includegraphics[height=6cm, keepaspectratio=true]
                  {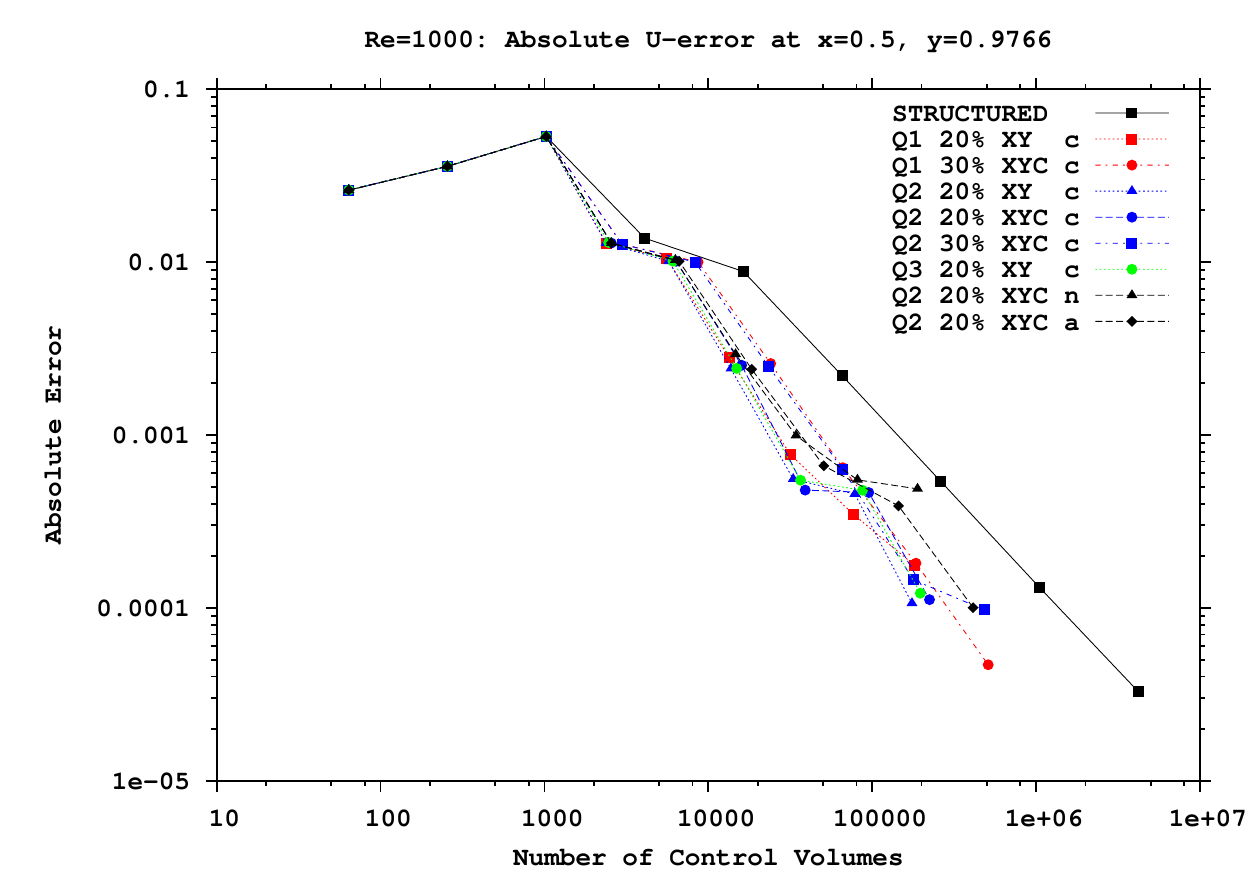}}
}
\noindent\makebox[\textwidth]{
 \subfigure {\label{cnv_Re_10000_coarse}
  \includegraphics[height=6cm, keepaspectratio=true]
                  {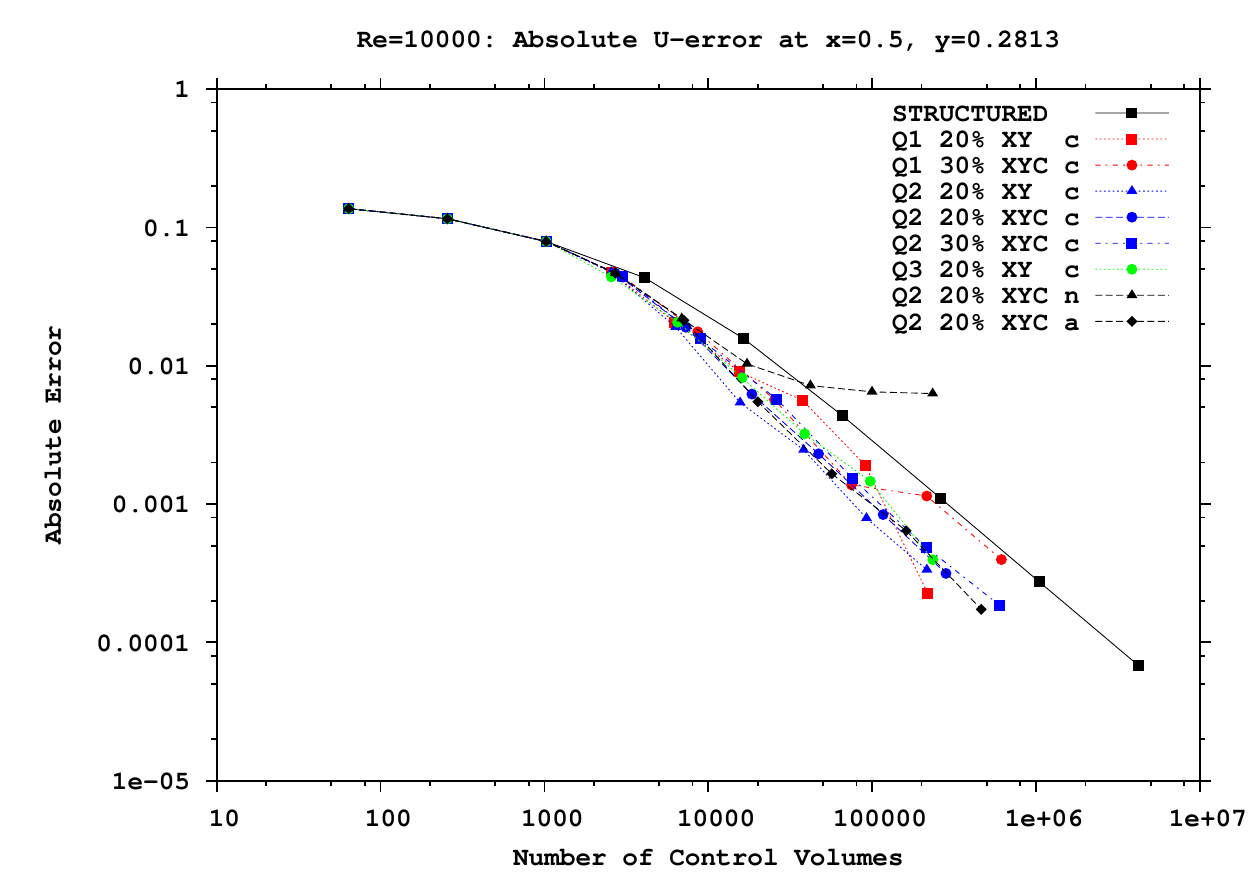}}
 \subfigure {\label{cnv_Re_10000_fine}
  \includegraphics[height=6cm, keepaspectratio=true]
                  {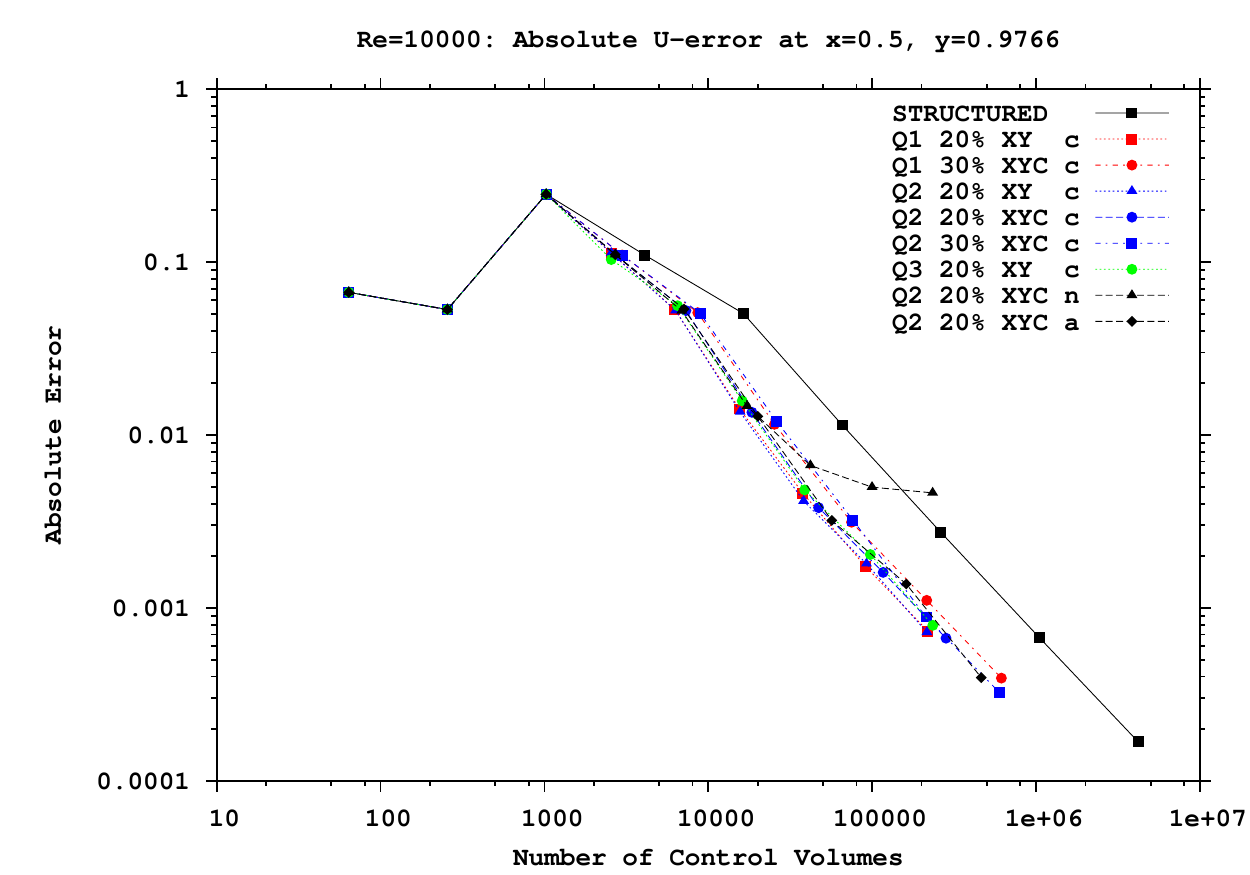}}
}
\caption{Convergence of $|\epsilon_u|$ at specific points, for various cases.
The left and right columns show results at points generally located at coarse
and fine regions of the composite grids respectively.}
\label{fig:convergence_at_points}
\end{figure}

Figures \ref{fig:cnv_Re_100_u} - \ref{fig:cnv_Re_10000_u} show the convergence
of $\lVert \epsilon_u \rVert_1$, the norm (\ref{eq:e_norm_1}) of the
discretisation error of $u$, with grid refinement for the selected Reynolds
numbers. The graphs for the other variables ($v$, $p$) are similar, but in
general mostly results on $u$ are presented here, due to space limitations.
Results for the other variables are included when appropriate. Figure
\ref{fig:convergence_at_points} shows similar graphs but for the absolute value
of the discretisation error of $u$ at specific points. For each Reynolds number
two points were selected among those listed in Table
\ref{table:vert_centreline}: One where the grids are coarse in general (left
graph) and one where the grids are fine in general (the top-most point in fact
- right graph). The results of Figure \ref{fig:convergence_at_points} must be
interpreted with some caution because the reduction of the discretisation error
at a specific point during a refinement cycle may be strongly influenced by
whether the grid was refined or not at that particular location during that
cycle.

Table \ref{table:err_norms_finest} contains results on the norms of the
discretisation errors of the other variables as well, for a limited number of
cases. These cases differ only in the quantity used as a refinement criterion
(Q1, Q2 or Q3) while the other parameters of the refinement schemes are kept the
same, to allow a direct comparison among these quantities.

\begin{table}[pb]
\caption{Error norms on the finest grid for various cases.}
\label{table:err_norms_finest}
\begin{center}
\noindent\makebox[\textwidth]{
\begin{scriptsize}   
\rowcolors{3}{black!20}{white}
\begin{tabular}{ r | c c c | c c c | c c c }
 \hline
    \multicolumn{1}{|c|}{}
  & \multicolumn{3}{ c|}{\textbf{Re = 100}} 
  & \multicolumn{3}{ c|}{\textbf{Re = 1000}}
  & \multicolumn{3}{ c|}{\textbf{Re = 10000}} \\
    \multicolumn{1}{|c|}{}
  & \multicolumn{1}{ c }{$\lVert \epsilon_u \rVert_1$}
  & \multicolumn{1}{ c }{$\lVert \epsilon_v \rVert_1$}
  & \multicolumn{1}{ c|}{$\lVert \epsilon_p \rVert_1$}
  & \multicolumn{1}{ c }{$\lVert \epsilon_u \rVert_1$}
  & \multicolumn{1}{ c }{$\lVert \epsilon_v \rVert_1$}
  & \multicolumn{1}{ c|}{$\lVert \epsilon_p \rVert_1$}
  & \multicolumn{1}{ c }{$\lVert \epsilon_u \rVert_1$}
  & \multicolumn{1}{ c }{$\lVert \epsilon_v \rVert_1$}
  & \multicolumn{1}{ c|}{$\lVert \epsilon_p \rVert_1$} \\
 \hline
Q1 20\% XY c & $2.80\text{E-5}$ & $3.05\text{E-5}$ & $1.28\text{E-5}$ &
               $8.98\text{E-5}$ & $8.55\text{E-5}$ & $3.92\text{E-5}$ &
               $1.98\text{E-4}$ & $2.06\text{E-4}$ & $1.25\text{E-4}$
\\
Q2 20\% XY c & $1.05\text{E-5}$ & $1.05\text{E-5}$ & $1.33\text{E-5}$ &
               $4.16\text{E-5}$ & $3.98\text{E-5}$ & $2.09\text{E-5}$ &
               $1.78\text{E-4}$ & $1.78\text{E-4}$ & $1.20\text{E-4}$
\\
Q3 20\% XY c & $1.00\text{E-5}$ & $1.00\text{E-5}$ & $5.14\text{E-6}$ &
               $4.07\text{E-5}$ & $3.84\text{E-5}$ & $2.04\text{E-5}$ &
               $2.05\text{E-4}$ & $2.04\text{E-4}$ & $1.55\text{E-4}$
\\ \hline
\end{tabular}
\end{scriptsize}}
\end{center}
\end{table}
\bigskip

\begin{figure}[tpb]
 \centering
\noindent\makebox[\textwidth]{
\includegraphics[scale=1]{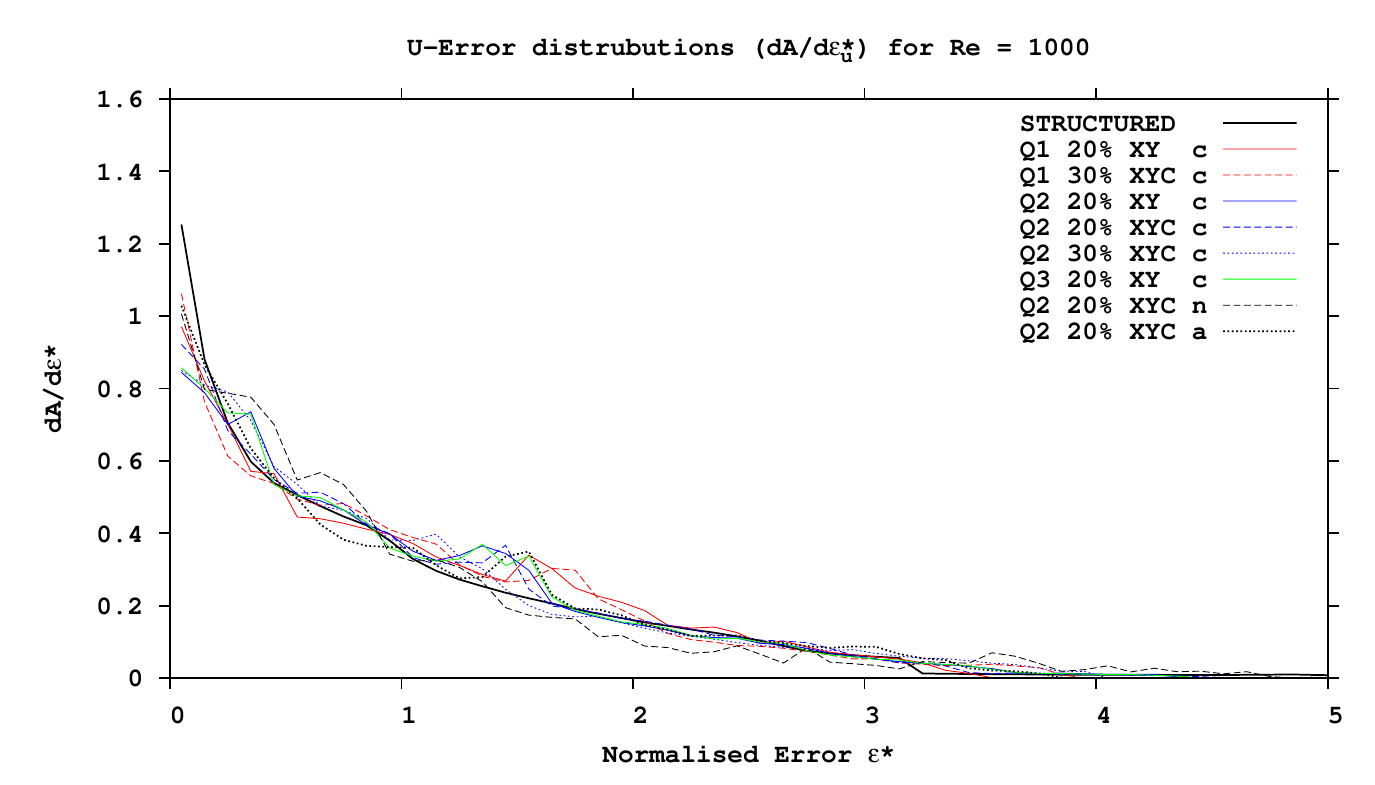}}
 \caption{Distributions of absolute normalised $u$-error on the finest grids, Re
= 1000.}
 \label{fig:eun_distributions_1000}
\end{figure}

\begin{figure}[tpb]
 \centering
\noindent\makebox[\textwidth]{
\includegraphics[scale=1]{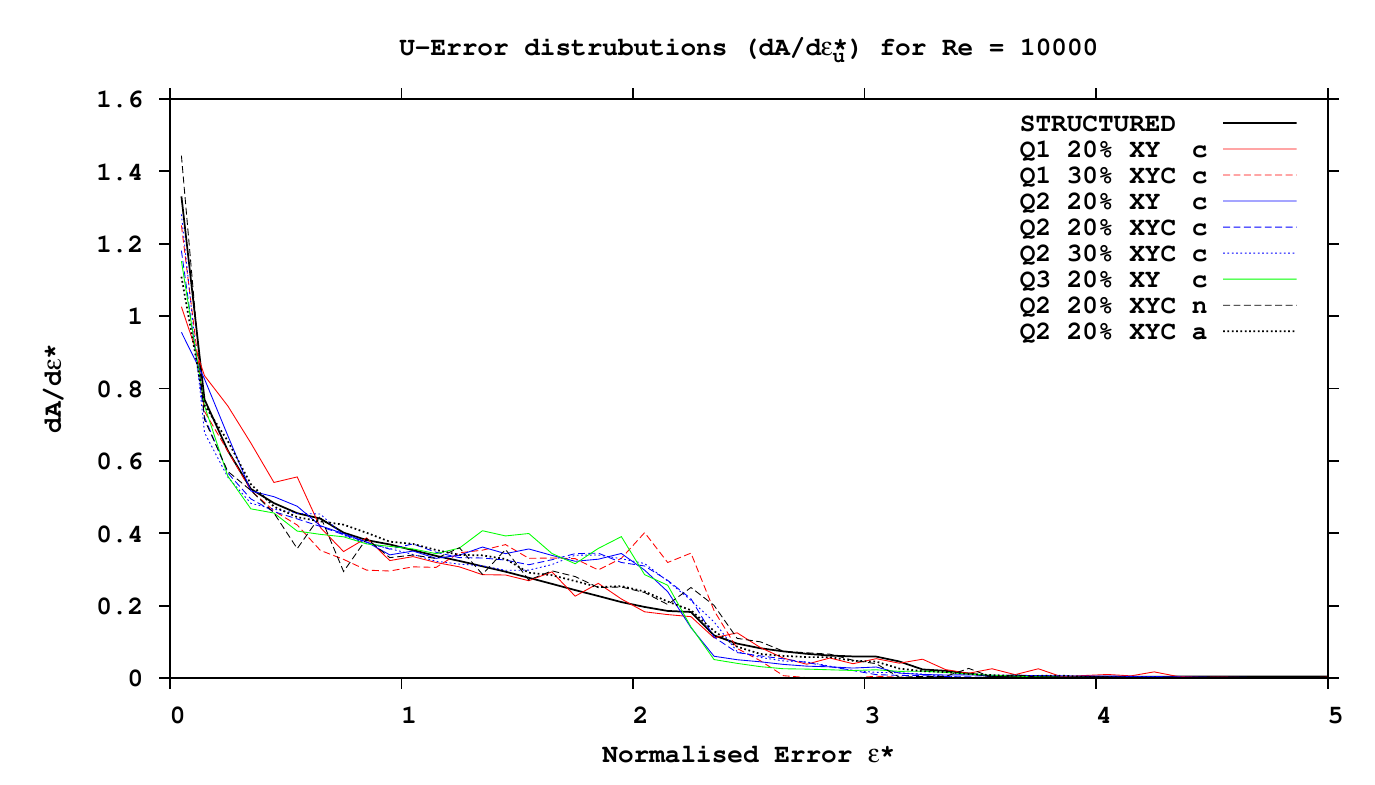}}
 \caption{Distributions of absolute normalised $u$-error on the finest grids, Re
= 10000.}
 \label{fig:eun_distributions_10000}
\end{figure}

To investigate the effect of local refinement on the distribution of
discretisation error across the domain, Figures \ref{fig:errors_un_Re_100},
\ref{fig:errors_un_Re_1000} and \ref{fig:errors_un_Re_10000} depict the
normalised $u$-errors $\epsilon_u^*$ across the domain for Re = 100, 1000 and
10000 respectively, for various cases. Also, Figure \ref{fig:errors_pn} shows
the normalised discretisation error of pressure $\epsilon_p^*$ on the structured
grids (top row) and on composite grids Q2 20\% XYC c (bottom row) for the range
of Reynolds numbers. Then, Figures \ref{fig:eun_distributions_100} -
\ref{fig:eun_distributions_10000} show the statistical distributions of the
$u$-errors, and Tables \ref{table:statistics_100} - \ref{table:statistics_10000}
contain associated metrics.

From this large amount of data, the following observations can be made
concerning the influence of various parameters on the efficiency of local
refinement:

\subsubsection*{Effect of Reynolds number}

Figure \ref{fig:tau} shows that as the Reynolds number increases, the variation
of truncation error becomes greater: Both the regions of high $\tau$ and the
regions of low $\tau$ become larger. The truncation error varies by several
orders of magnitude. For Re = 100 it is high near the lid and low everywhere
else. As the Reynolds number increases, it becomes high also at the outer part
of the main vortex, while in the interior of the vortex it becomes very small.
These variations of $\tau$ reflect on the variations of grid density as seen in
Figures \ref{fig:grids_criteria} and \ref{fig:grids_continuity}.

\begin{figure}[tpb]
\centering
\noindent\makebox[\textwidth]{
 \subfigure[{1024 $\times$ 1024}]{\label{e_structured_9_100}
  \includegraphics{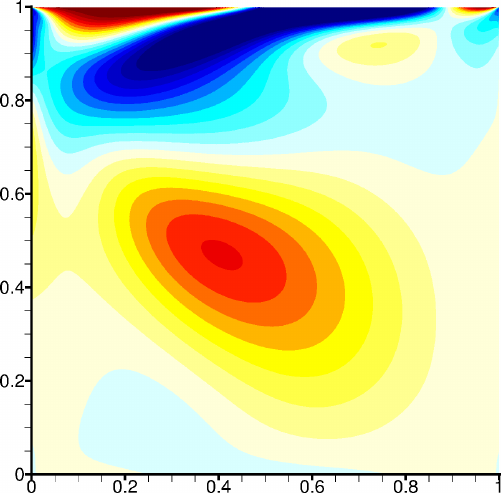}}
 \subfigure[{Q2 20\% XYCc}]{\label{e_Q2_20_XYCc_100}
  \includegraphics{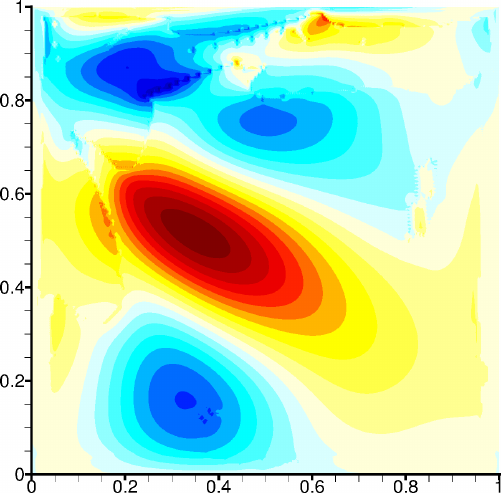}}
 \subfigure[{Q2 20\% XYCa}]{\label{e_Q2_20_XYCa_100}
  \includegraphics{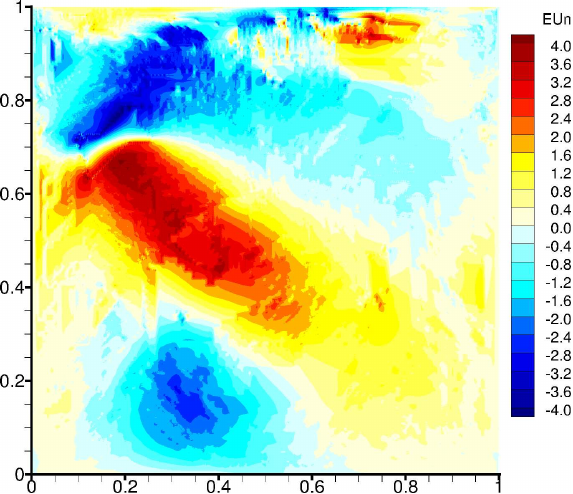}}
}
\noindent\makebox[\textwidth]{
 \subfigure[{Q1 20\% XY c}]{\label{e_Q1_20_XYc_100}
  \includegraphics{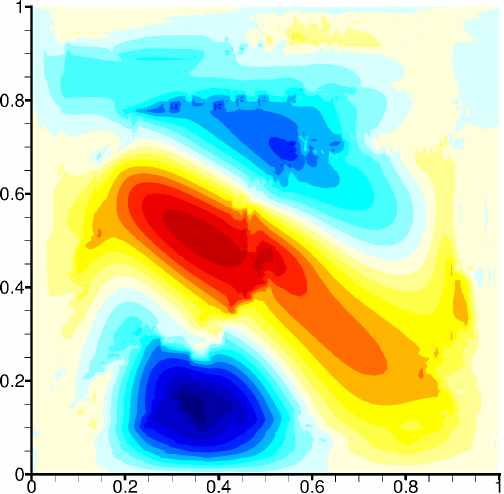}}
 \subfigure[{Q2 20\% XY c}]{\label{e_Q2_20_XYc_100}
  \includegraphics{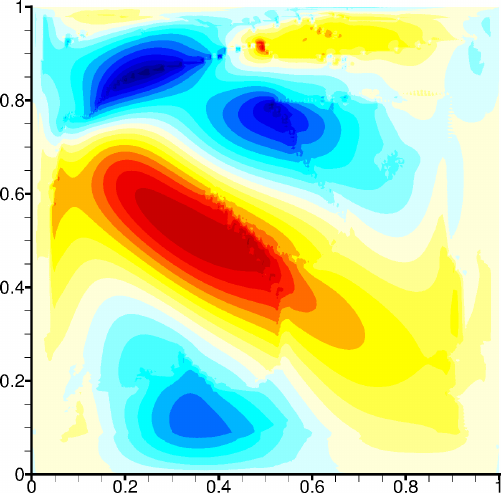}}
 \subfigure[{Q3 20\% XY c}]{\label{e_Q3_20_XYc_100}
  \includegraphics{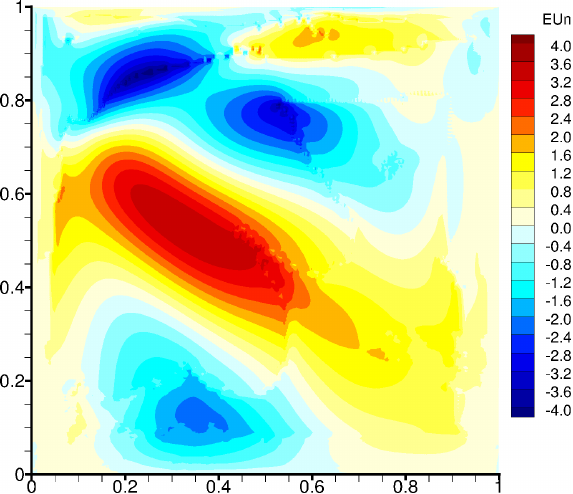}}
}
\caption{$\epsilon_u^*$ for various cases, Re = 100.}
\label{fig:errors_un_Re_100}
\end{figure}

Figure \ref{fig:cnv_Re_100_u} shows that for Re=100 the benefits from local
refinement are minimal concerning the total discretisation error in the domain,
no matter what refinement scheme is used (Q2 30\% XYC c appears to be the most
efficient - some schemes even appear to have a negative impact on the
efficiency). However, Figure \ref{fig:convergence_at_points} shows that local
refinement can be quite efficient at reducing the discretisation error locally
where the grid is dense. The efficiency is sometimes very poor at coarse regions
of the grid as also shown in Figure \ref{fig:convergence_at_points}. Figure
\ref{fig:errors_un_Re_100} shows that on structured grids there are very high
discretisation errors near the lid, which are greatly reduced by local refinement.
These errors concern only the $u$ variable and are quite high relative to the
mean error (Table \ref{table:statistics_100}: ($\epsilon_u^*)_{99\%} = 5.8$ on
the structured grid, ($\epsilon_u^*)_{99\%} \approx 3.5$ on composite grids). On
the other hand away from the lid the errors are not smoothed out, but, as shown in
Figure \ref{fig:errors_un_Re_100}, there are regions where $\epsilon_u^*$ is high
no matter what refinement scheme is used. This is an indication of the
complexity of the dependence of $\epsilon$ on $\tau$ (equation
(\ref{eq:eps_tau})), which cannot be tackled by the simple refinement criteria
tested here. Figure \ref{fig:eun_distributions_100} shows that indeed there is
some smoothing of $\epsilon_u^*$ caused by local refinement, as
$dA/d\epsilon_u^*$ is higher for the composite grids than for the structured
grid near $\epsilon_u^*=1$, and smaller for the composite grids than for the
structured grid near $\epsilon_u^*=0$ and for large values of $\epsilon_u^*$.
But the difference is not dramatic. This mild smoothing is also reflected in the
value of $\sigma_u^*$ in Table \ref{table:statistics_100} ($\approx$ 1.3 on
structured grid, $\approx$ 0.87 on composite grids). No smoothing takes place
for $v$ though, as seen in the same Table. This is because there are no high
$\epsilon_v$ regions near the lid.

\begin{table}[tpb]
\caption{Statistical indices of the distributions of normalised error, for
Re=100.}
\label{table:statistics_100}
\begin{center}
\rowcolors{2}{black!20}{white}
\noindent\makebox[\textwidth]{\begin{scriptsize}   
\begin{tabular}{ r | c c c | c c c | c c c }
 \hline \textbf{Re = 100}
 & $(\epsilon_u^*)_{\max}$ & $(\epsilon_u^*)_{99\%}$ & $\sigma_u^*$
 & $(\epsilon_v^*)_{\max}$ & $(\epsilon_v^*)_{99\%}$ & $\sigma_v^*$
 & $(\epsilon_p^*)_{\max}$ & $(\epsilon_p^*)_{99\%}$ & $\sigma_p^*$ \\
 \hline
Structured &
 $15.8$ & $5.8$ & $1.293$ & $5.52$ & $3.4$ & $0.897$ & $244$ & $3.1$ & $3.851$
\\ Q1 20\% XY c  & 
 $4.27$ & $3.6$ & $0.932$ & $3.85$ & $3.7$ & $1.015$ & $12.5$ & $4.0$ & $0.793$
\\ Q1 30\% XYCc  & 
 $4.00$ & $3.5$ & $0.878$ & $4.90$ & $3.8$ & $0.972$ & $34.3$ & $7.9$ & $1.622$
\\ Q2 20\% XY c  & 
 $4.73$ & $3.5$ & $0.873$ & $4.76$ & $3.5$ & $0.895$ & $13.0$ & $3.3$ & $0.585$
\\ Q2 20\% XYCc  & 
 $4.31$ & $4.0$ & $0.901$ & $4.30$ & $3.4$ & $0.875$ & $49.3$ & $6.1$ & $1.465$
\\ Q2 30\% XYCc  & 
 $5.01$ & $3.7$ & $0.868$ & $5.46$ & $3.7$ & $0.854$ & $66.2$ & $8.9$ & $2.150$
\\ Q3 20\% XY c  & 
 $4.76$ & $3.5$ & $0.870$ & $4.83$ & $3.5$ & $0.895$ & $33.1$ & $6.6$ & $1.331$
\\ Q2 20\% XYCn  & 
 $5.23$ & $3.9$ & $0.933$ & $5.39$ & $4.2$ & $0.966$ & $7.75$ & $3.8$ & $0.786$
\\ Q2 20\% XYCa  & 
 $6.41$ & $3.6$ & $0.874$ & $7.62$ & $3.9$ & $0.876$ & $51.3$ & $9.1$ & $1.787$
\\ \hline
\end{tabular}
\end{scriptsize}}
\end{center}
\end{table}

\begin{figure}[tpb]
\centering
\noindent\makebox[\textwidth]{
 \subfigure[{Re=100: 1024 $\times$ 1024}]{\label{ep_structured_9_100}
  \includegraphics{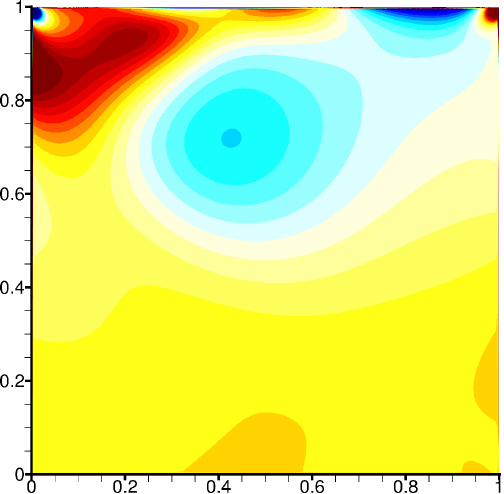}}
 \subfigure[{Re=1000: 1024 $\times$ 1024}]{\label{ep_structured_9_1000}
  \includegraphics{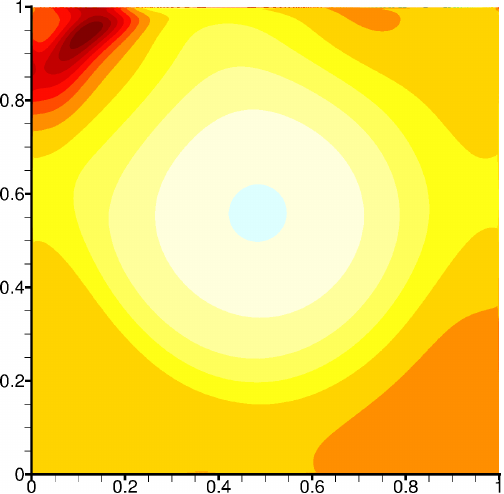}}
 \subfigure[{Re=10000: 1024 $\times$ 1024}]{\label{ep_structured_9_10000}
  \includegraphics{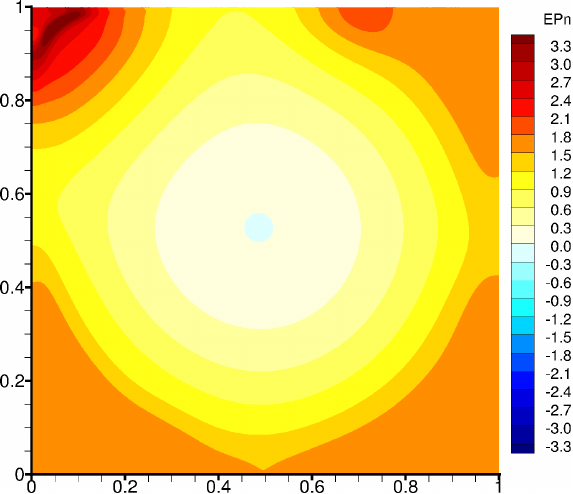}}
}
\noindent\makebox[\textwidth]{
 \subfigure[{Re=100: Q2 20\% XYCc}]{\label{ep_Q2_20_XYCc_100}
  \includegraphics{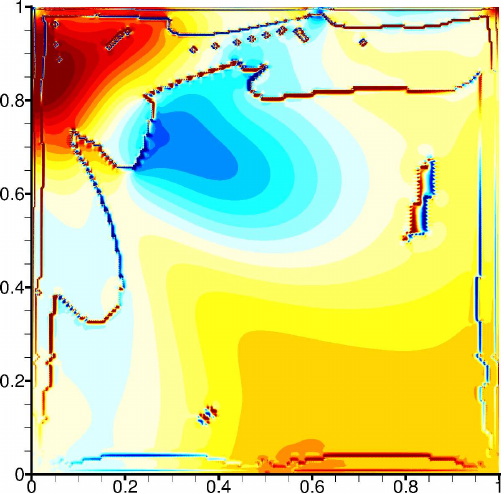}}
 \subfigure[{Re=1000: Q2 20\% XYCc}]{\label{ep_Q2_20_XYCc_1000}
  \includegraphics{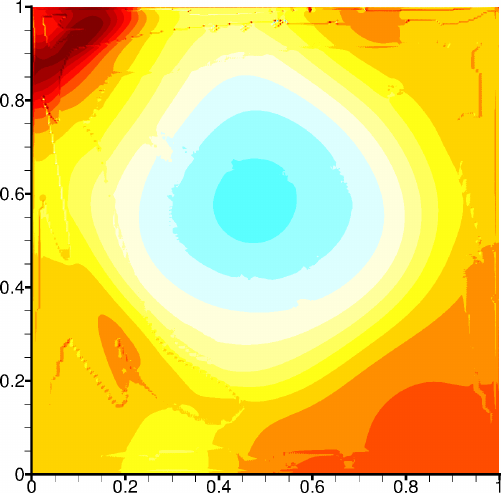}}
 \subfigure[{Re=10000: Q2 20\% XYCc}]{\label{ep_Q2_20_XYCc_10000}
  \includegraphics{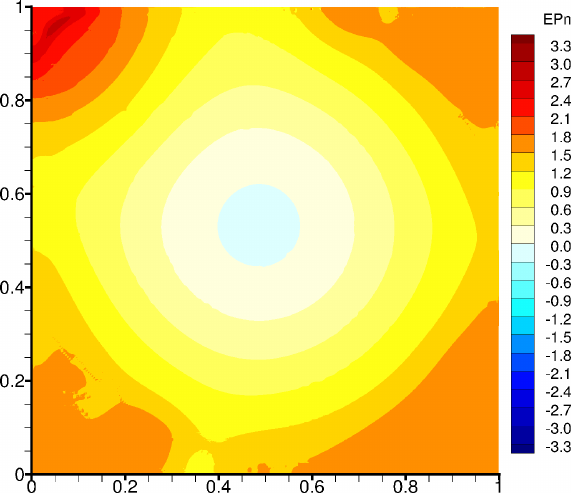}}
}
\caption{$\epsilon_p^*$ for various cases.}
\label{fig:errors_pn}
\end{figure}

The pressure variable also deserves some attention for Re=100. The corresponding
results of Table \ref{table:statistics_100} are difficult to interpret, but it
must be taken into account that due to linear extrapolation to the boundaries
the interpolation operator (\ref{eq:interpolation_o3}) may loose its accuracy
and large errors may appear there. This is indeed the case since Table
\ref{table:statistics_100} lists a value of $(\epsilon_p^*)_{max}=244$ but
merely a value of $(\epsilon_p^*)_{99\%}=3.1$ which excludes the high errors
at the boundaries. On the other hand, concerning the composite grids,
relatively high errors occur also at level interfaces as may be seen in Figure
\ref{fig:errors_pn}, something which is not observed with $u$ or $v$. In fact it
is also not observed with $p$ at higher Reynolds numbers. It is not known
whether these high $p$ errors have a significant effect on the overall
efficiency of local refinement.

\begin{figure}[tpb]
\centering
\noindent\makebox[\textwidth]{
 \subfigure[{1024 $\times$ 1024}]{\label{e_structured_9_1000}
  \includegraphics{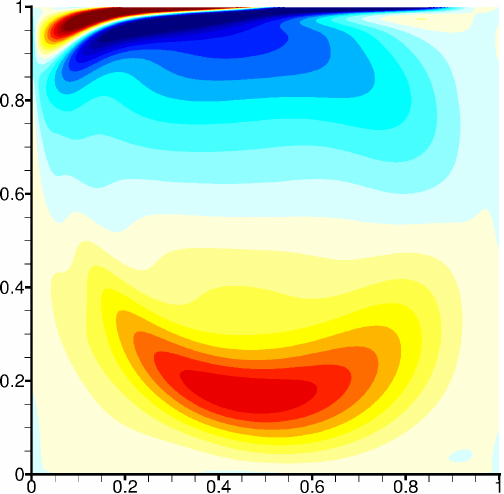}}
 \subfigure[{Q2 20\% XYCc}]{\label{e_Q2_20_XYCc_1000}
  \includegraphics{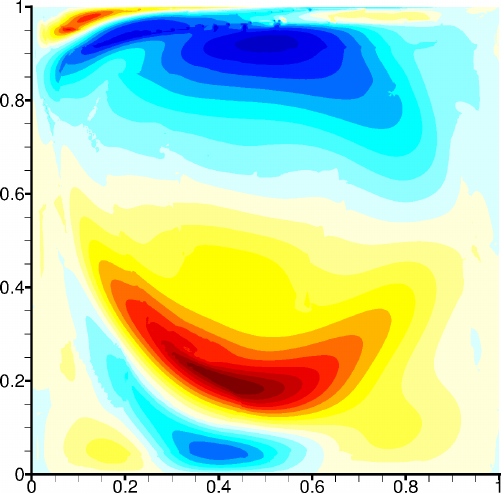}}
 \subfigure[{Q2 20\% XYCa}]{\label{e_Q2_20_XYCa_1000}
  \includegraphics{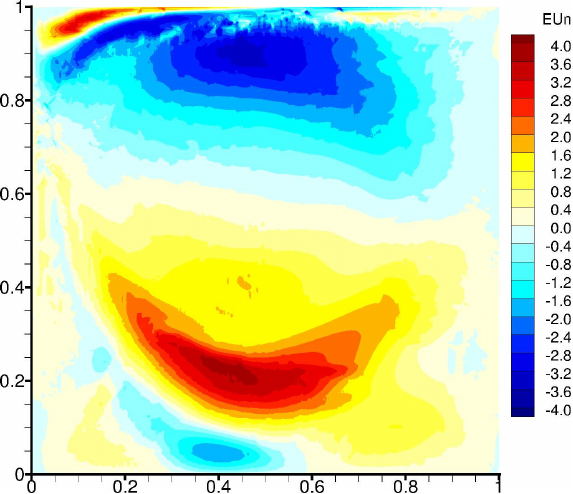}}
}
\noindent\makebox[\textwidth]{
 \subfigure[{Q1 20\% XY c}]{\label{e_Q1_20_XYc_1000}
  \includegraphics{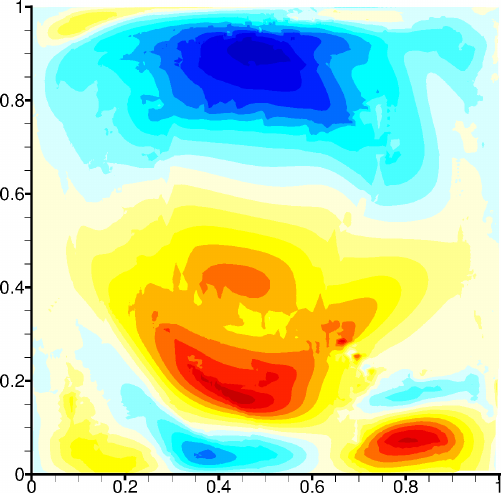}}
 \subfigure[{Q2 20\% XY c}]{\label{e_Q2_20_XYc_1000}
  \includegraphics{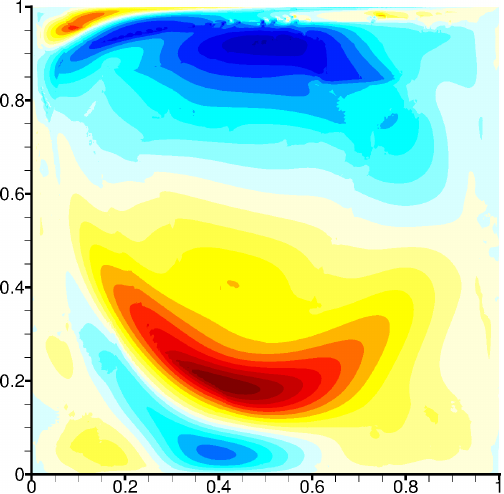}}
 \subfigure[{Q3 20\% XY c}]{\label{e_Q3_20_XYc_1000}
  \includegraphics{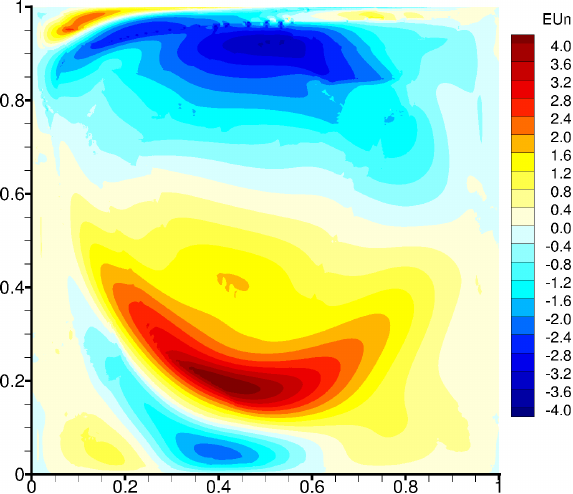}}
}
\caption{$\epsilon_u^*$ for various cases, Re = 1000.}
\label{fig:errors_un_Re_1000}
\end{figure}

As the Reynolds number increases, the efficiency of local refinement also
increases - Figures \ref{fig:cnv_Re_1000_u} - \ref{fig:cnv_Re_10000_u}. As
mentioned in Section \ref{sec:tau}, the discretisation error is transported
by convection and diffusion with the truncation error acting as source.
Therefore as the Reynolds number increases and convection becomes stronger, the
discretisation error is transported further away from its source regions of
high truncation error, because convection tends to transport a quantity
without reducing its magnitude, while diffusion attenuates the quantity that it
transports. This means that reducing the source (truncation error) by local
refinement will have a larger impact on the discretisation error of the whole
domain when the Reynolds number is larger. This is also seen in Figure
\ref{fig:convergence_at_points}, which shows that for Re=1000 and especially
for Re=10000 the discretisation error reduces faster with local refinement even
at coarse regions of the grid. Figures \ref{fig:errors_un_Re_1000},
\ref{fig:errors_un_Re_10000} and \ref{fig:errors_pn} show that the numerical
solution underestimates the strength of the main vortex on all grids. Local
refinement again makes $\epsilon_u^*$ a bit smoother due its reduction near the
lid, but $\epsilon_v^*$ and $\epsilon_p^*$ are not made smoother.
$\epsilon_p^*$ in particular is a lot smoother even on structured grids as the
Reynolds number increases.

\begin{table}[tpb]
\caption{Statistical indices of the distributions of normalised error, for
Re=1000.}
\label{table:statistics_1000}
\begin{center}
\rowcolors{2}{black!20}{white}
\noindent\makebox[\textwidth]{\begin{scriptsize}   
\begin{tabular}{ r | c c c | c c c | c c c }
 \hline \textbf{Re = 1000}
 & $(\epsilon_u^*)_{\max}$ & $(\epsilon_u^*)_{99\%}$ & $\sigma_u^*$
 & $(\epsilon_v^*)_{\max}$ & $(\epsilon_v^*)_{99\%}$ & $\sigma_v^*$
 & $(\epsilon_p^*)_{\max}$ & $(\epsilon_p^*)_{99\%}$ & $\sigma_p^*$ \\
 \hline
Structured &
 $10.5$ & $4.8$ & $1.007$ & $4.56$ & $3.1$ & $0.811$ & $10.6$ & $2.7$ & $0.568$
\\ Q1 20\% XY c  & 
 $3.61$ & $3.2$ & $0.814$ & $3.28$ & $3.0$ & $0.762$ & $3.81$ & $2.7$ & $0.548$
\\ Q1 30\% XYCc  & 
 $3.88$ & $3.6$ & $0.840$ & $4.03$ & $3.1$ & $0.783$ & $4.61$ & $2.7$ & $0.567$
\\ Q2 20\% XY c  & 
 $4.42$ & $3.6$ & $0.851$ & $3.83$ & $3.4$ & $0.816$ & $5.03$ & $2.8$ & $0.639$
\\ Q2 20\% XYCc  & 
 $4.57$ & $3.5$ & $0.855$ & $4.18$ & $3.5$ & $0.836$ & $5.67$ & $3.2$ & $0.649$
\\ Q2 30\% XYCc  & 
 $4.48$ & $3.7$ & $0.880$ & $3.84$ & $3.3$ & $0.808$ & $9.61$ & $2.7$ & $0.657$
\\ Q3 20\% XY c  & 
 $4.44$ & $3.6$ & $0.853$ & $3.83$ & $3.3$ & $0.819$ & $4.88$ & $3.0$ & $0.633$
\\ Q2 20\% XYCn  & 
 $7.02$ & $4.7$ & $1.053$ & $5.08$ & $4.0$ & $0.861$ & $3.49$ & $3.0$ & $0.644$
\\ Q2 20\% XYCa  & 
 $4.44$ & $3.4$ & $0.866$ & $4.29$ & $3.2$ & $0.839$ & $8.58$ & $3.0$ & $0.644$
\\ \hline
\end{tabular}
\end{scriptsize}}
\end{center}
\end{table}

\subsubsection*{Effect of refinement quantity}

\begin{figure}[tpb]
\centering
\noindent\makebox[\textwidth]{
 \subfigure[{1024 $\times$ 1024}]{\label{e_structured_9_10000}
  \includegraphics{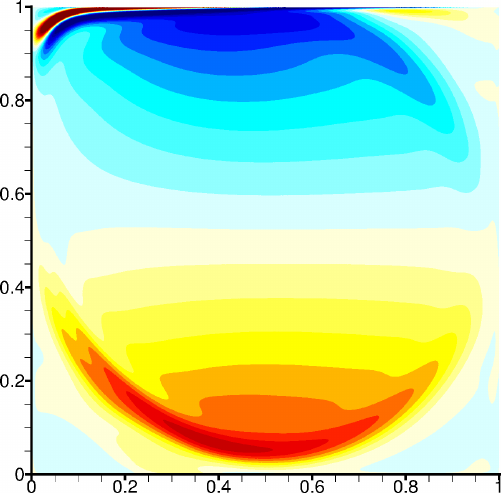}}
 \subfigure[{Q2 20\% XYCc}]{\label{e_Q2_20_XYCc_10000}
  \includegraphics{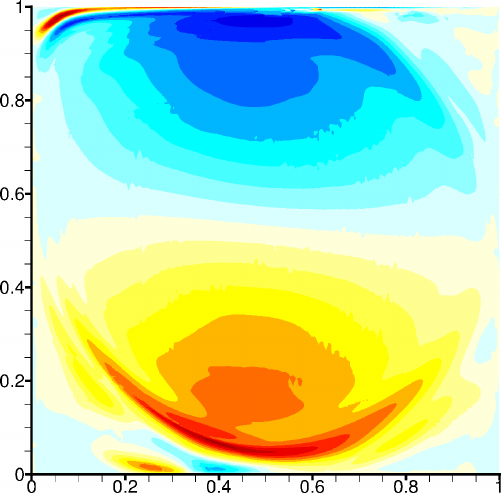}}
 \subfigure[{Q2 20\% XYCa}]{\label{e_Q2_20_XYCa_10000}
  \includegraphics{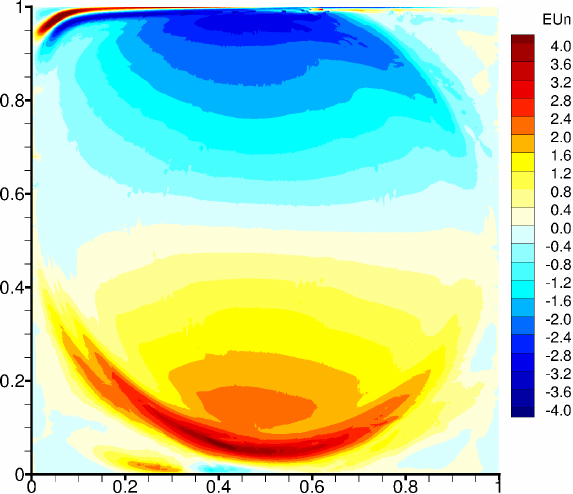}}
}
\noindent\makebox[\textwidth]{
 \subfigure[{Q1 20\% XY c}]{\label{e_Q1_20_XYc_10000}
  \includegraphics{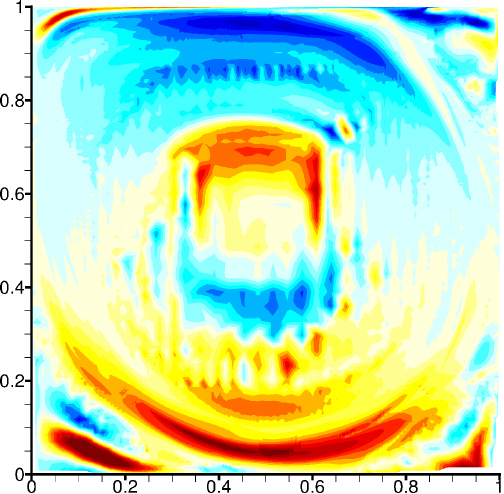}}
 \subfigure[{Q2 20\% XY c}]{\label{e_Q2_20_XYc_10000}
  \includegraphics{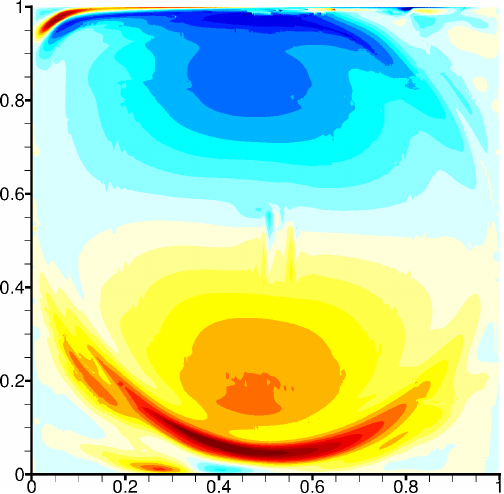}}
 \subfigure[{Q3 20\% XY c}]{\label{e_Q3_20_XYc_10000}
  \includegraphics{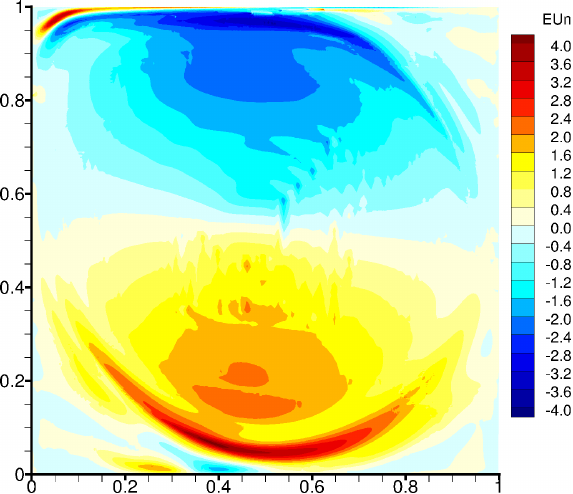}}
}
\caption{$\epsilon_u^*$ for various cases, Re = 10000.}
\label{fig:errors_un_Re_10000}
\end{figure}

To compare the criteria Q1, Q2 and Q3 it is best to compare the results for
schemes Q1 20\% XY c, Q2 20\% XY c and Q3 20\% XY c. A decrease in CV volume
reduces Q2 and Q3 both directly and indirectly (through the reduction of the
truncation error), but it only reduces Q1 indirectly. The result, as Figure
\ref{fig:grids_criteria} shows, is that Q1 produces grids that are very fine
at regions oh high $\tau$ and very coarse in regions of low $\tau$, greatly
reducing $\epsilon^*$ near the lid but increasing it further away.
Figure \ref{fig:errors_un_Re_10000} shows that Q1 20\% XY c is the only
scheme which allows the development of errors in the middle of the domain, due
to the coarseness of the grid there. Q2 and Q3 produce smoother grids and
Figures \ref{fig:cnv_Re_100_u} - \ref{fig:cnv_Re_10000_u} show that this makes
them more efficient than Q1. An exception is Q1 20\% XY c for Re = 10000 which
although it is less efficient than Q2 and Q3 at coarser grids, it catches up in
efficiency as the grids become finer, and in fact it has the steepest slope
suggesting that if even more refinement cycles were carried out it could surpass
the other quantities in efficiency.  Figure \ref{fig:convergence_at_points}
shows that Q1 may be quite efficient at particular fine-grid points.

Figure \ref{fig:grids_criteria} shows that Q2 and Q3 produce very similar grids,
as expected according to the discussion in Section \ref{sec:criteria}, since all
CVs of the grid have the same aspect ratio. The differences in the grids they
produce are minimal at Re = 100 but are slightly more noticable at Re = 10000,
which is not surprising since the convection terms of the main diagonal
coefficients are stronger in the latter case. Figures \ref{fig:cnv_Re_10000_u}
and \ref{fig:convergence_at_points} show that for Re = 10000 Q2 is slightly more
efficient which is interesting, since it suggests that the activation of
the convection terms of $A_{P,P}$ in quantity 3, which reduces the chance of a
CV being refined if the velocity is high, makes things worse rather than
better. Figures \ref{fig:cnv_Re_100_u}, \ref{fig:cnv_Re_1000_u} and
\ref{fig:convergence_at_points} show that both quantities produce very similar
results for Re = 100 and 1000.

\begin{table}[tpb]
\caption{Statistical indices of the distributions of normalised error, for
Re=10000.}
\label{table:statistics_10000}
\begin{center}
\rowcolors{2}{black!20}{white}
\noindent\makebox[\textwidth]{\begin{scriptsize}   
\begin{tabular}{ r | c c c | c c c | c c c }
 \hline \textbf{Re = 10000}
 & $(\epsilon_u^*)_{\max}$ & $(\epsilon_u^*)_{99\%}$ & $\sigma_u^*$
 & $(\epsilon_v^*)_{\max}$ & $(\epsilon_v^*)_{99\%}$ & $\sigma_v^*$
 & $(\epsilon_p^*)_{\max}$ & $(\epsilon_p^*)_{99\%}$ & $\sigma_p^*$ \\
 \hline
Structured &
 $9.80$ & $3.5$ & $0.901$ & $6.97$ & $3.5$ & $0.862$ & $4.01$ & $2.6$ & $0.592$
\\ Q1 20\% XY c  & 
 $5.62$ & $3.8$ & $0.875$ & $6.24$ & $3.9$ & $0.858$ & $1.67$ & $1.6$ & $0.395$
\\ Q1 30\% XYCc  & 
 $3.19$ & $2.5$ & $0.758$ & $3.15$ & $2.5$ & $0.753$ & $2.03$ & $1.8$ & $0.385$
\\ Q2 20\% XY c  & 
 $8.91$ & $3.2$ & $0.779$ & $3.56$ & $3.1$ & $0.757$ & $2.97$ & $2.2$ & $0.531$
\\ Q2 20\% XYCc  & 
 $3.80$ & $2.8$ & $0.758$ & $3.07$ & $2.8$ & $0.733$ & $2.93$ & $2.3$ & $0.531$
\\ Q2 30\% XYCc  & 
 $5.83$ & $2.9$ & $0.778$ & $4.21$ & $2.7$ & $0.751$ & $3.00$ & $2.2$ & $0.521$
\\ Q3 20\% XY c  & 
 $4.34$ & $3.1$ & $0.754$ & $3.40$ & $2.8$ & $0.720$ & $2.67$ & $2.1$ & $0.474$
\\ Q2 20\% XYCn  & 
 $3.72$ & $3.0$ & $0.817$ & $3.68$ & $2.8$ & $0.767$ & $2.16$ & $2.1$ & $0.483$
\\ Q2 20\% XYCa  & 
 $5.38$ & $3.2$ & $0.804$ & $4.25$ & $3.2$ & $0.803$ & $3.38$ & $2.4$ & $0.555$
\\ \hline
\end{tabular}
\end{scriptsize}}
\end{center}
\end{table}

\subsubsection*{Effect of continuity equation}

To investigate the effect of whether the continuity equation is used in the
refinement scheme we compare the schemes Q2 20\% XY c and Q2 20\% XYC c.
Figures \ref{fig:cnv_Re_100_u} - \ref{fig:cnv_Re_10000_u} reveal that the use of
the continuity equation makes local refinement slightly more efficient for Re =
100 and slightly less efficient for Re = 10000, compared to the case where the
refinement decision is based only on the momentum equations. For Re = 1000 the
difference in efficiency is marginal. The increase in the number of CVs of the
grid caused by the inclusion of the continuity equation in the refinement
criterion is 36\% for Re = 100, 28\% for Re = 1000 and 31\% for Re = 10000
(compare also Figure \ref{fig:grids_continuity} against the middle column of
Figure \ref{fig:grids_criteria}). Also noteworthy is that Table
\ref{table:statistics_100} suggests that the inclusion of the continuity
equation worsens the problem of high $\epsilon_p^*$ at level interfaces.

\subsubsection*{Effect of refinement ratio}

Comparing schemes Q2 20\% XYC c and Q2 30\% XYC c in Figure
\ref{fig:cnv_Re_100_u} reveals that increasing the ratio form 20\% to 30\%
improves the efficiency slightly, just like including the continuity equation
which also increases the number of CVs of the grid. A much greater improvement
is observed among schemes Q1 20\% XY c and Q1 30\% XYC c which includes the
combined effect of including the continuity equation and increasing the
refinement ratio. For Re = 1000 there appears to be no benefit from increasing
the refinement ratio as far as quantity Q2 is concerned, but there is an
important benefit combined with the use of the continuity equation for Q1. For
Re = 10000 the increase of the refinement ratio reduces the efficiency,
just like including the continuity equation. This shows that for high Reynolds
numbers it is more efficient to restrict local refinement to small regions of
high momentum truncation error.

\subsubsection*{Effect of level interface treatment}

Comparing schemes Q2 20\% XYCc, Q2 20\% XYCn and Q2 20\% XYCa it is clear that
it is most efficient to allow refinement only at the coarse side of a level
interface. Not allowing refinement at all produces terrible results at all
Reynolds numbers, as Figures \ref{fig:cnv_Re_100_u} - \ref{fig:cnv_Re_10000_u}
show. The reason is clear from Figure \ref{grid_Q2_20_XYCn_Re_1000}: Refinement
is prevented from occurring in large areas of the domain. On the other hand,
not posing any restrictions at all produces grids like the one shown in Figure
\ref{grid_Q2_20_XYCa_Re_1000} where level interfaces themselves constitute
refinement nuclei and consume many of the available CV refinements. This is
expected to become worse as refinement cycles progress, because the truncation
error reduces at the rest of the domain due to refinement, but not at level
interfaces where $\tau \in O(1)$. An advantage of criteria Q2 and Q3 in this
respect is that the size of the CV alone plays a role in the selection of the
CVs to be refined. Therefore even though the truncation error does not decrease
with grid refinement at level interfaces, the quantities Q2 and Q3 do decrease,
thereby decreasing the likelihood of refinement in these interface CVs.
Figures \ref{fig:cnv_Re_100_u} - \ref{fig:cnv_Re_10000_u} show the rather
surprising result that scheme Q2 20\% XYCa is only slightly less efficient than
Q2 20\% XYCc. The $\epsilon_u^*$ distributions in space of Q2 20\% XYCa are very
similar to those of the other refinement schemes as shown in Figures
\ref{fig:errors_un_Re_100} - \ref{fig:errors_un_Re_10000} only that they are
rugged, but the ruggedness is smoothed out as the Reynolds number increases.

\section{Conclusions} \label{sec:conclusions}

For the present study a number of numerical experiments were conducted involving
the solution of the regularised lid-driven cavity problem using various local
grid refinement schemes for a range of Reynolds numbers. All local refinement
schemes are based on an estimate of the truncation error.The goal was to assess
the effect of varying various local refinement parameters on the efficiency of
the finite volume method. To this end the regularised lid-driven cavity problems
were solved to a great level of accuracy. Several conclusions can be drawn from
the results presented in this paper.

First of all, truncation error-based local refinement appears to offer little
benefit on a global scale when the Reynolds number is low, although it can be
efficient at a local scale. However, local refinement becomes quite efficient as
the Reynolds number increases. It is suggested that this is due to the
discretisation error being convected farther away from the source locations (of
high truncation error) at higher Reynolds numbers, so reducing the sources
causes a greater reduction in the global discretisation error.

Refinement criteria that are proportional to the CV volume in addition to the
truncation error are found to be more efficient than simply using the truncation
error by itself. Using the product of the truncation error times the CV volume
has been found to be better than using the truncation error divided by the main
diagonal coefficient of the linearised operator.

Using the truncation error of the continuity equation in addition to those of
the momentum equations, and increasing the percentage of CVs that are refined in
each refinement cycle increases the efficiency when the Reynolds number is low,
but decreases it when it is large.

Despite the fact that the present local refinement schemes attempt to smooth out
the truncation error distribution in the domain, it was observed that they only
slightly smooth the distribution of discretisation error. This is an indication
of the complexity of the $\tau-\epsilon$ relationship.

Also, an effort was made to deal with the issue that at level interfaces
$\tau \in O(1)$. Although the exact effect of these high truncation errors on
the discretisation error is not completely clear, the present results show that
this issue is best dealt with by allowing refinement to occur only at the coarse
side of a level interface rather than either not allowing refinement at all
there or not imposing any restrictions. However, in the latter case of not
restricting refinement at level interfaces, rather surprisingly, the efficiency
does not appear to suffer severely by the pile-up of refinement levels on top of
the original level interface.

Finally, we would like to make some general comments on the refinement strategy
adopted and to discuss some implications in the case that a transient problem was
solved instead. Most of the previous researchers that based their refinement
criterion on either the truncation error or the residual used a threshold to select
the CVs to be refined; a CV is refined if the refinement quantity there is above the
threshold. The refinement cycles continue until the refinement quantity is below the
threshold at all CVs, or until a maximum allowed number of levels has been reached.
We did not follow this procedure because of two reasons: The first is that the
truncation error does not decrease at certain regions of the grid such as at level
interfaces and at domain boundaries, and therefore it is impossible for the refinement
criterion quantity to drop below the threshold everywhere. The second is that due to
the complexity of $\tau-\epsilon$ relationship it is very difficult to relate a
desired accuracy to a certain threshold. Therefore, in our finite volume scheme such
a procedure would cause the following: The refinement procedure would stop in the
interior of grid levels when the quantity dropped below the threshold, but the
refinement cycles would continue since the quantity would be above the threshold at
level interfaces and at domain boundaries. Eventually the cycles would stop when the
maximum allowed level was reached. One would then have to decide on choosing a
particular maximum allowed level. By allowing more and more levels the number of CVs
of the grid would continually increase (at level interfaces and domain boundaries),
but the discretisation error would not converge to zero since CVs would not be refined
in the interior of the levels where the refinement quantity is below the threshold.
The discretisation error would converge to a certain non-zero value, as observed e.g.
in \cite{Syrakos06b}.

On the other hand, the aim of the present study is to compare a number of refinement
criterion quantities, and therefore the most natural approach would be to refine
the CVs where this quantity is largest. The comparison would be most accurate if
it happened on a CV-by-CV basis, i.e. if during a refinement cycle only one CV was
refined, the one where the refinement quantity is largest. Then by observing how
fast the discretisation error reduces as the number of refinement cycles increases,
one could evaluate the various refinement quantities. Of course, such a procedure
is completely impractical due to the overhead involved in each refinement cycle
(the data structures have to change, the equations have to be solved on the new
composite grid, the truncation error has to be calculated afresh etc.). However, the
next best thing that one can do which is also practical, is to not refine a single
CV in a refinement cycle but to refine many CVs selected on the basis that their
refinement quantity is larger than at the rest of the CVs of the grid. So in the
present study the number of these CVs was selected to be a percentage $R$ of the
total number of CVs of the grid. This allows a clearer comparison between the various
refinement quantities than using a threshold, because each refinement quantity would
have its own threshold and it is not straightforward to make these thresholds equivalent.
Also, by allowing more and more refinement cycles to occur this procedure (which uses
the percentage $R$) causes the discretisation error to converge to zero, if the refinement
quantity is proportional to the CV size, (as the present results verify), since there is
no threshold below which a CV will not be refined. Furthermore, the percentage $R$
combined with a chosen fixed number of refinement cycles makes the number of CVs of the
final composite grid to be approximately known a priori; it approximately equals $K\cdot
(1+3R)^n$ where $K$ is the number of CVs of the initial grid, and $n$ the number of
allowed refinement cycles (when a CV is refined it is replaced by 4 child CVs, so the
total number of CVs increases by 3), so we have two parameters, $R$ and $n$ which can
be varied simultaneously to get the desired grid size. Therefore, the refinement strategy
adopted here can be described as aiming to produce the best possible accuracy for a given
selected grid size. This is achieved better if $R$ is small, but $n$ must not become so
large that the overhead of the refinement procedure causes the overall efficiency to degrade.
A disadvantage of the method is that it does not provide much control over the discretisation
error, but the same also holds even when a threshold is used.

The present study is concerned only with the steady-state case. Transient problems would
have additional complexities. First of all the truncation error would contain also components
which would be a function of the time step size $\Delta t$, in addition to those which are
functions of the grid spacing $h$. The present truncation error estimation method which
uses the coarse grid $2h$ would only identify the part of the truncation error which is due
solely to the spatial discretisation. In order that the whole truncation error be calculated
the coarse grid operator should also use a time step of $2\Delta t$. In this case local
refinement should also be performed in time, by using smaller time steps where appropriate.
To reduce the complexity one can limit the local refinement to the spatial grid only,
trying to reduce only the spatial components of the truncation error, but this procedure
cannot be expected to reduce the discretisation error below a certain point, as the temporal
components of the truncation error do not decrease. The refinement procedure described here
could be used within each time step if an implicit temporal discretisation scheme is chosen,
but not without modification. Indeed, using a refinement percentage $R$ time step after time
step would cause the grid to become huge in terms of number of CVs very soon. Since the aim
of the present method is to achieve the best accuracy with a given grid size, the procedure
could be adapted as follows: During the first time step an initial coarse grid would be
adapted using selected values of $R$ and $n$, so that the desired grid size is reached,
and during the following time steps there would be as many CV refinements as there are
unrefinements, so that the total number of CVs of the composite grid remains constant, with
the aim that the refinement quantity is as low as possible in the domain. The increase in
truncation error that an unrefinement would cause can be estimated from the truncation error
estimate at the child CVs assuming $\tau \in O(h^p)$. Therefore in this case it is necessary
that the method has the ability to perform unrefinement, that is to remove 4 child CVs to
recover their parent. This is an issue with its own complexities that is beyond the scope of
the present work; implementation details can be found e.g. in \cite{Jasak_00, Rosenberg_06}.





\section*{References}

\bibliographystyle{ieeetr}
\bibliography{2012_Syrakos_JCP}

\begin{thebibliography}{10}

\bibitem{Yamaleev_01}
N.~K. Yamaleev, ``Minimization of the truncation error by grid adaptation,''
  {\em Journal of Computational Physics}, vol.~170, pp.~459--497, 2001.

\bibitem{Vilsmeier_93}
R.~Vilsmeier and D.~H{\"a}nel, ``Adaptive methods on unstructured grids for
  {E}uler and {N}avier-{S}tokes equations,'' {\em Computers \& Fluids},
  vol.~22, no.~4-5, pp.~485 -- 499, 1993.

\bibitem{Chen_97}
W.~L. Chen, F.~S. Lien, and M.~A. Leschziner, ``Local mesh refinement within a
  multi-block structured-grid scheme for general flows,'' {\em Computer Methods
  in Applied Mechanics and Engineering}, vol.~144, pp.~327--369, 1997.

\bibitem{Brandt}
A.~Brandt, ``Multi-level adaptive solutions to boundary-value problems,'' {\em
  Mathematics of Computation}, vol.~31, pp.~333--390, 1977.

\bibitem{Berger_84}
M.~J. Berger and J.~Oliger, ``Adaptive mesh refinement for hyperbolic partial
  differential equations,'' {\em Journal of Computational Physics}, vol.~53,
  no.~3, pp.~484 -- 512, 1984.

\bibitem{Berger_89}
M.~Berger and P.~Colella, ``Local adaptive mesh refinement for shock
  hydrodynamics,'' {\em Journal of Computational Physics}, vol.~82, no.~1,
  pp.~64 -- 84, 1989.

\bibitem{Thompson_89}
M.~C. Thompson and J.~H. Ferziger, ``An adaptive multigrid technique for the
  incompressible {N}avier-{S}tokes equations,'' {\em Journal of Computational
  Physics}, vol.~82, pp.~94--121, 1989.

\bibitem{Trottenberg}
U.~Trottenberg, C.~Oosterlee, and A.~Schuller, {\em Multigrid}.
\newblock Academic Press, 2001.

\bibitem{Aftosmis02}
M.~J. Aftosmis and M.~J. Berger, ``Multilevel error estimation and adaptive
  h-refinement for cartesian meshes with embedded boundaries,'' {\em AIAA Paper
  2002-0863,40-th AIAA Aerospace Sciences Meeting, Reno, Nevada}, January 2002.

\bibitem{Brown_05}
D.~J. Brown and L.~L. Lowe, ``Multigrid elliptic equation solver with adaptive
  mesh refinement,'' {\em Journal of Computational Physics}, vol.~209,
  pp.~582--598, 2005.

\bibitem{Syrakos06b}
A.~Syrakos and A.~Goulas, ``Finite volume adaptive solutions using {SIMPLE} as
  smoother,'' {\em International Journal for Numerical Methods in Fluids},
  vol.~52, pp.~1215--1245, 2006.

\bibitem{Lee_07}
Y.~Lee, H.~Thompson, and P.~Gaskell, ``An efficient adaptive multigrid
  algorithm for predicting thin film flow on surfaces containing localised
  topographic features,'' {\em Computers \& Fluids}, vol.~36, no.~5, pp.~838 --
  855, 2007.

\bibitem{Muzaferija_97}
S.~Muzaferija and D.~A. Gosman, ``Finite-volume {CFD} procedure and adaptive
  error control strategy for grids of arbitrary topology,'' {\em Journal of
  Computational Physics}, vol.~138, pp.~766--787, 1997.

\bibitem{Jasak_01}
H.~Jasak and A.~D. Gosman, ``Residual error estimate for the finite volume
  method,'' {\em Numerical Heat Transfer, Part B Fundamentals}, vol.~39,
  pp.~1--19, 2001.

\bibitem{Jasak_03}
H.~Jasak and A.~D. Gosman, ``Element residual error estimate for the finite
  volume method,'' {\em Computers \& Fluids}, vol.~32, pp.~223--248, 2003.

\bibitem{Hay_06}
A.~Hay and M.~Visonneau, ``Error estimation using the error transport equation
  for finite-volume methods and arbitrary meshes,'' {\em International Journal
  of Computational Fluid Dynamics}, vol.~20, pp.~463--479, 2006.

\bibitem{Hay_07}
A.~Hay and M.~Visonneau, ``Adaptive finite-volume solution of complex turbulent
  flows,'' {\em Computers \& Fluids}, vol.~36, no.~8, pp.~1347 -- 1363, 2007.

\bibitem{Ganesh_09}
N.~Ganesh, N.~V. Shende, and N.~Balakrishnan, ``{R}-parameter: A local
  truncation error based adaptive framework for finite volume compressible flow
  solvers,'' {\em Computers \& Fluids}, vol.~38, no.~9, pp.~1799 -- 1822, 2009.

\bibitem{Roy_09}
C.~J. Roy and A.~J. Sinclair, ``On the generation of exact solutions for
  evaluating numerical schemes and estimating discretization error,'' {\em
  Journal of Computational Physics}, vol.~228, no.~5, pp.~1790 -- 1802, 2009.

\bibitem{Pretorius_06}
F.~Pretorius and M.~W. Choptuik, ``Adaptive mesh refinement for coupled
  elliptic-hyperbolic systems,'' {\em Journal of Computational Physics},
  vol.~218, no.~1, pp.~246 -- 274, 2006.

\bibitem{Shen_11}
C.~Shen, J.-M. Qiu, and A.~Christlieb, ``Adaptive mesh refinement based on high
  order finite difference {WENO} scheme for multi-scale simulations,'' {\em
  Journal of Computational Physics}, vol.~230, no.~10, pp.~3780 -- 3802, 2011.

\bibitem{Peric}
J.~H. Ferziger and M.~Peric, {\em Computational methods for fluid dynamics}.
\newblock Springer, 3rd~ed., 2002.

\bibitem{Manteuffel86}
T.~A. Manteuffel and A.~B. White, ``The numerical solution of second-order
  boundary value problems on nonuniform meshes,'' {\em Mathematics of
  Computation}, vol.~47, pp.~511--535, 1986.

\bibitem{Jones2000}
W.~P. Jones and K.~R. Menzies, ``{Analysis of the Cell-Centred Finite Volume
  Method for the Diffusion Equation},'' {\em Journal of Computational Physics},
  vol.~165, pp.~45--68, Nov. 2000.

\bibitem{Syrakos06a}
A.~Syrakos and A.~Goulas, ``Estimate of the truncation error of finite volume
  discretization of the {N}avier-{S}tokes equations on colocated grids,'' {\em
  International Journal for Numerical Methods in Fluids}, vol.~50,
  pp.~103--130, 2006.

\bibitem{diskin_nia_07_08}
B.~Diskin and J.~L. Thomas, ``Accuracy analysis for mixed-element finite-volume
  discretization schemes,'' tech. rep., National Institute of Aerospace, TR
  2007-08, Hampton, Virginia, 2007.

\bibitem{Thomas_08}
J.~L. Thomas, B.~Diskin, and C.~L. Rumsey, ``Toward verification of
  unstructured-grid solvers,'' {\em AIAA Journal}, vol.~46, pp.~3070--3079,
  2008.

\bibitem{Eriksson09}
S.~Eriksson and J.~Nordstr{\"o}m, ``Analysis of the order of accuracy for
  node-centered finite volume schemes,'' {\em Applied Numerical Mathematics},
  vol.~59, pp.~2659--2676, 2009.

\bibitem{johansen98}
H.~Johansen and P.~Colella, ``A cartesian grid embedded boundary method for
  poisson's equation on irregular domains.,'' {\em Journal of Computational
  Physics}, vol.~147, pp.~60--85, 1998.

\bibitem{Matsunaga2000}
N.~Matsunaga and T.~Yamamoto, ``Superconvergence of the {S}hortley-{W}eller
  approximation for {D}irichlet problems,'' {\em Journal of Computational and
  Applied Mathematics}, vol.~116, no.~2, pp.~263--273, 2000.

\bibitem{Svard_06}
M.~Sv{\"a}rd and J.~Nordstr{\"o}m, ``On the order of accuracy for difference
  approximations of initial-boundary value problems,'' {\em Journal of
  Computational Physics}, vol.~218, no.~1, pp.~333 -- 352, 2006.

\bibitem{Bernert97}
K.~Bernert, ``$\tau$--extrapolation. theoretical foundation, numerical
  experiment,and application to {N}avier--{S}tokes equations,'' {\em SIAM
  Journal on Scientific Computing}, vol.~18, no.~2, pp.~460--478, 1997.

\bibitem{Fulton03}
S.~R. Fulton, ``On the accuracy of multigrid truncation error estimates,'' {\em
  Electronic Transactions on Numerical Analysis}, vol.~15, pp.~29--37, 2003.

\bibitem{Briggs}
W.~L. Briggs, V.~E. Henson, and S.~F. McCormick, {\em A multigrid tutorial -
  2nd ed.}
\newblock SIAM, 2000.

\bibitem{Syrakos}
A.~Syrakos, {\em Analysis of a finite volume method for the incompressible
  {N}avier-{S}tokes equations}.
\newblock PhD thesis, Aristotle University of Thessaloniki, 2006.
\newblock (in Greek).

\bibitem{Rhie_Chow}
C.~M. Rhie and W.~L. Chow, ``Numerical study of the turbulent flow past an
  airfoil with trailing edge separation,'' {\em AIAA Journal}, vol.~21,
  pp.~1525--1532, 1983.

\bibitem{Patankar_72}
S.~V. Patankar and D.~B. Spalding, ``A calculation procedure for heat, mass and
  momentum transfer in three-dimensional parabolic flows,'' {\em International
  Journal of Heat and Mass Transfer}, vol.~15, pp.~1787--1806, 1972.

\bibitem{Botella_98}
O.~Botella and R.~Peyret, ``Benchmark spectral results on the lid-driven cavity
  flow,'' {\em Computers \& Fluids}, vol.~27, no.~4, pp.~421 -- 433, 1998.

\bibitem{Fraysse_12}
F.~Fraysse, J.~de~Vicente, and E.~Valero, ``The estimation of truncation error
  by $\tau$-estimation revisited,'' {\em Journal of Computational Physics},
  vol.~231, no.~9, pp.~3457 -- 3482, 2012.

\bibitem{Lewis}
H.~R. Lewis and L.~Denenberg, {\em Data structures {\&} their algorithms}.
\newblock Addison Wesley, 1991.

\bibitem{Khosla_74}
P.~K. Khosla and S.~G. Rubin, ``A diagonally dominant second-order accurate
  implicit scheme,'' {\em Computers \& Fluids}, vol.~2, no.~2, pp.~207 -- 209,
  1974.

\bibitem{Ghia_82}
U.~Ghia, K.~N. Ghia, and C.~T. Shin, ``High-{R}e solutions for incompressible
  flow using the {N}avier-{S}tokes equations and a multigrid method,'' {\em
  Journal of Computational Physics}, vol.~48, no.~3, pp.~387 -- 411, 1982.

\bibitem{Bruneau_06}
C.-H. Bruneau and M.~Saad, ``The 2{D} lid-driven cavity problem revisited,''
  {\em Computers \& Fluids}, vol.~35, no.~3, pp.~326 -- 348, 2006.

\bibitem{Shen_89}
J.~Shen, ``Dynamics of regularized cavity flow at high reynolds numbers,'' {\em
  Applied Mathematics Letters}, vol.~2, no.~4, pp.~381 -- 384, 1989.

\bibitem{Shen_91}
J.~Shen, ``Hopf bifurcation of the unsteady regularized driven cavity flow,''
  {\em Journal of Computational Physics}, vol.~95, no.~1, pp.~228 -- 245, 1991.

\bibitem{Trefethen}
L.~N. Trefethen and D.~{Bau III}, {\em Numerical Linear Algebra}.
\newblock SIAM, 1997.

\bibitem{Jasak_00}
H.~Jasak and A.~D. Gosman, ``Automatic resolution control for the finite-volume
  method, part 2: Adaptive mesh refinement and coarsening,'' {\em Numerical
  Heat Transfer, Part B: Fundamentals}, vol.~38, no.~3, pp.~257--271, 2000.

\bibitem{Rosenberg_06}
D.~Rosenberg, A.~Fournier, P.~Fischer, and A.~Pouquet,
  ``Geophysical-astrophysical spectral-element adaptive refinement ({GASpAR}):
  Object-oriented $h$-adaptive fluid dynamics simulation,'' {\em Journal of
  Computational Physics}, vol.~215, no.~1, pp.~59 -- 80, 2006.

\end{thebibliography}







\end{document}